\documentclass[aps,amsfonts,prb,twocolumn,superscriptaddress]{revtex4-2}
\usepackage{epsfig,amsmath,amssymb,bm,epsf,graphicx,bbold,dsfont,cancel}
\usepackage[all]{xy}
\usepackage{color}
\usepackage{physics}
\usepackage{tikz}
\usetikzlibrary{positioning, calc, patterns, topaths, shapes.geometric, shapes.misc}
\usepackage{pgfmath}
\usepackage{algorithm}
\usepackage{algpseudocode}
\usepackage{mathtools}
\usepackage{hyperref}
\usepackage{amsmath}
\usepackage{caption}
\usepackage{multirow}
\usepackage{subcaption}

\newpage

\newenvironment{algorithmnofloat}
  {
     \refstepcounter{algorithm}
     \kern12pt\hrule height.8pt depth0pt \kern2pt
     \renewcommand{\caption}[2][\relax]{
       {\raggedright\textbf{Algorithm~\thealgorithm} ##2\par}
       \ifx\relax##1\relax 
         \addcontentsline{loa}{algorithm}{\protect\numberline{\thealgorithm}##2}
       \else 
         \addcontentsline{loa}{algorithm}{\protect\numberline{\thealgorithm}##1}
       \fi
       \kern2pt\hrule\kern2pt
     }
  }{
     \kern12pt\hrule\kern12pt
  }

\tikzstyle{myNodeStyle}=[draw, inner sep=0, fill=white]
\tikzstyle{triangleNode}=[myNodeStyle, isosceles triangle, isosceles triangle apex angle=90, shape border rotate=#1, minimum width=1.2cm]
\tikzstyle{widetriangleNode}=[myNodeStyle, isosceles triangle, isosceles triangle apex angle=120, shape border rotate=#1, minimum width=0.5cm]
\tikzstyle{circleNode}=[myNodeStyle, circle, inner sep=.1mm, minimum width=.3cm]
\tikzstyle{circleNodeSmall}=[myNodeStyle, circle, inner sep=.1mm, minimum width=.15cm]
\tikzstyle{squareNode}=[myNodeStyle, regular polygon, regular polygon sides=4, minimum width=.7cm]
\tikzstyle{rectangleNode}=[myNodeStyle, rectangle, inner sep=1mm, minimum width=.7cm]
\tikzstyle{squareNodeSmall}=[myNodeStyle, regular polygon, regular polygon sides=4, minimum width=.5cm, text width=0, fill=white]
\tikzstyle{trapeziumNode}=[myNodeStyle, trapezium, shape border rotate=#1, inner sep=.1mm, minimum width=.7cm]
\tikzstyle{semicircleNode}=[myNodeStyle, semicircle, shape border rotate=#1, inner sep=.3mm, minimum width=.7cm]
\tikzstyle{roundedrectangleNode}=[myNodeStyle, rounded rectangle, rounded rectangle west arc=0pt, shape border rotate=#1, inner sep=.3mm, minimum width=.3cm]

\newcommand\svdeq{\stackrel{\text{SVD}}{=}}

\newcommand\danglinglegbias[4]{
    \ifnum\pdfstrcmp{#2}{north}=0 ($(#1.north)+(#4,0)$)--($(#1.north)+(#4,#3)$) \fi
    \ifnum\pdfstrcmp{#2}{south}=0 ($(#1.south)+(#4,0)$)--($(#1.south)+(#4,-#3)$) \fi
    \ifnum\pdfstrcmp{#2}{east} =0 ($(#1.east)+(0,#4)$)--($(#1.east)+(#3,#4)$) \fi
    \ifnum\pdfstrcmp{#2}{west} =0 ($(#1.west)+(0,#4)$)--($(#1.west)+(-#3,#4)$) \fi
}

\DeclareMathOperator{\argmin}{argmin}
\DeclareMathOperator{\sgn}{sgn}

\renewcommand{\eqref}[1]{Eq.~(\ref{#1})}
\newcommand{\figref}[1]{Fig.~\ref{#1}}
\renewcommand{\algref}[1]{Alg.~\ref{#1}}
\newcommand{\secref}[1]{Sec.~\ref{#1}}
\newcommand{\appendixref}[1]{App.~\ref{#1}}
\newcommand{\tabref}[1]{Table.~\ref{#1}}

\usetikzlibrary{external}
\begin{document}

\title{Tensor Network Methods for Extracting CFT Data from Fixed-Point Tensors and Defect Coarse Graining}
\author{Wenhan Guo}
\affiliation{C. N. Yang Institute for Theoretical Physics and Department of Physics and Astronomy, State University of New York at Stony Brook, Stony Brook, NY 11794-3840, USA}
\author{Tzu-Chieh Wei}
\affiliation{C. N. Yang Institute for Theoretical Physics and Department of Physics and Astronomy, State University of New York at Stony Brook, Stony Brook, NY 11794-3840, USA}

\begin{abstract}
We present a comprehensive study on extracting CFT data using tensor network methods, especially, from the fixed point tensor of the linearized tensor renormalization group
(lTRG) for the classical 2D Ising model near the critical temperature. Utilizing two different methods, we extract operator scaling dimensions
and operator product expansion (OPE) coefficients by introducing defects on the lattice and by employing the fixed-point tensor. We also explore the effects of point-like defects in the lattice on the coarse-graining process. We find that there is a correspondence between coarse-grained defect tensors and conformal states obtained from the lTRG fixed point equation. We also analyze the capabilities and
limitations of our proposed coarse-graining scheme for tensor networks with point-like defects, including graph-independent local truncation (GILT) and higher-order tensor renormalization group (HOTRG). Our results provide a better understanding of the capacity and limitations
of the tenor renormalization group scheme in coarse-graining defect tensors, and we show that GILT+HOTRG can be used
to give accurate two- and four-point functions under specific conditions. We also find that employing the minimal canonical form further improves the stability of the RG flow.

\end{abstract}

\date \today
 \maketitle

\section{Introduction}

The study of the renormalization group (RG) is of central importance in the study of critical phenomena in statistical mechanics and many-body quantum systems~\cite{kadanoff1966scaling,wilson1975renormalization}. The basic idea is to remove the short-range information from a given system while keeping its long-range behavior intact. For critical systems, there is a similarity for behaviors at different length scales and, often, they can be described by conformal field theory (CFT)~\cite{francesco2012conformal}, which emerges at fixed points of the RG. By studying the perturbative behavior of the RG near fixed points, one can extract useful information about the CFT, such as the central charge, the scaling dimensions of conformal operators, and the operator product expansion (OPE) coefficients between them.

The tensor network (TN) is a versatile framework for describing strongly correlated systems by encoding the correlation and/or the entanglement structure into mutually connected tensors. It can be used to describe partition functions of classical statistical models and wave functions of quantum many-body ground states. It can also be used to study topological order~\cite{cirac2021matrix} and, as toy models, to study conformal field theory and holography~\cite{vidal2007entanglementRenormalization,swingle2012entanglementHolography,bao2015consistency}. 
In particular, tensor network renormalization provides a new perspective of the RG flow, where the properties of the coarse-grained system are parametrized by a tensor. Given the comparison that CFTs are fixed points of the RG flow when `integrating out' the UV (short-distance) details, it is a natural question whether one can build a correspondence or relationship between the fixed-point tensor and the CFT. One particular direction of the above question is whether it is possible to determine the CFT data of a critical lattice system when given the fixed-point tensor from coarse-graining the lattice model. This question has previously been asked and discussed using tensor network methods~\cite{milsted2018MeraGeometric, evenbly2016TNRScale, haegeman2013ERRS, zou2020SpinChainOPE, verstraete2008TNSReview, evenbly2013MERA}.

Tensor-based real-space RG methods have been developed over the past several decades, offering certain advantages over traditional numerical methods, such as Monte Carlo methods. The density matrix renormalization group (DMRG)~\cite{white1992DMRG.69.2863,white1993DMRGalgorithm,schollwock2005DMRG,schollwock2011DMRGMPS} method, which can be understood in terms of the matrix product state (MPS)~\cite{verstraete2004density,klumper1992MPS_VBS,ostlund1995MPS_DMRGLimit,vidal2003MPSQC} ansatz, has been the most successful numerical method for gapped one-dimensional many-body systems. The multiscale entanglement renormalization ansatz (MERA)~\cite{evenbly2013MERA, vidal2007MERA,evenbly2009MERAalgorithm} is a tree-like tensor network that provides reliable and accurate results for studying one-dimensional gapless systems.

Projected entangled pair states (PEPS)~\cite{verstraete2004PEPS,verstraete2004PEPS_VBSQC,verstraete2006PEPSAreaLaw}, proposed by Cirac and Verstraete, are a natural extension of the MPS for quantum systems on a lattice of two or higher dimensions that obey the area entanglement law. However, contracting the tensor network in two and higher dimensions is a computationally difficult task. Approximations are necessary to make this computation efficient. One of the early coarse-graining methods is the corner transfer matrix, originally developed by Baxter~\cite{baxter1981corner,nishino1996corner}, which has since been applied to quantum systems~\cite{orus2012exploring}. Another approach for the real-space coarse-graining scheme in two-dimensional systems is the tensor renormalization group (TRG) by Levin and Nave~\cite{levin2007TRG}, which led to several variants and further ideas~\cite{xie2009SRG,chen2020SRGAD,xie2012HOTRG,evenbly2015TNR,yang2017loopTNR}, including the higher-order TRG (HOTRG)~\cite{xie2012HOTRG} that we will use in this work.

Although these methods give an accurate estimate of the free-energy density and local observables, they still do not generate a correct RG flow in the tensor space. Gu and Wen~\cite{gu2009TEFR_TMscD} discovered that real-space TN RG methods do not remove all short-distance correlations, and local entanglement filtering was needed to achieve a fixed point tensor. In particular, during the real-space RG, the local correlations residing across the boundary of the coarse-graining blocks survive, which can be illustrated by a toy model called corner double-line (CDL) tensors. Various methods have been proposed to deal with the issue of CDL tensors~\cite{gu2009TEFR_TMscD,evenbly2015TNR,yang2017loopTNR,hauru2018GILT}. With these methods, the fixed-point tensors of the RG flow can be found approximately~\cite{goldenfeld2018lectures}.

Among these entanglement filtering methods, graph-independent local truncation (GILT), proposed by Hauru, Delcamp, and Mizera~\cite{hauru2018GILT}, has the advantage of simplicity, especially since it is compatible with several coarse graining methods, such as TRG and HOTRG. It also offers a clear geometric meaning of how CDLs are removed in coarse-graining. Using HOTRG + GILT, as well as a $Z_2$ gauge fixing procedure, Lyu, Xu, and Kawashima demonstrated that the scaling dimensions can be extracted by linearizing the equation that relates successive coarse-graining tensors around the fixed point~\cite{lyu2021lTRG},  which is a procedure referred to as the linearized TRG (lTRG) and is analogous to the traditional RG equation. Scaling dimensions were also extracted from the transfer matrix at the fixed point~\cite{gu2009TEFR_TMscD}, or from the eigenvalues of the local-scale transformation~\cite{evenbly2015TNR,evenbly2017TNR_detail,evenbly2016TNRScale}. 

It was emphasized in~\cite{lyu2021lTRG} that a good gauge fixing is essential in making the notion of a fixed-point tensor well defined and in getting the correct scaling dimensions from lTRG. Here, in this work, in addition to the $Z_2$ gauge fixing employed in Ref.~\cite{lyu2021lTRG}, we also employ the minimal canonical form (MCF), proposed recently by Acuaviva et al.~\cite{acuaviva2022MCF}, who showed how to identify gauge-equivalent tensors by an iterative scheme.  We adopt the MCF into the lTRG and obtain a more consistent RG flow of coarse-grained tensors, allowing better characterization of how the system approaches and flows away from the fixed point. This allows us to study the 2D classical Ising model and reproduce the CFT data of the Ising CFT, including the scaling dimensions $\Delta_\sigma$, $\Delta_\varepsilon$, and the OPE coefficients $C_{\sigma\sigma\varepsilon}$, from the fixed point tensor. 

Our results of the scaling dimensions and OPE coefficients are consistent with calculations of the two-point and four-point correlation functions. However, we encounter and resolve issues in computing the latter to achieve the comparison. The correlation functions calculated from HOTRG suffer from a smearing effect.  Despite removing CDL local correlations, we find that GILT suffers from inaccuracy in evaluating correlation functions at longer distances.  We provide a solution through an averaging procedure (described in detail in the following). Furthermore,
 we compare the coarse-grained lattice-level defect tensors with the tensors extracted from the lTRG spectrum, confirming that both represent conformal states.
 We also explore whether our coarse-graining method can accommodate excited states in addition to the ground state.

The remainder of this paper is as follows.
In \secref{sec:Background}, we review the Tensor Renormalization Group concept and the coarse graining techniques that we employ, which include HOTRG, MCF, and GILT. We also review the concept of lTRG.
In \secref{sec:Results}, we present our main numerical findings.
In \secref{sec:Results_TdiffFlow}, we demonstrate that our coarse-graining scheme successfully obtains a fixed point tensor, and this allows us to extract the full CFT data of the Ising CFT, including scaling dimensions $\Delta_\sigma$, $\Delta_\varepsilon$, and OPEs $C_{\sigma\sigma\varepsilon}$, respectively in \secref{sec:Results_ScDim} and \secref{sec:Results_OPEFromlTRG}.
For comparison, in \secref{sec:Results_2pt} and \secref{sec:Results_4pt}, we calculated the two- and four-point correlation functions to extract the full CFT data. We also compute the two-point function on a finite-size torus in \secref{sec:Results_2pt_torus}.
In \secref{sec:Results_eigvecCompare} and \secref{sec:Results_encoding_position}, we compare the coarse-grained lattice-level defect tensors and the tensors extracted from the lTRG spectrum, confirming that both represent conformal states.
In \secref{sec:Discussion}, we explore whether our coarse-graining method can accommodate excited states in addition to the ground state. We identify the primary error of HOTRG and GILT as resulting from the "smearing" of point-like defects and suppression of edge modes. To address this, we recommend sampling points randomly and avoiding placing impurity tensors at corners and edges of coarse-grained blocks.
Moreover, we demonstrate that the advanced gauge-fixing method MCF improves TRG flow stability and propose a method for detecting CDL tensors along the RG flow. We also provide evidence that coarse graining introduces an error that breaks conformal symmetry and ultimately drives the system away from the fixed point.
In~\secref{sec:Conclusion}, we summarize our work and suggest potential research directions. In the appendix, we present further details to supplement the main text, such as the algorithmic procedures that we use in this paper. In addition, we also review some related CFT background and our proposed TN approach to relevant CFT data.

\section{Background}
\label{sec:Background}

In this section, we review the connection between the conventional RG flow and their tensor network counterpart. We start by introducing the tensor network and the concept of TRG in Sec.~\ref{subsec:TRG}. Then we briefly review the concept of RG flow, their fixed points, and CFT in Sec.~\ref{subsec:RGCFT}.
After that, we describe some of the previous studies on TRG, including HOTRG, local entanglement filtering, and GILT, in Secs.~\ref{subsec:RGCFT} and~\ref{sec:GILT}, as these are the main tools we will use in this work. Further details are presented in the Appendix. We also discuss the gauge redundancy in the tensor network and the newly proposed gauge fixing method that we will use, that is, MCF in Sec.~\ref{subsec:MCF}. We then discuss the linearized tensor renormalization group proposed in Ref.~\cite{lyu2021lTRG} to extract scaling dimensions of operators in Sec~\ref{subsec:lTRG}.
Finally, we describe our coarse-graining scheme, which involves a combination of HOTRG, GILT, MCF, and a residual $U(\chi)^{\otimes m}$ gauge fixing procedure in Sec.~\ref{sec:CoarseGrainingScheme}. In addition, we describe how we produce coarse-grain defect tensors, which can be used to compute correlation functions.

\subsection{Tensor Network and Tensor Renormalization Group}\label{subsec:TRG}

The translation invariant tensor network is a natural way to describe the partition function of many statistical models and the wave functions of quantum systems, such as the Ising model on a square lattice~\cite{mccoy2009advanced}  and the ground state wave function of the Affleck-Kennedy-Lieb-Tasaki model~\cite{affleck1987rigorous}. By replacing certain tensors with a ``defect tensor'', one can further compute the $n$-point expectation value or correlation function.

\eqref{eq:TNSquareLattice} below shows the tensor network representation of the vacuum partition function (or, in turn, the free energy via $F=-\frac{1}{\beta}\log Z$) of a lattice model on a square lattice with periodic boundary condition,
\begin{equation}
    Z=\tr
    \vcenter{\hbox{\includegraphics[page=1]{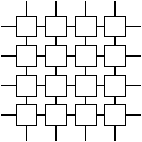}}},
    \label{eq:TNSquareLattice}
\end{equation}

\eqref{eq:TNSquareLatticeExpVal} below shows how to evaluate the two-point correlator,
\begin{equation}
    \expval{\mathcal{O}_{1}(x_1)\mathcal{O}_{2}(x_2)}=
    \frac{1}{Z}\tr
    \vcenter{\hbox{\includegraphics[page=2]{TRGCFTGraphics.pdf}}},
    \label{eq:TNSquareLatticeExpVal}
\end{equation}
where $\tr T =\vcenter{\hbox{\includegraphics[page=57,scale=0.5]{TRGCFTGraphics.pdf}}}$ means contracting opposite pairs of edges, according to the periodic boundary condition. 

We denote the ``virtual'' Hilbert space as $\mathcal{H}_v=\mathbb{C}^{\chi}$ associated with tensor indices on one leg, where $\chi$ is the corresponding bond dimension of the tensor leg. A tensor $T$ on the lattice has $2m$ legs, where $m$ is the spacetime dimension of the lattice used to describe the system. For example, $m=2$ in the classical Ising model on the square lattice. Thus, the vector space of all the $2m$-legged tensors is $T \in \mathcal{H}_v^{\otimes 2m}$. 

To evaluate the contraction of the tensor network in \eqref{eq:TNSquareLattice} on an infinitely large lattice, which yields the partition function of the system at the continuum limit, one iteratively transforms the tensor network into a new tensor network on a coarse-grained lattice until the boundary effect and the finite-size effect are negligible. This is generally called the real-space tensor renormalization procedure, as illustrated below. (We note that in the infinite system instead of the value of the partition function, we obtain the free-energy density instead.) 
\begin{equation}
\label{eq:HOTRG_concept}
    \vcenter{\hbox{\includegraphics[page=3]{TRGCFTGraphics.pdf}}}
    \approx
    \vcenter{\hbox{\includegraphics[page=4]{TRGCFTGraphics.pdf}}}
\end{equation}

The bond dimension of the coarse-grained tensor under exact merging will grow exponentially. To make it feasible for numerical computation, one needs to truncate the Hilbert space of the virtual degrees of freedom (i.e., bond dimensions) in each leg. Various numerical methods such as TRG~\cite{lyu2021lTRG}, HOTRG~\cite{xie2012HOTRG}, TEFR~\cite{gu2009TEFR_TMscD}, TNR~\cite{evenbly2015TNR,evenbly2017TNR_detail}, SRG~\cite{xie2009SRG}, and GILT~\cite{hauru2018GILT} are proposed to properly truncate the Hilbert space in the coarse-graining process.

\smallskip\noindent\textbf{Truncation to finite bond dimension}. When coarse-graining the tensor toward a fixed point, scale invariance requires that $\mathcal{H}_v\otimes\mathcal{H}_v\cong \mathcal{H}_v$, which implies that $\text{dim} \mathcal{H}_v=1$ or $\infty$ strictly. So in principle, one needs infinitely many iterations so that the coarse-grained tensor network can reach either a trivial or an infinite-dimensional tensor. In practice, we assume that the system approximately reaches the thermodynamic limit after a finite but large number of steps, and the higher-index components of the coarse-grained tensors are negligible, and we truncate the Hilbert space on each leg to retain only its most significant $\chi$-dimensional subspace.

One has to bear in mind that the truncation error introduces an artificial length scale to the coarse-grained system, which breaks the conformal symmetry. This issue will be demonstrated in \secref{sec:Results_ScDim}; for example, see \figref{fig:flow_scd}.
In this paper, we focus on a combination of methods using HOTRG and GILT~\cite{lyu2021lTRG}. The description of our coarse-grain scheme is given in \secref{sec:CoarseGrainingScheme}.

\subsection{Renormalization Group Flow and Conformal Field Theory}\label{subsec:RGCFT}

\begin{figure}[h]
    \centering
    \includegraphics[width=.8\hsize]{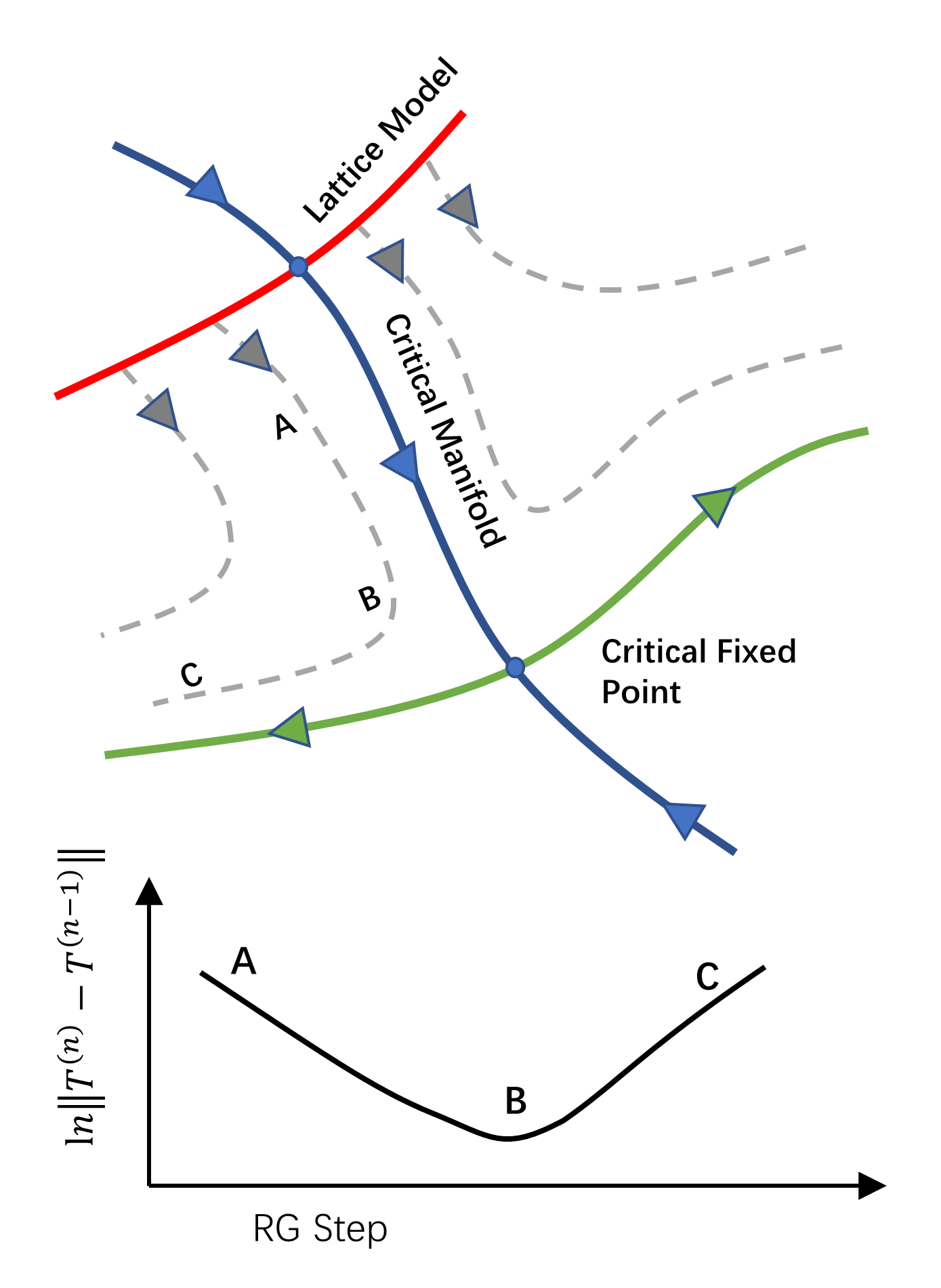}
    \caption{(color online) Illustration of the renormalization group flow (top) and the changes in tensors during the RG process. }
    \label{fig:RGFlow}
\end{figure}

The behavior of statistical and quantum systems at different length scales can be described using the renormalization group flow in the system parameter space, as illustrated in \figref{fig:RGFlow}.
At criticality, the system flows into a certain type of fixed point, which sometimes can be described by a CFT, where there is an emergence of scale invariance at long distances. As a result, the two-point functions obey a power law
\begin{equation}
    \expval{\mathcal{O}(x_1)\mathcal{O}(x_2)} \propto \frac{1}{(x_1-x_2)^{2\Delta_{\mathcal{O}}}},
\end{equation}
where $\Delta_{\mathcal{O}}$ is the scaling dimension of the operator $\mathcal{O}$.
Further information about CFT will be summarized in \appendixref{sec:Appendix_CFTStates} and \appendixref{sec:Appendix_CFTOPE}.

When the system is near the critical manifold of the system parameters, it can firstly converge to someplace near the critical manifold and then deviate from it exponentially, moving towards perhaps another fixed point. On the logarithmic scale, the change in the tensor is represented by a V-shaped curve after each RG step, as demonstrated in Fig.~\ref{fig:RGFlow}.

In TRG, the difference between coarse-grained tensors from successive RG steps also follows a curve schematically shown in~\figref{fig:RGFlow},  which was previously reported in Refs.~\cite{meurice2013SmallXExactRG,evenbly2015TNR,lyu2021lTRG} and can be used as a method to determine how close the flow is to the criticality~\footnote{See also~\figref{fig:Tdiff} from our numerical simulations}.  We expect the change of the system after each RG transformation to behave as a V-shaped curve on the logarithmic scale. For example, the location B in~\figref{fig:RGFlow} is the closest to a nearby critical manifold, and for length scales where the system is at the proximity of B, the system behaves like a CFT at the corresponding fixed point. As a result, the system can behave like different CFTs at corresponding length scales, depending on the detailed RG flow.

A CFT can be uniquely determined by its CFT data. The CFT data consists of the list of primary operators $\{\mathcal{O}_i\}$, their conformal dimensions $\{\Delta_i\}$, and the OPE coefficients $C_{ijk}$ between them. With these, one can compute all the $n$-point correlation functions. In the context of tensor networks, the conformal dimensions describe how a defect tensor scales under coarse graining, and the OPE coefficients describe how two defect tensors fuse into another defect tensor during this process.

\subsection{Higher Order Tensor Renormalization Group}\label{subsec:HOTRG}

Now we describe the key coarse-graining procedure that we shall use, i.e., the HOTRG. 
The idea is to find a coarse-grained tensor $T'$ and isometry $w$:  
$\mathcal{H}_v\rightarrow\mathcal{H}_v\otimes\mathcal{H}_v$, which specifies the subspace to keep after one step of coarse-graining, as illustrated in the equations below.
\begin{equation}
    \label{eq:HOTRG_steps_demo}
    \vcenter{\hbox{\includegraphics[page=20]{TRGCFTGraphics.pdf}}}
    =
    \vcenter{\hbox{\includegraphics[page=21]{TRGCFTGraphics.pdf}}}
\end{equation}
\begin{equation}
    \label{eq:HOTRG_steps2_demo}
    \vcenter{\hbox{\includegraphics[page=22]{TRGCFTGraphics.pdf}}}
    =
    \vcenter{\hbox{\includegraphics[page=23]{TRGCFTGraphics.pdf}}}
\end{equation}
One can repeat the coarse-graining procedure horizontally and then  vertically at alternative steps. 

The strategy of choosing the isometry tensor $w$ is to make the coarse-grained tensor be able to mimic the original tensor subnetwork as much as possible, as follows.
\begin{equation}
    \label{eq:HOTRG_equation}
    \vcenter{\hbox{\includegraphics[page=9]{TRGCFTGraphics.pdf}}}
    \approx
    \vcenter{\hbox{\includegraphics[page=10]{TRGCFTGraphics.pdf}}}
\end{equation}
The detailed prescription of computing $w$ is described in \algref{alg:HOTRG} (\appendixref{sec:HOTRG_alg}).

\subsection{Local Entanglement Filtering and Graph Independent Local Truncation}\label{sec:GILT}

\begin{figure}[h]
    \centering
    \includegraphics[width=.8\hsize]{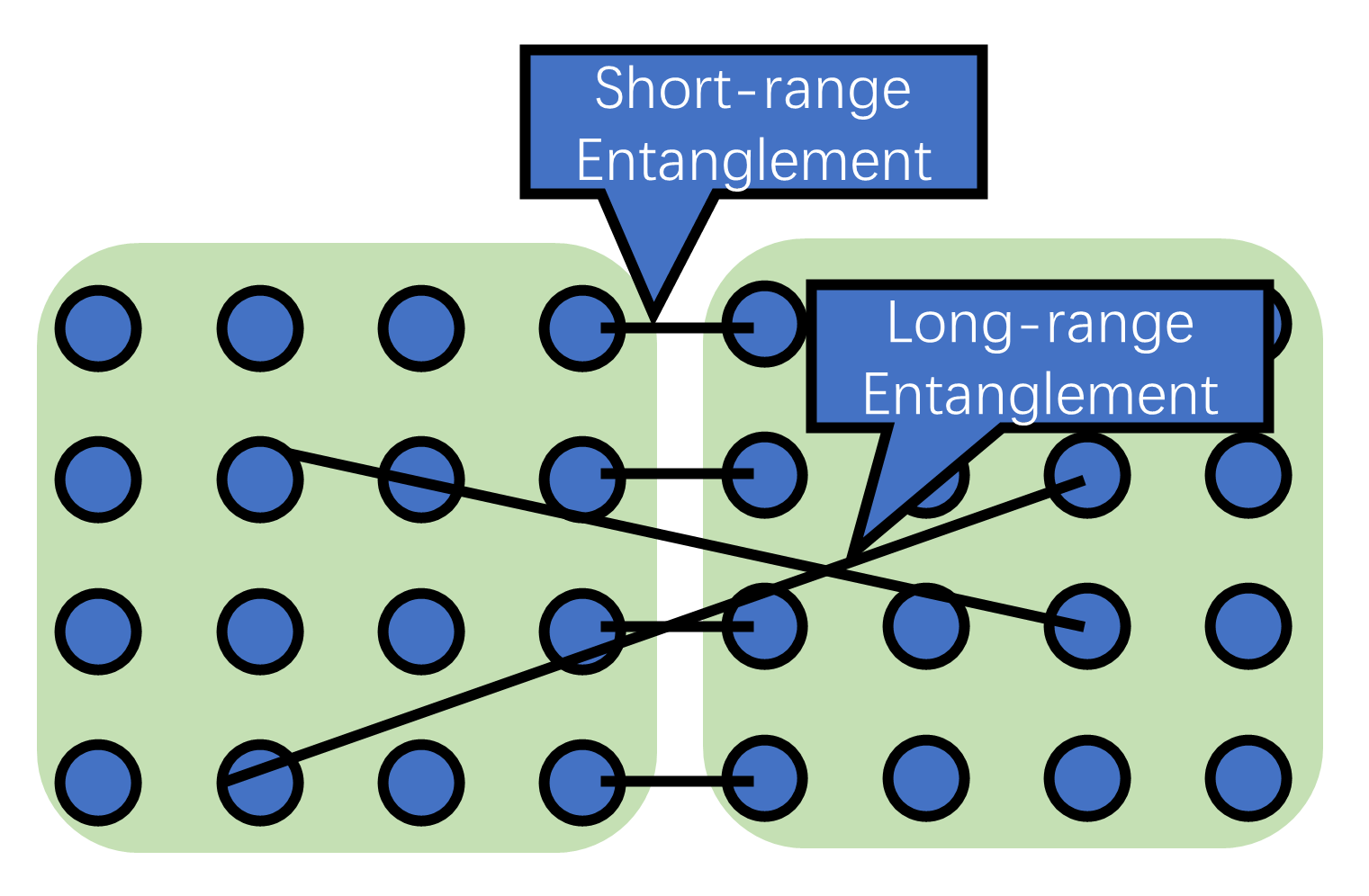}
    \caption{(color online) Local Entanglement resides on the boundary and Long Range Entanglement in the bulk between two rectangular blocks of lattice sites.}
    \label{fig:ShortRangeEntanglementIllust}
\end{figure}

Real-space RG methods suffer from accumulating microscopic entanglement between the pair of sites adjacent to the boundary of coarse-grained regions. As illustrated in \figref{fig:ShortRangeEntanglementIllust}, the entanglement between two blocks, which determines the necessary bond dimension required to faithfully represent the system, arises from the contribution of short-distance pairs across the boundary of the two blocks.

On a square lattice, local entanglement can be parameterized by corner-double-line tensors~\cite{gu2009TEFR_TMscD},
\begin{equation}
\label{eq:cdl}
    \vcenter{\hbox{\includegraphics[page=11]{TRGCFTGraphics.pdf}}}
    =
    \vcenter{\hbox{\includegraphics[page=12]{TRGCFTGraphics.pdf}}}
\end{equation}
To effectively capture the long-wavelength behavior of the system using a tensor with a finite bond dimension, local entanglement removal must be done. The basic idea of local entanglement removal involves gluing two tensors together and then slicing them apart again.

CDL tensors can be removed by graph-independent local truncation~\cite{hauru2018GILT}. The idea of GILT is to modify a subset of a tensor network such as a plaquette, so that the modified sub-network still looks the same from external legs, but the entanglement through a certain internal leg is reduced, \begin{equation}
    \vcenter{\hbox{\includegraphics[page=13]{TRGCFTGraphics.pdf}}}
    \approx
    \vcenter{\hbox{\includegraphics[page=14]{TRGCFTGraphics.pdf}}}
    \svdeq
    \vcenter{\hbox{\includegraphics[page=15]{TRGCFTGraphics.pdf}}}
    \label{eq:GILT}
\end{equation}

\subsection{Gauge Fixing and Minimal Canonical Form}
\label{subsec:MCF}
The ``virtual'' Hillbert space $\mathcal{H}_v$ on each leg is unphysical, suffering from a $GL(\chi)^{\otimes m}$ gauge redundancy~\cite{acuaviva2022MCF}, as shown in \eqref{eq:gauge_transform}. We recall that $\chi$ is the bond dimension and $m$ is the `spacetime' dimension of the lattice. It is essential to fix the gauge to compare the tensors of different RG steps component-wise; a gauge transformation on the four legs is illustrated in the equation below.

 \begin{equation}
     \vcenter{\hbox{\includegraphics[page=5]{TRGCFTGraphics.pdf}}}
     \sim
     \vcenter{\hbox{\includegraphics[page=6]{TRGCFTGraphics.pdf}}}
     \label{eq:gauge_transform}
 \end{equation}

Ref.~\cite{lyu2021lTRG} introduced a rudimentary gauge-fixing procedure, which is based on the eigenvalue decomposition during the standard HOTRG procedure \eqref{eq:HOTRG_equation1}. Their method successfully yields a V-shaped curve~\footnote{ The Figure~9 in \cite{lyu2021lTRG} does not show a clear V-shaped curve, indicating that the gauge is not completely fixed when comparing tensors at successive RG steps. However, the GitHub code they provide yields a clear V-shaped curve, as shown in \figref{fig:CompareGilt}.}.

Recently, it has been proposed in Ref.~\cite{acuaviva2022MCF} that the tensor $T$ can be iteratively gauge-transformed to a ``minimal canonical form'' (MCF)~\cite{acuaviva2022MCF}, which is unique among gauge-equivalent tensors up to a $U(\chi)^{\otimes m}$ gauge ambiguity, 
\begin{equation}
    T_{min}=\argmin\{||\tilde{T}||_2:\tilde{T}\in \overline{G \cdot T} \},
\end{equation}
where $G\cdot T=\{g_1\otimes g_2 \otimes g_1^{-1} \otimes g_2^{-1} \cdot T | g_1, g_2 \in GL(\chi) \}$. The overline means that we take the closure of the gauge orbit, i.e., including the limits of any sequence of group actions. (The consideration of closure is for mathematical rigor and does not need to be explicitly considered in our scenario.)

We summarize the prescription for obtaining the MFC in \algref{alg:MCF} (\appendixref{sec:MCF_alg}). The residual $U(\chi)^{\otimes m}$ gauge can be robustly fixed using an improved method based on the preliminary method in Ref.~\cite{lyu2021lTRG}. The method is summarized in \algref{alg:Fix_U1} (\appendixref{sec:Fix_U1_alg}).
With the implementation of MCF, we obtain a more consistent gauge fixing when comparing coarse-grained tensors between successive RG steps, and it produces a V-shaped curve; see the discussion in Sec.~\ref{sec:Discussion}.

\subsection{Linearized Tensor Renormalization Group}\label{subsec:lTRG}

The lTRG~\cite{lyu2021lTRG} studies how the perturbation in the input tensor changes the flow of the coarse-grained tensor. It is defined as follows,
\begin{equation}
    \label{eq:lTRG}
    M^{i'j'k'l'}_{ijkl}=\frac{\partial T'_{i'j'k'l'}}{\partial T_{ijkl}},
\end{equation}
where $T$ and $T'$ are tensors before and after coarse-graining.
At the critical point, the eigenvalues $\lambda_\alpha$ of $M$ give the conformal dimensions $\Delta_{ \mathcal{O}_\alpha}$ of the CFT operators,
\begin{equation}
    \label{eq:lTRG_eigvecs}
    M\ket{\mathcal{O}_\alpha}=\lambda_\alpha \ket{\mathcal{O}_\alpha},
\end{equation}
where
\begin{equation}
    \label{eq:scdim_lTRG}
    \frac{\lambda_\alpha}{\lambda_0}=s^{-\Delta_{\mathcal{O}_\alpha}},
\end{equation}
and  $s=2$ is the linear scaling of the system after two RG steps (which merge four sites by a vertical and a horizontal coarse-graining steps). Note that proper normalization is required; see \appendixref{sec:Appendix_RenormalizeTensorStates}. 

Ref.~\cite{lyu2021lTRG} shows that the scaling dimensions of the first few operators in the Ising CFT can be obtained from the linearized HOTRG scheme with GILT and a preliminary $Z_2$ ($U(1)$ for the complex case) gauge-fixing procedure.
One can also compare the lTRG result with the scaling dimensions obtained from the transfer matrix method~\cite{gu2009TEFR_TMscD}, as well as the known analytical results from conformal field theory~\cite{francesco2012conformal}. For example, in Ref.~\cite{evenbly2017TNR_detail}, the first 101 scaling dimensions of local and non-local operators in the Ising CFT are calculated using the transfer matrix method on the coarse-grained tensor obtained from the TNR method, which confirms well with the expected analytical result.

In this paper, we will further demonstrate that the eigenvectors $\ket{\mathcal{O}_\alpha}$ of lTRG in \eqref{eq:lTRG} have a clear physical meaning. They are the conformal states associated with the corresponding operators $\mathcal{O}_\alpha$. 
The conformal states are obtained by inserting conformal primaries and descendants into the vacuum state (see \appendixref{sec:Appendix_CFTStates}). In this paper, the states are defined on the boundary of a coarse-graining block. Thus, we place the operator at the center of the block. At the lattice level, these states can be achieved by coarse-graining the tensor network \eqref{eq:TNSquareLattice} with the operator insertion as defect tensors:
\begin{equation}
    \label{eq:state-operator-correspondence}
    \ket{\mathcal{O}_\alpha}=\lim_{r\rightarrow 0}\mathcal{O}_\alpha(r)\ket{0}
    \approx
    \sigma(x_1)\sigma(x_2)...\ket{0}_{\text{lattice}}.
\end{equation}
By also calculating the two-point correlation function of the Ising model on the square lattice using the TRG method, we find that the position of the operator in the coarse-grained defect tensor $T_{\sigma_{x_1}}$ is encoded by its projections onto the primary and descendant tensors or lTRG.

\subsection{Coarse Graining using HOTRG, GILT and MCF}
\label{sec:CoarseGrainingScheme}

Here, we integrate GILT and MCF into the HOTRG steps,

\begin{minipage}{\columnwidth}
\begin{equation}
\label{eq:HOTRGStepFull}
    \vcenter{\hbox{\includegraphics[page=7]{TRGCFTGraphics.pdf}}}
    =
    \vcenter{\hbox{\includegraphics[page=8]{TRGCFTGraphics.pdf}}}
\end{equation}
\end{minipage}

In the above, at each coarse-graining layer, $w$, $w^\dagger$ are the isometry tensors from HOTRG, and 
$g_{12}$, $g_{13}$, $g_{22}$, and $g_{23}$ are the entanglement truncation matrices from GILT. They filter out certain local correlations that cannot be removed by $w$ and $w^\dagger$.
Moreover, $h_0$ and $h_2$ are the gauge-fixing matrices, determined by MCF, and the residual $U(\chi)^{\otimes m}$ gauge redundancy is fixed by the sign-fixing procedure in Ref.~\cite{lyu2021lTRG}.

\smallskip\noindent\textbf{Adapting GILT to HOTRG}. In GILT, one has to repeat the process on all the possible plaquette-leg combinations to remove all the local entanglements. In practice, as suggested in~\cite{lyu2021lTRG}, it is sufficient only to apply GILT on the following plaquette-legs:
\begin{equation}
    \vcenter{\hbox{\includegraphics[page=16]{TRGCFTGraphics.pdf}}}
    \approx
    \vcenter{\hbox{\includegraphics[page=17]{TRGCFTGraphics.pdf}}}
\end{equation}
The tensor in the gray shade is to be coarse-grained, with circle-shaped GILT tensors being split using the SVD. We only truncate half of the CDL loops across the legs to be coarse-grained.

\subsection{Coarse-graining the Defect Tensors}
\label{sec:coarse_graining_defect_tensors}

To compute the $n$-point correlation function in \eqref{eq:TNSquareLatticeExpVal}, we iteratively apply the coarse-graining step \eqref{eq:HOTRGStepFull} to each pair of tensors in all coarse-graining blocks, as shown in \eqref{eq:CGDefectLarge} below:
\begin{widetext}
\begin{equation}
    \label{eq:CGDefectLarge}
     \tr\vcenter{\hbox{\includegraphics[page=59]{TRGCFTGraphics.pdf}}}
     =
     \tr\vcenter{\hbox{\includegraphics[page=60]{TRGCFTGraphics.pdf}}}
     =
     \tr\vcenter{\hbox{\includegraphics[page=61]{TRGCFTGraphics.pdf}}}
     =
     \tr\vcenter{\hbox{\includegraphics[page=62]{TRGCFTGraphics.pdf}}}
     =
     \tr\vcenter{\hbox{\includegraphics[page=63]{TRGCFTGraphics.pdf}}},
\end{equation}
\end{widetext}

where $\sigma_1$, $\sigma_2$ are two defect operators, and $:\sigma_1 \sigma_2:$ means the two defects are fused together.

The coarse-graining of tensors is performed according to \eqref{eq:HOTRGStepFull}, in which we use the same set of coarse-graining tensors $w$, $g$, $h$ as when coarse-graining vacuum tensors. The symbol $\tr$ indicates contracting opposite pairs of edges, according to periodic boundary conditions. Note that the coarse-grained tensor is dependent on not only the type of defect but also the position of the defect relative to the coarse-graining block.

Similarly, when calculating the linearized TRG \eqref{eq:lTRG}, on the other hand, we also keep the coarse-graining tensors $w$, $g$, $h$ as constants, that is, their variation with respect to the change of the input tensor is discarded. 

As a result, the coarse-grained tensor is quad-linear in the four input tensors:

\begin{widetext}
\begin{equation}
    \label{eq:CGVacuumVariation}
     \vcenter{\hbox{\includegraphics[page=18]{TRGCFTGraphics.pdf}}}
     =
     \vcenter{\hbox{\includegraphics[page=64]{TRGCFTGraphics.pdf}}}
     +
     \vcenter{\hbox{\includegraphics[page=65]{TRGCFTGraphics.pdf}}}
     +
     \vcenter{\hbox{\includegraphics[page=66]{TRGCFTGraphics.pdf}}}
     +
     \vcenter{\hbox{\includegraphics[page=67]{TRGCFTGraphics.pdf}}}
\end{equation}
\end{widetext} \eqref{eq:CGVacuumVariation} above shows the variation of the coarse-grained tensor with the change in the input tensor, which is used to obtain the lTRG equation \eqref{eq:lTRG}.

Faithful scaling dimensions can be computed from the linearized TRG with fixed $g$ and $w$, as shown in \secref{sec:Discussion}. However, in terms of $n$-point correlation functions, sometimes undesirable results are obtained if GILT is applied, which will be discussed in more detail later in \secref{sec:Discussion_GILTProblemDefect}.

\section{Main Results}
\label{sec:Results}

Here we present our results of the CFT data from the fixed point tensor of the linearized Tensor Renormalization Group~\cite{lyu2021lTRG}, using the improved gauging fixing via MCF.  We obtain the CFT data by two distinct approaches: one involving the introduction of defects on the lattice and the other relying solely on the fixed-point tensor, without any information about the lattice. We also compare the tensors corresponding to conformal operators extracted from the two methods.

We first follow the work of Lyu et al. \cite{lyu2021lTRG}, who extracted the conformal dimension spectrum from the eigenvalues of lTRG. Our first contribution is to improve the stability of the RG flow by introducing MCF for gauge fixing. The results of the RG flow and scaling dimensions are summarized in Sections \ref{sec:Results_TdiffFlow} and \ref{sec:Results_ScDim}.

Next, we go beyond reproducing scaling dimensions and propose and implement how to extract the OPE coefficients, such as $C_{\sigma\sigma\varepsilon}$, through the fusion of eigenvectors and lTRG fixed point equations (note that these eigenvectors are themselves tensors). The results agree with known values and are summarized in Section \ref{sec:Results_OPEFromlTRG}.

Additionally, we verify that eigenvectors of lTRG serve as a useful representation of conformal states at low-energy sectors.  To do this, specifically, we also construct these conformal states by placing lattice-level operators on the lattice (see Table~\ref{tab:scdim}) and then coarse-graining them. We compare these
with the eigenvectors of lTRG and find that the latter align well with the coarse-grained tensors in the low-energy subspace (i.e., the subspace consists of states of small scaling), as demonstrated in Section~\ref{sec:Results_eigvecCompare}. Moreover, by expanding the coarse-grained tensor using the basis of primary and descendant states, it becomes evident that the position of the primary operator inserted in the lattice is encoded in the components of the descendant basis, similar to the standard Taylor expansion used for operators. This observation is discussed in Section \ref{sec:Results_encoding_position}.

However, it remains unclear whether coarse graining introduces significant truncation errors in such states with higher excited levels. To characterize the quality of the coarse-graining on the defect tensors, we calculate their two-point and four-point correlation functions using HOTRG+GILT.  They agree well with the analytical result, and we can also use them to extract the CFT data $\Delta_\sigma$, $\Delta_\varepsilon$ and $C_{\sigma\sigma\varepsilon}$ by fitting the correlation functions. However, the error is noticeably larger than the one we obtained from the fusion method. Additionally, we calculate the two-point function on a finite-size torus and find that the numerical result agrees well with the analytical one and that the correlation function respects the periodic boundary condition of the torus geometry. These results are summarized in sections \ref{sec:Results_2pt}, \ref{sec:Results_4pt}, and \ref{sec:Results_2pt_torus}.

By studying the $N$-point correlation functions, we can better understand the capability and limitations of our real-space RG scheme in coarse-graining defect tensors. A technical discussion is provided in section \ref{sec:Discussion}. With this knowledge, we are confident that the coarse-grained tensor can serve as a reference for conformal states.

In summary, we demonstrate two different methods for extracting CFT data from a critical lattice model. The first method involves extracting operator scaling dimensions from two-point correlation functions and OPE coefficients from four-point correlation functions by introducing defects on the lattice. The second method utilizes the fixed-point tensor to extract the CFT data without relying on lattice-level knowledge (i.e., only from the fixed-point tensor and equation). Specifically, we obtain the scaling dimensions from the eigenvalues of lTRG and the OPE coefficients from the fusion rules of the eigenvectors of lTRG, which correspond to conformal states. Our findings demonstrate a good agreement between the conformal states obtained from lTRG and those obtained from coarse-graining lattice-level defects.

\subsection{Convergence onto the Conformal Fixed Point in Higher-order Tensor Renormalization Group}
\label{sec:Results_TdiffFlow}

After applying GILT~\cite{hauru2018GILT} to remove local entanglement and performing the MCF~\cite{acuaviva2022MCF} for gauge-fixing, we can identify a non-trivial fixed point in HOTRG~\cite{xie2012HOTRG}, where the tensor remains almost the same after coarse-graining. This fixed point from the 2D classical Ising model corresponds to the Ising CFT $\mathcal{M}(4/3)$, with a central charge of $c=1/2$~\cite{francesco2012conformal}.

\begin{figure}[h]
    \centering
    \includegraphics[width=\hsize]{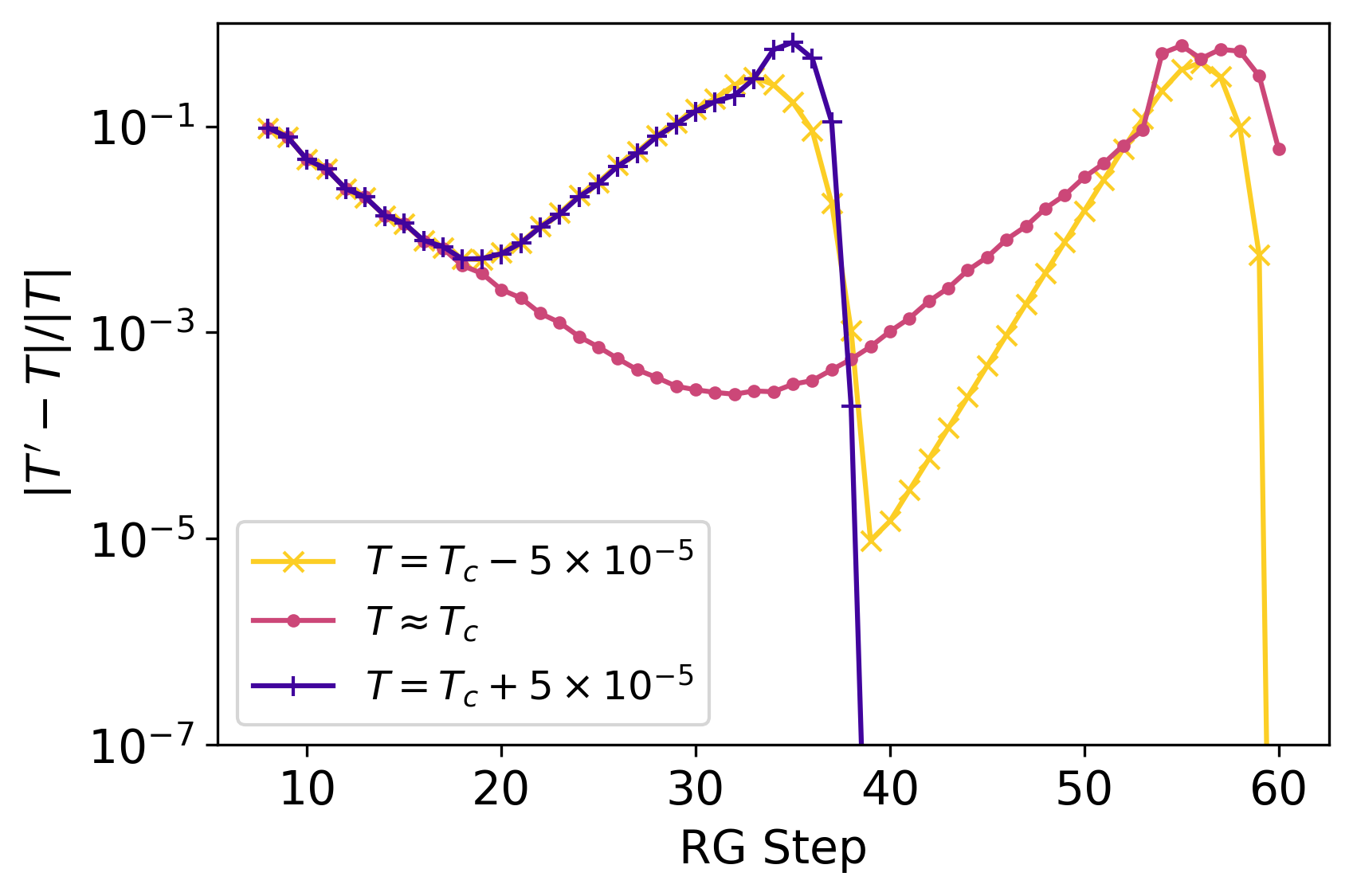}
    \caption{(color online) Difference of coarse-grained tensor $||T^{(l)}-T^{(l-2)}||$ before and after every two steps of HOTRG coarse-graining for the 2D square lattice Ising model near the critical temperature $T_c$. The bond dimension is $\chi=24$. GILT is applied with $\epsilon_{\text{GILT}}=8\times 10^{-7}$, $\text{nIter}_{\text{GILT}}=1$. MCF is applied. Three temperatures are considered: $T = T_c(\chi=24)$, $T=T_c(\chi=24)-5\times 10^{-5}$ and $T=T_c(\chi=24)+5\times 10^{-5}$.  The numerical scheme specified critical temperature $T_c(\chi=24)=2.26920063397$ is chosen such that the system has the most stable HOTRG flow (see \appendixref{sec:Find_Tc}).}
    \label{fig:Tdiff}
\end{figure}

\begin{figure}[h]
    \centering
    \includegraphics[width=\hsize]{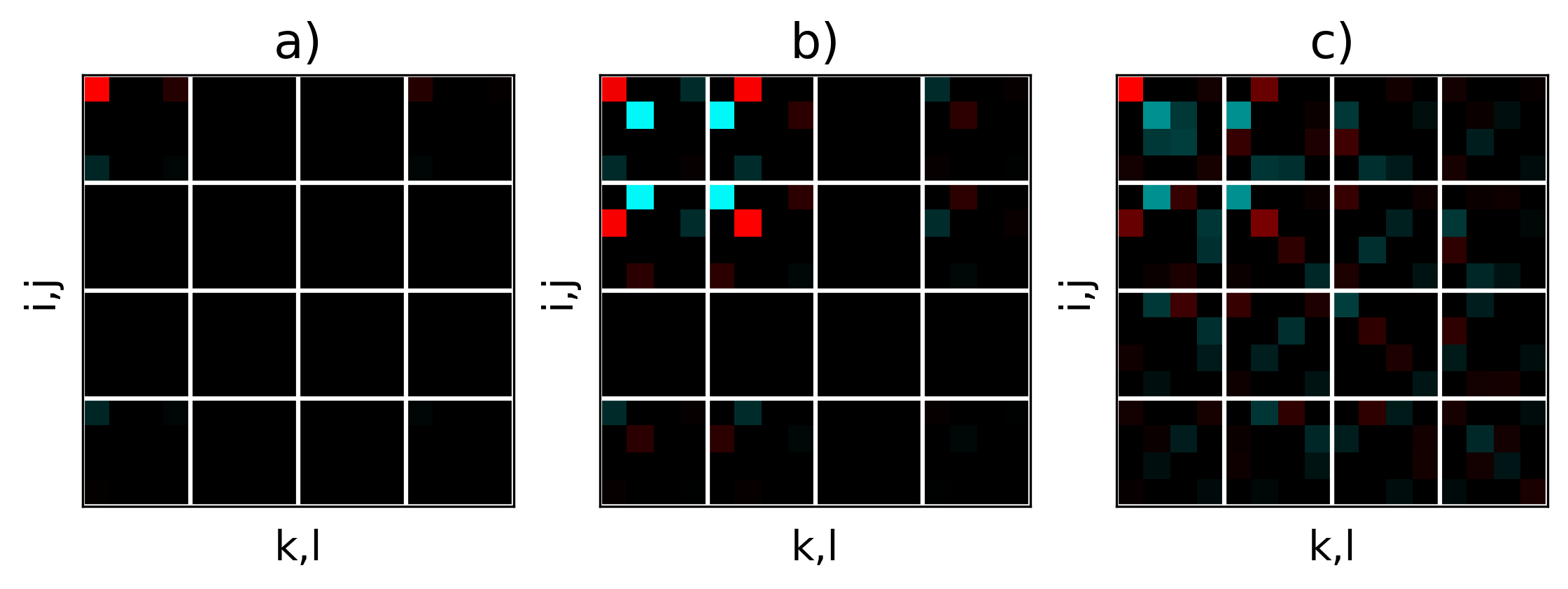}
    \caption{(color online) The first few components of fixed-point tensors $T_{ijkl}^{(fix)}$ for the a) high temperature, b) low temperature and c) critical fixed point of the RG flow demonstrated in \figref{fig:Tdiff}. The same parameters are used as in \figref{fig:Tdiff}.  The components $T_{ijkl}^{(fix)}$'s are shown as  pixels at row $4\times i + j$ and at column $4\times k + l$. The pixel's brightness stands for the component's absolute value, and the pixel's color stands for the component's argument (i.e., red is positive, blue is negative, and black is zero). One can see that a) is the trivial fixed point of HOTRG, b) is the tensor product between two trivial tensors, indicating an unbroken $Z_2$ symmetry, and c) is a more complicated fixed point tensor. The same parameters are used as in \figref{fig:Tdiff}.  }
    \label{fig:fpTensors}
\end{figure}

In \figref{fig:Tdiff}, we present the tensor difference $||T^{(l)}-T^{(l-2)}||$ of the coarse-grained Ising model on a 2D square lattice as a function of the HOTRG iterations near the critical temperature $T_c$ (see purple circles in the figure). As the system approaches the critical point around step 30, the coarse-grained tensor approximately converges to a fixed point, where the tensor remains almost the same after the coarse-graining steps (with the change in the tensors being less than $3\times 10^{-4}$). After step 30, the tensor deviates exponentially,  because the critical fixed point is an unstable fixed point, and any error, such as the numerical error, truncation error, or fluctuation of initial values, will result in a spontaneous symmetry broken after RG iterations.

This behavior of first approaching and then deviating from the fixed point is a hallmark of the RG flow near a critical point that can be identified from a V-shaped curve in the logarithmic plot.
Figure~\ref{fig:Tdiff} also illustrates how the system evolves into distinct fixed points as the temperature is slightly perturbed. When starting at a temperature slightly away from $T_c$ (above and below), the system does not converge as closely to the critical fixed point as the one at the critical temperature. The flows in both cases approach their ``perihelions'' of the fixed point around step 20. 

In the low-temperature phase, the system first flows to a degenerate low-temperature fixed point (with the only two nonzero elements being $T_{0,0,0,0}=T_{1,1,1,1}$) before eventually flowing to a tensor of the form $T_{0,0,0,0}=0$, due to spontaneous symmetry breaking caused by perturbations from numerical errors, since we did not enforce the $Z_2$ symmetry during coarse graining. 
In the high temperature phase, the system converges directly to the high-temperature fixed point, with one nonzero element being $T_{+,+,+,+}$ (whose tensor form is nevertheless equivalent to $T_{0,0,0,0}$ by a basis transformation in the virtual indices~\eqref{eq:gauge_transform}).
The tensor components at the three fixed points are shown in~\figref{fig:fpTensors}.

The numerical scheme for the critical temperature $T_c(\chi)$ is obtained using a bisection algorithm, as suggested in \cite{lyu2021lTRG}. We assume that there is an abrupt change in the coarse-grained tensors across the phase transition. The critical point can be identified by comparing the gauge-fixed coarse-grained tensors $T^{(l_M)}$. The detailed procedure is described in \algref{alg:Find_Tc} in \appendixref{sec:Find_Tc}.

The curves in Fig.~\ref{fig:Tdiff} exhibits more fluctuation and discontinuities in the range where $|T'-T|/|T|$ is large. It is because when two tensors are more different from each other, it is more likely that the MCF fails to find the correct gauge to compare the two tensors. More specifically, a gauge fixing scheme could introduce discontinuities; that is, two physically nearly identical tensors are separated by the ``cut'' of a gauge choice, resulting in an O(1) inconsistent in their gauge-fixed form. This problem might be diagnosed by inspecting the level crossing incident in the SVD spectrum.

\subsection{Scaling Dimension from Linearized Tensor Renormalization group}
\label{sec:Results_ScDim}

In Table~\ref{tab:scdim}, we compare the results of the scaling dimensions obtained by two different methods: (1) diagonalizing the transfer matrix on a cylinder and (2) solving the eigenvalues of lTRG~\eqref{eq:lTRG}. We also compare for each case how the results depend on whether GILT is used or not. First, we confirm that GILT generally improves the numerical results for scaling dimensions~\cite{hauru2018GILT}, especially for higher values. However, we observe that the transfer matrix spectrum achieves more accurate information on higher conformal dimensions compared to the lTRG method (see their values for operators having $\Delta=3$ or higher).

\begin{table}[h]
    \centering
    \begin{tabular}{|c|c|c|c|c|c|c|c|}
        \hline
         & 
         & \multicolumn{2}{|c|}{with GILT} 
         & \multicolumn{2}{|c|}{without GILT} 
         & \multicolumn{2}{|c|}{Ref.\cite{lyu2021lTRG}} \\
        \hline
        exact &
        $\Delta$
        & $\Delta_{\text{cyl}}$ & $\Delta_{\text{RG}}$
        & $\Delta_{\text{cyl}}$ & $\Delta_{\text{RG}}$ 
        & $\Delta_{\text{cyl}}$ & $\Delta_{\text{RG}}$\\
        \hline
        $\mathbb{1}$            & 0               &         &         &          &         &       &  \\ \hline
        $\sigma$                & $\frac{1}{8}$   & 0.1259  & 0.1275  &  0.1101  & 0.1065  & 0.125 & 0.127 \\ \hline
        $\varepsilon$              & 1            & 1.0116  & 1.0021  &  1.0041  & 1.2844  & 1.002 & 1.009 \\ \hline
        $\partial_x \sigma$,    & $1+\frac{1}{8}$ & 1.1395  & 1.1264  &  1.1600  & 1.4137  & 1.128 & 1.125 \\
        $\partial_y \sigma$     &                 & 1.1399  & 1.1446  &  1.2380  & 1.4269  & 1.128 & 1.128 \\ \hline
        $\partial_x \varepsilon$,  & $2$          & 2.0354  & 1.9981  &  1.2586  & 2.0431  & 2.014 & 2.002 \\
        $\partial_y \varepsilon$,  &              & 2.0357  & 2.0008  &  1.2587  & 2.0431  & 2.014 & 2.004 \\
        $T_{xx}-T_{yy}$,        &                 & 2.0382  & 2.0033  &  1.3253  & 2.0462  & 2.016 & 2.068 \\
        $T_{xy}+T_{yx}$,        &                 & 2.0428  & 2.0447  &  1.3525  & 2.0471  & 2.016 & 2.073 \\ \hline
        $\partial_{xx}\sigma$,  & $2+\frac{1}{8}$ & 2.1607  & 2.1197  &  1.4917  & 2.0512  & & \\
        $\partial_{xy}\sigma$,  &                 & 2.1831  & 2.1987  &  1.5193  & 2.0651  & & \\
        $\partial_{yy}\sigma$,  &                 & 2.1897  & 2.2031  &  1.6390  & 2.0719  & & \\ \hline
        $\partial_{xx}\varepsilon$ & 3            & 3.0145  & 2.5010  &  1.6434  & 2.0720  & & \\
        $\partial_{xy}\varepsilon$ &              & 3.0548  & 2.6845  &  1.6479  & 2.0931  & & \\
        $\partial_{yy}\varepsilon$ &              & 3.1142  & 2.6845  &  1.6512  & 2.0931  & & \\
        $\partial_{x}T_{xx}$    &                 & 3.1145  & 2.7911  &  1.6582  & 2.0993  & & \\
        $\partial_{y}T_{yy}$    &                 & 3.1306  & 2.8228  &  1.6639  & 2.0993  & & \\ \hline
        ...                     & $3+\frac{1}{8}$ & 3.1359  & 2.8264  &  1.6641  & 2.1025  & & \\
        \hline
        
    \end{tabular}
    \caption{The central charge and the first few scaling dimensions at RG step 30 obtained using two different methods: (1) transfer matrix on a cylinder\cite{gu2009TEFR_TMscD} with scaling dimensions denoted as $\Delta_{\text{cyl}}$, and (2) linearized Tensor Renormalization Group with scaling dimensions denoted as $\Delta_{\text{RG}}$. The cases of whether GILT is applied are compared, where the RG steps for both cases is 30. The bond dimension is set to $\chi=24$, and GILT is applied with $\epsilon_{\text{GILT}}=8\times 10^{-7}$ and $\text{nIter}_{\text{GILT}}=1$. Note that Ref.~\cite{lyu2021lTRG} reports the results at step = 22 and with $\chi=30$. Moreover, the conformal operators in each group are degenerate. }
    \label{tab:scdim}
\end{table}

The spectrum of the transfer matrix, $\Delta_{cyl}$, can be used to represent the quality of the coarse-grained tensor. The method with GILT enabled produces the correct scaling dimensions and their degeneracy number up to $\Delta=3$. Without GILT, the number of degeneracy errors starts from $\Delta=2$, and the error in the scaling dimensions is much larger.

The lTRG scaling dimension spectrum, $\Delta_\text{RG}$, is of greater interest to us, since we shall extend the approach to OPE coefficients. The method with GILT enabled does not provide the correct scaling dimensions from $\Delta=3$ and above, but it does show a gap between $\Delta=2+\frac18$ and $\Delta=3$, with the correct number of degeneracy at $\Delta=2+\frac18$. On the other hand, the case without GILT has the last meaningful gap between $\Delta=1$ and $\Delta=1+\frac 18$. Furthermore, the GILT case has fewer errors at $\Delta=\frac 18, 1, 1+\frac 18$.

For further comparison, we present the number of degeneracy in each scaling dimension in \tabref{tab:CFT_degeneracy}.
 The error of scaling dimensions compared to the analytic results at each RG step is further discussed in \ref{sec:Discussion_ErrorFlow},

\subsection{OPE Coefficient from Linearized Tensor Renormalization Group}
\label{sec:Results_OPEFromlTRG}

Despite that lTRG performs worse than the transfer matrix in scaling dimensions, we demonstrate in this subsection that using lTRG one can extract the OPE coefficients $C_{ijk}$ only from its fixed point tensor $T_{\mathbb{1}}$, without any knowledge about the lattice-level details.
In particular, what one needs is the coarse-graining $\mathcal{M}: T_1\otimes T_2\otimes T_3\otimes T_4 \mapsto T'$, and the conformal states $T_\mathbb{1}$, $T_\sigma$,  and $T_\varepsilon$, which are the first few eigenvectors of the lTRG transformation $M: T\mapsto T\otimes T\otimes T\otimes T \mapsto T'$.

We recall the OPE expansion and its coefficients in terms of  the conformal states rather than operators, 
\begin{equation}
    \label{eq:CFTOPE_demo}
    \ket{\mathcal{O}_2\mathcal{O}_3}=\frac{C_{231}}{r_{23}^{\Delta_2+\Delta_3-\Delta_1}}\ket{\mathcal{O}_1}+...,
\end{equation}
where $...$ denotes states orthogonal to $\ket{\mathcal{O}_1}$.

From our coarse-graining scheme, we have that

\begin{equation}
    \bra{T_\varepsilon}\mathcal{M}\ket{\begin{array}{cc}
    T_\sigma & T_\sigma	\\
    T_\mathbb{1} & T_\mathbb{1}
    \end{array}}=\frac{C_{\sigma\sigma\varepsilon}}{a_0^{2\Delta_\sigma-\Delta_\varepsilon}},
\end{equation}

where $\ket{\begin{array}{cc}
    T_\sigma & T_\sigma	\\
    T_\mathbb{1} & T_\mathbb{1}
    \end{array}}$ is the 8-legged tensor from contracting the corresponding tensor sub-network, $\mathcal{M}$ is the coarse-graining procedure \eqref{eq:HOTRGStepFull} that converts the 8-legged tensor to the 4-legged coarse-grained tensor. The inner product $\bra{T_2}\ket{T_1}=\vcenter{\hbox{\includegraphics[page=58,scale=0.7]{TRGCFTGraphics.pdf}}}$.   $\ket{T_{\mathbb{1}}}$ and $\ket{T_{\sigma}}$ are renormalized so that:
\begin{equation}
    \mathcal{M}\ket{\begin{array}{cc}
    T_\mathbb{1} & T_\mathbb{1}	\\
    T_\mathbb{1} & T_\mathbb{1}
    \end{array}}=\ket{T_\mathbb{1}}.
\end{equation}

\begin{equation}
    \mathcal{M}\ket{\begin{array}{cc}
    T_\sigma & T_\sigma	\\
    T_\mathbb{1} & T_\mathbb{1}
    \end{array}}=\frac{C_{\sigma\sigma\mathbb{1}}}{a_0^{2\Delta_\sigma}}\ket{T_\mathbb{1}}+...
\end{equation}

The detailed scheme is further explained in  Appendix~\ref{sec:Appendix_RenormalizeTensorStates}. Note that in practice, we place defect tensors on different combinations of sites in the coarse-graining block and average over the resulting renormalization coefficients and the OPE coefficients.

The result of the OPE coefficients of Ising CFT is shown in Table~\ref{tab:OPEFromlTRG}. The scaling dimensions and the fusion between two $\sigma$'s are fairly accurate; however, the values that incur slightly larger errors are $C_{\sigma\varepsilon\sigma}$ and $C_{\varepsilon\varepsilon\varepsilon}$. This is because as an operator of higher scaling dimensions, $\epsilon$, suffers more error than $\sigma$ from the coarse-graining process.
We remark that OPE coefficients (and scaling dimensions) can also be extracted from four-point correlation functions, as will be discussed in Sec.~\ref{sec:Results_4pt}. However, the results there are not as accurate.

\begin{table}[h]
    \centering
    \begin{tabular}{|c|c|c|}
        \hline
         & exact & lTRG\\
         \hline
        $\Delta_\sigma$ & 1/8 & 0.127 \\
        $\Delta_\varepsilon$ & 1 & 1.002 \\
        \hline
        $C_{\sigma\sigma\sigma}$ & 0 & $2.4\times10^{-7}$ \\
        $C_{\sigma\sigma\varepsilon}$ & 1/2 & 0.512 \\
        $C_{\varepsilon\varepsilon\sigma}$ & 0 & $1.1\times10^{-6}$ \\
        $C_{\varepsilon\varepsilon\varepsilon}$ & 0 & 0.168 \\
        $C_{\sigma\varepsilon\sigma}$ & 1/2 & 0.409 \\
        $C_{\sigma\varepsilon\varepsilon}$ & 0 & $1.5\times10^{-7}$ \\
        \hline
    \end{tabular}
    \caption{The CFT data extracted from the fusion rules of lTRG following the procedure detailed in \appendixref{sec:Appendix_RenormalizeTensorStates}.}
    \label{tab:OPEFromlTRG}
\end{table}

\subsection{Scaling Dimension from Two-point Correlators}
\label{sec:Results_2pt}

\begin{figure}[h]
    \centering
    \includegraphics[width=\hsize]{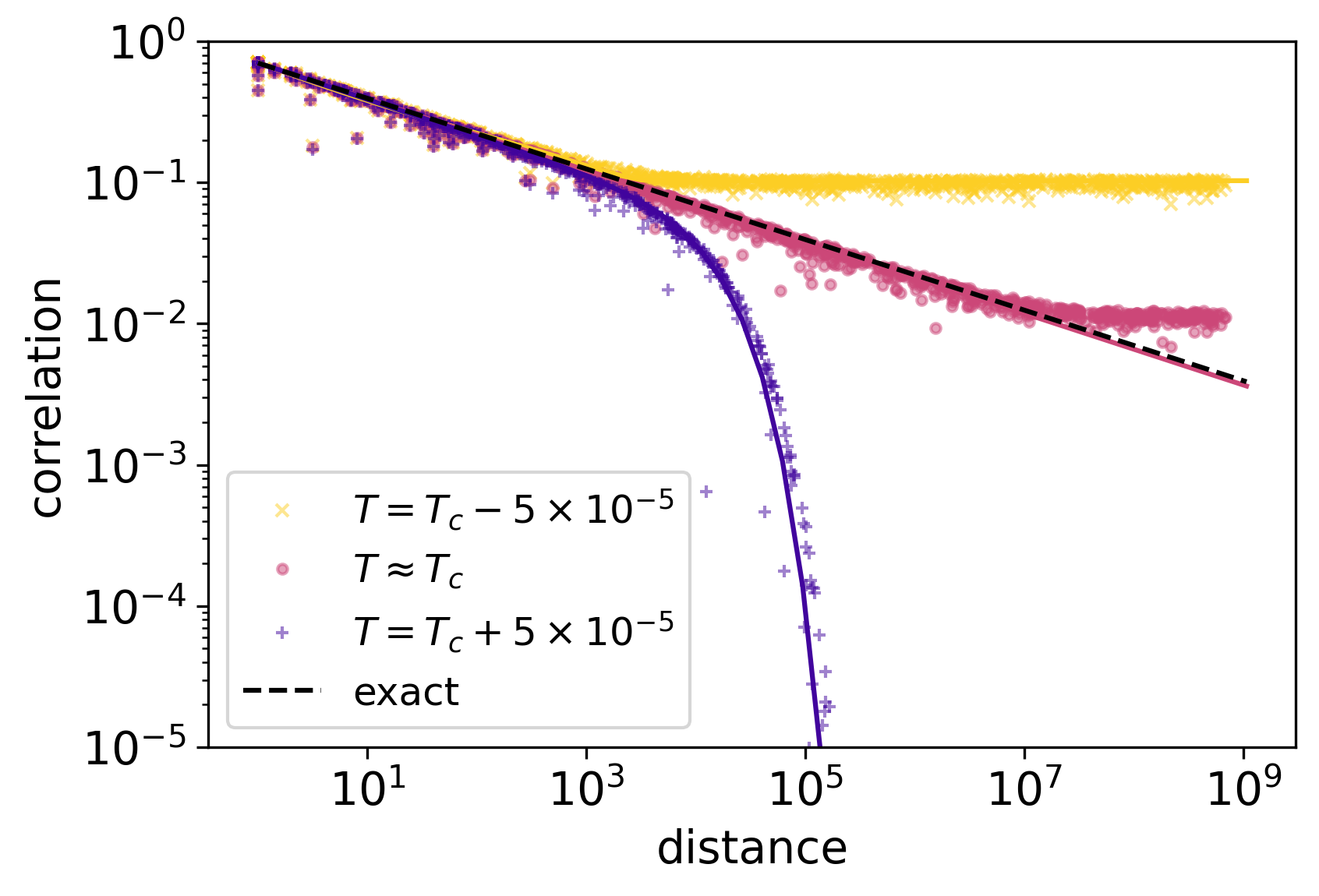}
    \caption{(color online) The two-point correlator $\langle\sigma(x_0,y_0)\sigma(x_1,y_1)\rangle$ of the Ising model on a $2^{30}\cross 2^{30}$ square lattice with periodic boundary condition. The locations of $(x_0,y_0)$ and $(x_1,y_1)$ were randomly chosen. The result is obtained from coarse graining defect tensors using HOTRG with bond dimension $\chi=24$. GILT is applied with $\epsilon_{\text{GILT}}=8\times 10^{-7}$ and $\text{nIter}_{\text{GILT}}=1$. MCF is also applied. Three temperatures are considered: $T = T_c(\chi=24)$, $T=T_c(\chi=24)-5\times 10^{-5}$ and $T=T_c(\chi=24)\times 10^{-5}$, where the numerical scheme specified the critical temperature $T_c(\chi=24)=2.26920063397$. }
    \label{fig:Corr}
\end{figure}

It is well known that the two-point correlation function $\langle\sigma(0)\sigma(r)\rangle$ at criticality scales as $1/r^{d-2+\eta}$, where the anomalous dimension $\eta$ is related to the scaling dimension $\Delta$, via $d-2+\eta=2\Delta$. Thus, it provides an alternative approach to extracting scaling dimensions. 
Figure~\ref{fig:Corr} shows the two-point correlation function $\langle\sigma(x_0,y_0)\sigma(x_1,y_1)\rangle$ of the square lattice Ising model near the critical temperature.

The points $(x_0,y_0)$ and $(x_1,y_1)$ were randomly sampled on the xy plane. The correlation function is determined by the distance between $(x_0,y_0)$ and $(x_1,y_1)$, irrespective of the relative orientation or absolute positions for most cases. This indicates that the translational symmetry is preserved and the $SO(2)$ rotational symmetry has emerged at large distances.

To calculate the two-point function using coarse-grain tensors. We place two defect tensors $T_{\sigma(x_1,y_1)}^{(0)}$, $T_{\sigma(x_2,y_2)}^{(0)}$ on a $L_x\times L_y$ square lattice tensor network with periodic boundary conditions, where the tensors at the other sites are the vacuum tensor $T^{(0)}$. Then we coarse-grain every adjacent pair of tensors as described in Sec.~\ref{sec:coarse_graining_defect_tensors}. Then the coarse-grained vacuum tensor $T^{(l+1)}$ and the coarse-grained defect tensors $T^{(l+1)}_{\sigma(x_1,y_1)}$, $T^{(l+1)}_{\sigma(x_2,y_2)}$ are obtained. Furthermore, when two defects are met in the same coarse grain block, they fuse into one defect tensor $T^{(l+1)}_{\sigma(x_1,y_1)\sigma(x_2,y_2)}$.

At the final coarse-graining step, all defects fuse into one tensor $T^{(l_M)}_{\sigma(x_1,y_1)\sigma(x_2,y_2)}$. The correlation function is computed using the following
\begin{equation}
\label{eq:tracing_defect_tensor}
    \expval{\sigma(x_1,y_1)\sigma(x_2,y_2)}=\frac{\tr T^{(l_M)}_{\sigma(x_1,y_1)\sigma(x_2,y_2)}}{\tr T^{(l_M)}},
\end{equation}
where $\tr T = \sum_{i,k}T_{iikk}=\vcenter{\hbox{\includegraphics[page=57,scale=0.5]{TRGCFTGraphics.pdf}}}$ indicates contracting the opposite pairs of legs of the tensor.

We used three different temperatures. (1) When $T<T_c$, the correlation function saturates at a finite distance, indicating a long-range order.
(2) When $T\approx T_c$, the correlation function follows a power-law behavior, with a slope in the log-log scale determined by the scaling dimension, indicating the critical behavior, discussed in more detail below.
(3) When $T>T_c$, the correlation function decays at a finite distance, indicating the absence of long-range order.

We now analyze the results of the two-point correlations in more detail and aim to extract the scaling dimension. To do this, we fit the correlation functions to the following ansatzes,  in the respective temperature ranges,
\begin{align}
    \label{eq:ansatz_lowT1}
        \langle\sigma(x_0,y_0)\sigma(x_1,y_1)\rangle&=
        A \frac{e^{-x/\zeta}}{x^{2\Delta'}} +m_0^2, 
 \, T<T_c, \\
    \label{eq:ansatz_critical}
        \langle\sigma(x_0,y_0)\sigma(x_1,y_1)\rangle&=
        A \frac{1}{x^{2\Delta}},  
 \, T=T_c, \\
    \label{eq:ansatz_highT}
        \langle\sigma(x_0,y_0)\sigma(x_1,y_1)\rangle&=
        A \frac{e^{-x/\zeta}}{x^{2\Delta}} , 
 \, T>T_c,
\end{align}
where $\Delta(T)$ is the scaling dimension of the magnetization operator fitted at the corresponding temperature.  $\zeta(T)$ is the correlation length, $m_0=\expval{\sigma}$ is the magnetization in the ordered phase, $\Delta'(T)$ is the ``scaling dimension'' of $\sigma-\expval{\sigma}$, where $\expval{\sigma}=(L_x L_y)^{-1}\sum_{x,y}\expval{\sigma(x,y)}$ is the one-point function averaged at every lattice site.

Regarding the outliers in the data, we adopt Huber loss \eqref{eq:HuberLoss} (explained in \appendixref{sec:Appendix_HuberLoss}) with $\epsilon=10^{-2}$ (in log-log space) to suppress their excessive effects. Since the fitting is done in the log-log space, we also give the data points a weight that is proportional to the magnitude of its correlation.

When $T\approx T_c$,  we obtain $\Delta=0.1266$; when $T=T_c+5\times 10^{-5}$,  we obtain $\Delta=0.1293$ and $\zeta=1.70\times 10^{4}$; when $T=T_c-5\times 10^{-5}$, we obtain $\Delta'=0.1652$, $\zeta=1.62\times 10^{3}$, and $m_0=0.313245$.

The three correlation functions behave in a similar way at short distance $x<10^3$. By fitting with \eqref{eq:ansatz_critical}, we get $\Delta=0.1251,0.1272,0.1300$ for low, critical, and high temperatures, respectively. 
The above results agree well with each other and with the analytic result $\Delta=0.125$.

We remark that, in order to get the correct scaling dimension, it is necessary to average over the positions of the correlated pairs, due to an issue in the coarse-graining of impurity operators. Also, Huber loss is used to take into account the outliers. We elaborate on these issues later in \secref{sec:Discussion}.

\subsection{OPE Coefficient from Four-point Correlators}
\label{sec:Results_4pt}

\begin{figure}[h]
    \centering
    \includegraphics[width=\hsize]{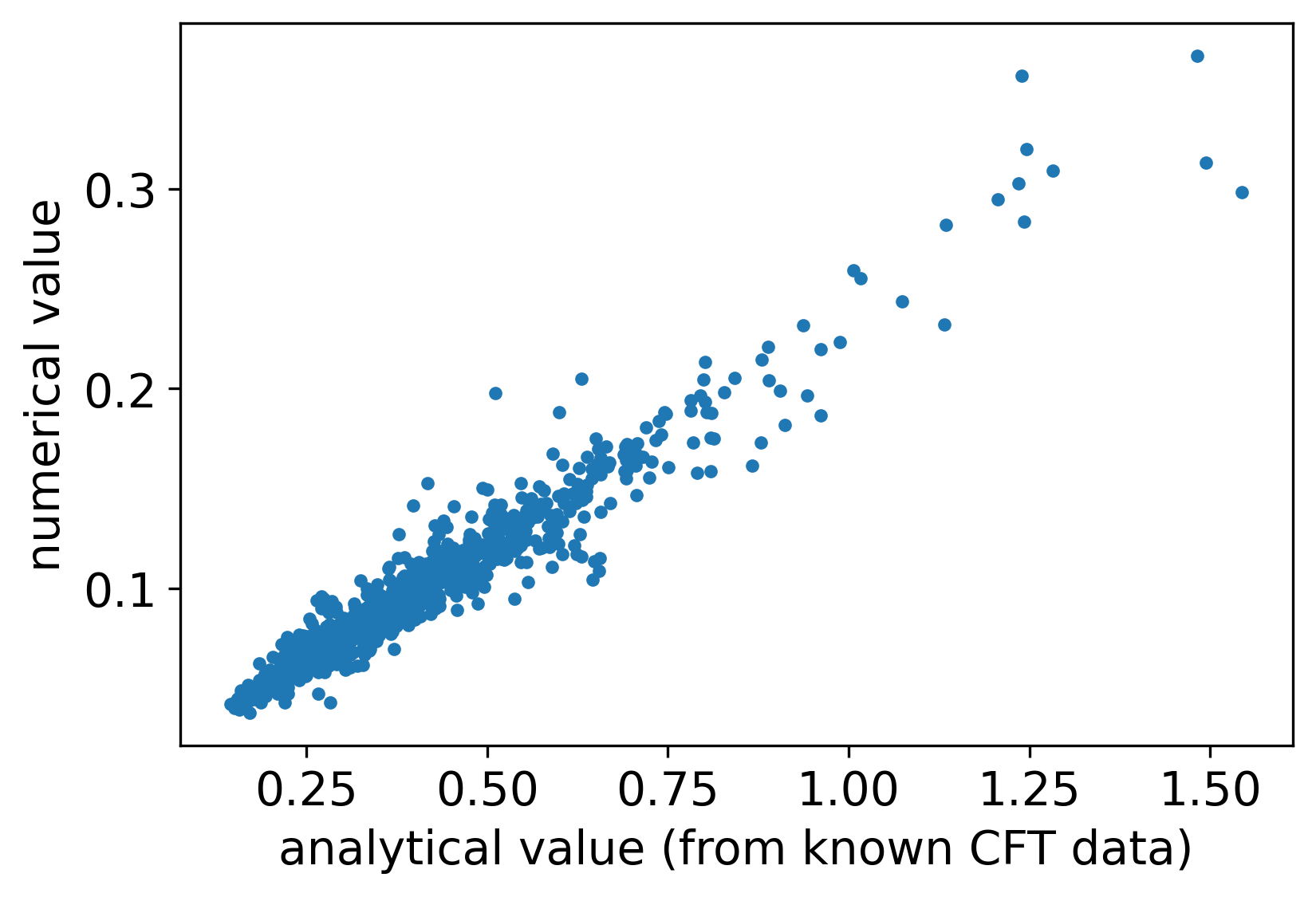}
    \caption{(color online) The four-point correlator $\langle\sigma(x_1,y_1)\sigma(x_2,y_2)\sigma(x_3,y_3)\sigma(x_4,y_4)\rangle$ of the Ising model on a $1024\cross 1024$ square lattice at the critical temperature $T \approx T_c$ for different choices of operator locations. The horizontal axis indicates the analytical value from \eqref{eq:4ptFunction} using the known CFT data $\Delta_\sigma=0.125$, $\Delta_\varepsilon=1$, $C_{\sigma \sigma \varepsilon}=0.5$.  The vertical axis indicates our numerical results. The four points were chosen randomly. The parameters used here are the same as in \figref{fig:Corr}.}
    \label{fig:Corr4PTAnalyticVsNumerical}
\end{figure}

 The essential data for Ising CFT are the scaling dimensions $\Delta_\sigma$, $\Delta_\varepsilon$, and the OPE coefficient $C_{\sigma \sigma \varepsilon}$.  In 2D CFT, the four-point function can be uniquely determined by the CFT data according to \eqref{eq:4ptFunction}.
Thus, we compute the four-point correlation function $\langle\sigma(x_1,y_1)\sigma(x_2,y_2)\sigma(x_3,y_3)\sigma(x_4,y_4)\rangle$ of the square lattice Ising model near the critical temperature. To confirm that we can extract the OPE coefficients, we first compare our numerical values of the four-point correlation function with the analytical CFT prediction, shown in \figref{fig:Corr4PTAnalyticVsNumerical}. The horizontal axis is the analytical value, and the vertical is the numerical value obtained from our TN procedure. One can see that, although the data is noisy, the trend indicates that they are proportional to each other up to some normalization factor,  resulting from the normalization of the spin operators. 

With the above comparison, we further use \eqref{eq:4ptFunction} for the fitting and thus extract the CFT data from our numerical results. We use the Huber loss \eqref{eq:HuberLoss} with $\epsilon=0.1$ to fit the curve in the logarithmic space, where the input data are the distances between pairs of points. The fitted values (vs. known values) are
$\Delta_\sigma=0.1120\, ({\rm vs.}\, 0.125)$, $\Delta_\varepsilon=0.9729\,({\rm vs.}\, 1)$, and $C_{\sigma \sigma \varepsilon}=0.4564\,({\rm vs.}\, 0.5)$, respectively. Thus,  four-point functions provide a crude estimation of all the essential data of Ising CFT and give another confirmation of our previous results using lTRG.

\subsection{Two-point Correlator on a Torus}
\label{sec:Results_2pt_torus}

\begin{figure}[h]
    \centering
    \includegraphics[width=\hsize]{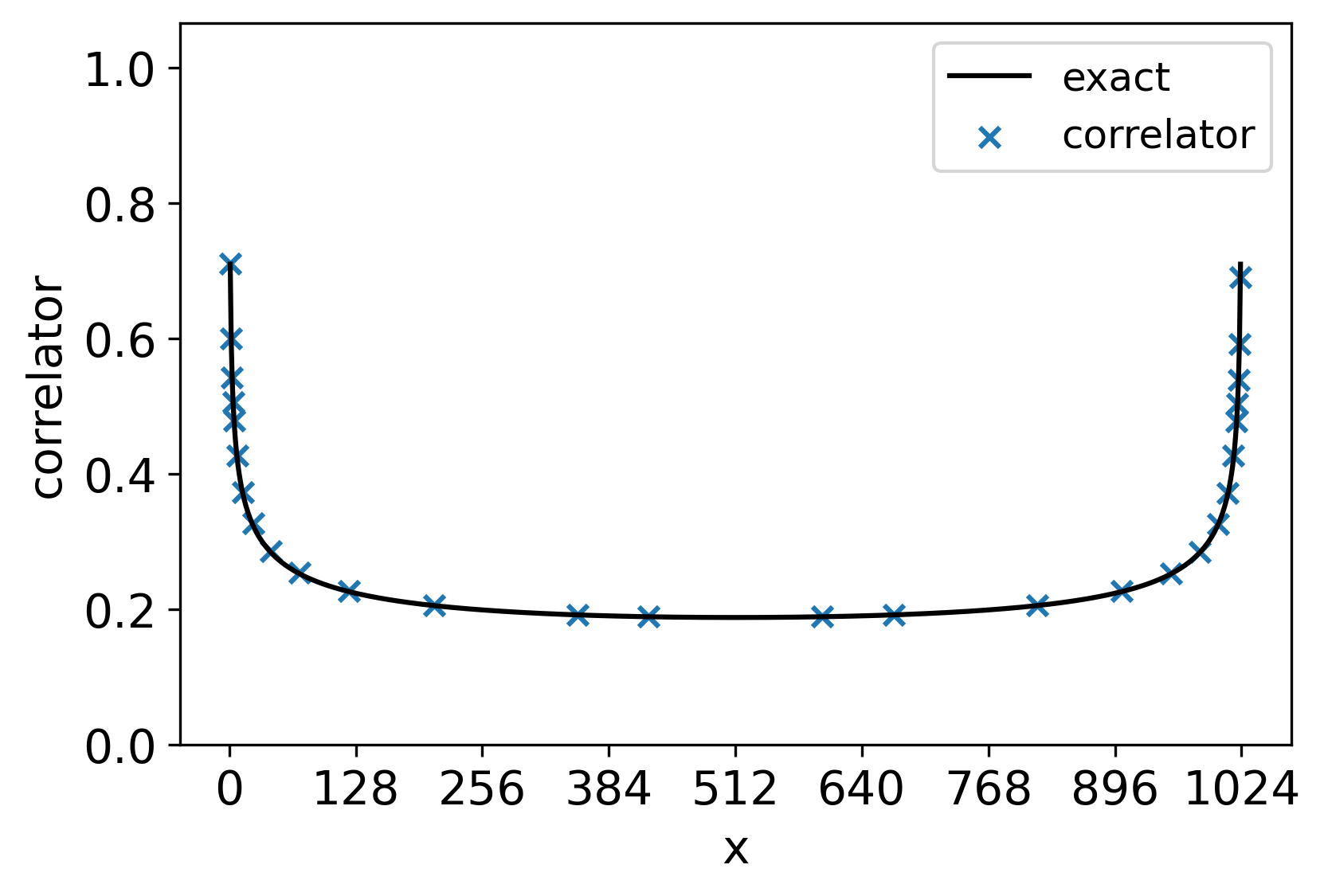}
    \caption{(color online) The two-point correlator $\langle\sigma(0,0)\sigma(x,0)\rangle$ of the Ising model at $T\approx T_c$ on a $1024\times1024$ torus. GILT is not used. The bond dimension is $\chi=24$. }
    \label{fig:Torus1024}
\end{figure}

We also compute the two-point correlation function $\langle\sigma(0,0)\sigma(x,0)\rangle$ on a $1024\times1024$ torus using the HOTRG method. The results are shown in Figure \ref{fig:Torus1024}.
\begin{figure}[h]
    \centering
    \includegraphics[width=\hsize]{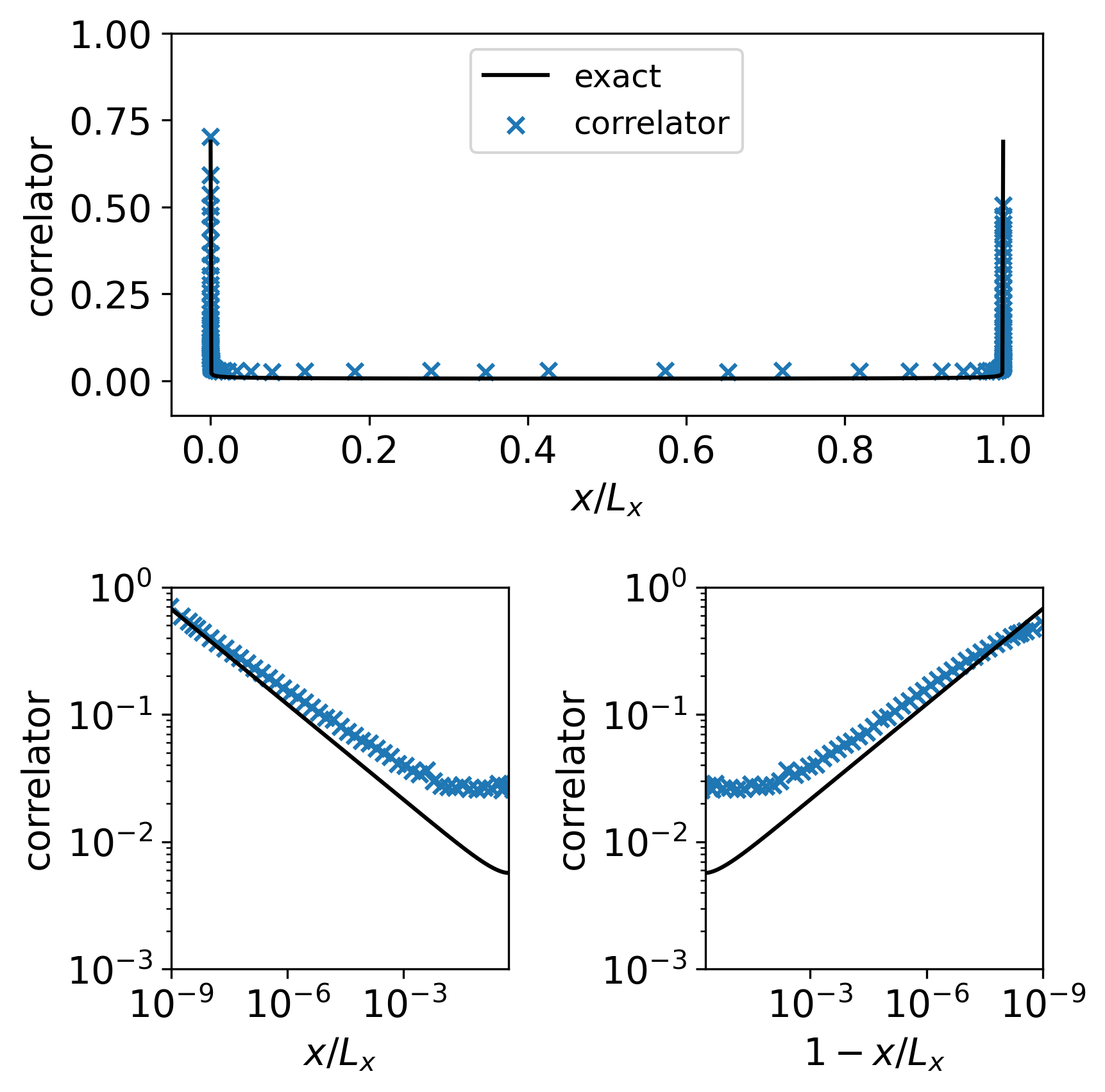}
    \caption{(color online) The two-point correlator $\langle\sigma(0,0)\sigma(x,0)\rangle$ of the Ising model at $T\approx T_c$ on a $2^{30}\times2^{30}$ torus. The data are plotted in linear and log-log space. Here, the same parameters are used as in \figref{fig:Torus1024}.}
    \label{fig:TorusBig}
\end{figure}
The result confirms well with the spin-spin correlation of the Ising CFT on a torus \cite{BigYellowBook} (12.108):
\begin{equation}
    \expval{\sigma(0,0)\sigma(x,y)}
    \propto
    \left| \frac{\partial_z \theta_1 (0|\tau) }{\theta_1(z|\tau)} \right|^{\frac{1}{4}}
    \frac{\sum_{\nu=1}^{4}|\theta_\nu(z/2|\tau)|}
    {\sum_{\nu=2}^{4}|\theta_\nu(0|\tau)|},
    \label{eq:torus_correction}
\end{equation}
where $z=\frac{1}{L_x}(x+iy)$ is the complex coordinate, $L_x$, $L_y$ are the size of the lattice,  $\tau=i L_y / L_x$, and  $\theta_\nu$'s are the elliptic theta functions, following the convention in \cite{mpmath-jacobi-theta}.  We only examine the case with $y=0$ here.

\figref{fig:TorusBig} shows the two-point function on a much larger torus, corresponding to 60 layers in the coarse graining. The correlator looks like the high-temperature correlation function as in \figref{fig:Corr} and \eqref{eq:ansatz_highT} from both sides. This is because the coarse-grained tensors we use had already flowed away from the conformal fixed point on the length scale $2^{15}$, as shown in \figref{fig:Tdiff} and \figref{fig:Corr}.

As expected by the periodic boundary condition, the curves in \figref{fig:Torus1024} and \figref{fig:TorusBig} both exhibit a peak when $x\rightarrow L_x$. However, the sharp peak in $x\rightarrow L_x$ seemingly contradicts the purpose of the coarse graining in HOTRG: removing microscopic details. The peak is the result of correlations between two point-like defects. However, in the case $x\rightarrow L_x$, the two defects meet in the very last step of the coarse-graining process. So, what was computed instead is the correlation function between two ``smeared'' operators, which gives a much smoother curve compared to that of the ``bare'' operators. Further discussions will be presented in \secref{sec:Discussion}.

The fact that a sharp peak can be obtained when $x\rightarrow L_x$ indicates that the truncated Hilbert space of the coarse-grained tensor still captures the modes in which the excitation is sharply localized on the boundary of the coarse-grained regions, as described in \secref{sec:Discussion}. Figure~\ref{fig:TorusBig} further illustrates that those point-like localized modes are kept throughout the entire RG flow.

\subsection{Comparison between Eigenvectors of linearized Tensor Renormalization Group with Coarse-Grained Defect Tensors}
\label{sec:Results_eigvecCompare}

\begin{figure}[h]
    \centering
    \includegraphics[width=1\hsize]{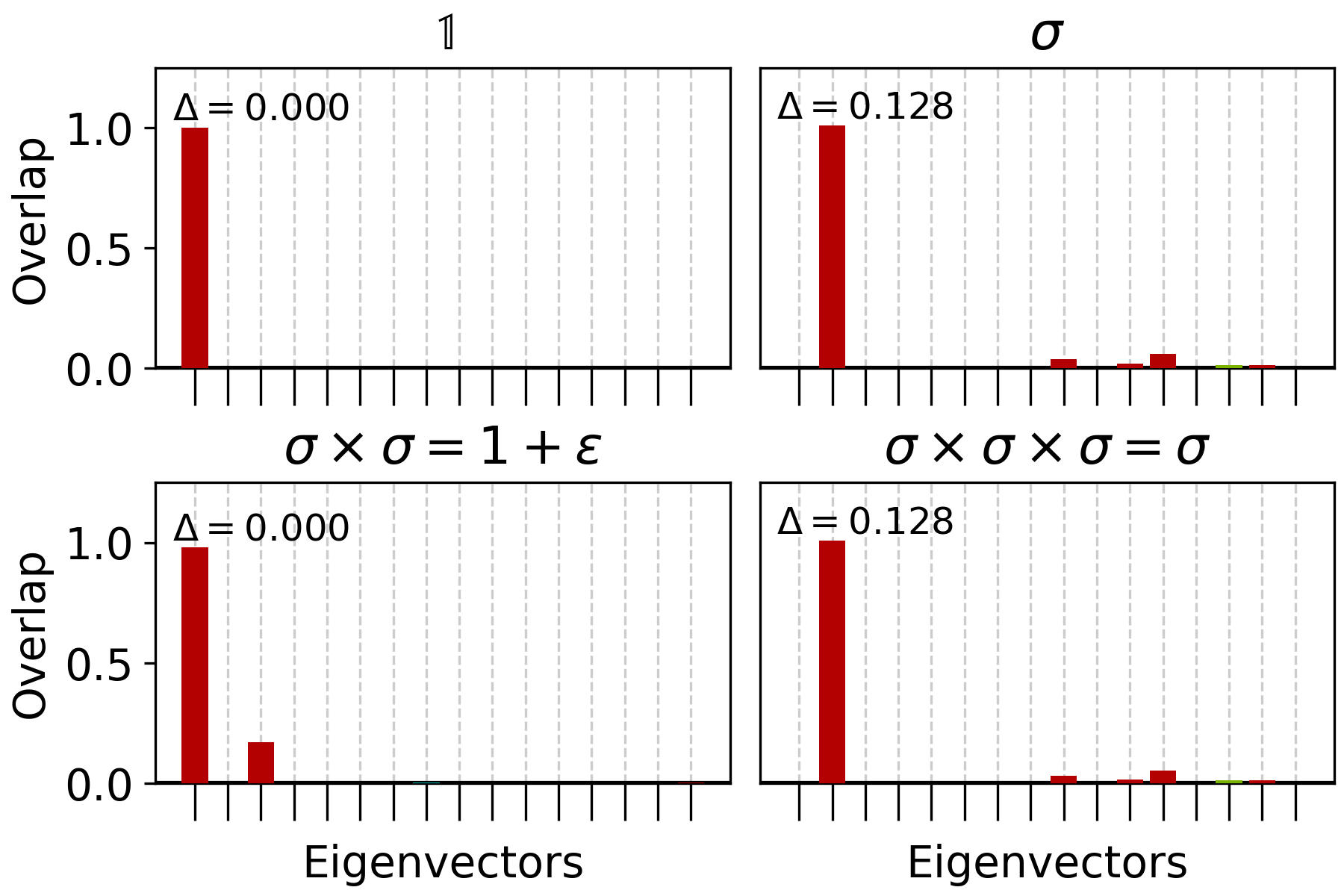}
    \caption{(color online) Projection of coarse-grained tensors with operator insertion onto the first few eigenvectors eigenvectors of lTRG at the RG step 30. The color indicates the argument $\theta$ of the overlapping $\braket{T_\text{lattice}}{T_\text{lTRG}}=re^{i\theta}$: red indicates positive values $\theta=0$ and blue indicates negative values $\theta=\pi$. The bond dimension used is $\chi=24$, and GILT is applied with $\epsilon_{\text{GILT}}=8\times 10^{-7}$, $\text{nIter}_{\text{GILT}}=1$. The results suggest that the eigenvectors correspond to states of conformal primaries and descendants. }
    \label{fig:NptFunction}
\end{figure}

As demonstrated in Sec.~\ref{sec:Results_ScDim} that lTRG via HOTRG+GILT+MCF gives good estimates of scaling dimensions, we now propose that additionally the eigenvectors of the lTRG coarse graining equation (\eqref{eq:lTRG}) at the fixed point correspond to the physical states of conformal primaries and their descendants. This claim is supported by a direct comparison between the coarse-grained tensors with operator insertion $T_{\sigma_{x_1}\sigma_{x_2}...}$ and the first few eigenvectors $v_i$ of the lTRG \eqref{eq:lTRG} near the conformal point, as shown in \figref{fig:NptFunction}.

From there we observe that the coarse-grained vacuum tensor $T_{\mathbb{1}}$ lies in the eigenspace with conformal dimension 0, which corresponds to the identity operator $\mathbb{1}$.
The coarse-grained tensor with an insertion of one operator $T_{\sigma_x}$ lies in the 0.125 conformal dimension eigenspace, which corresponds to the spin operator $\sigma$. However, as can be seen in Fig.~\ref{fig:NptFunction}, there are also significant projections on higher eigenvectors, which may be due to numerical errors and/or the operator insertion position not being exactly in the middle.
Next, the coarse-grained tensor with two operator insertions $T_{\sigma_{x_1}\sigma_{x_2}}$ lies in the joint subspace of $\Delta=0$ and $\Delta=1$. This is consistent with the fusion rule of the Ising CFT: $\sigma \times \sigma = \mathbb{1} + \varepsilon$, where $\varepsilon$ is the energy density operator with $\Delta_\varepsilon=1$.
Lastly,
the coarse-grained tensor with three operator insertions $T_{\sigma_{x_1}\sigma_{x_2}\sigma_{x_3}}$ further confirms the fusion rule $\varepsilon \times \sigma = \sigma$ and $\mathbb{1} \times \sigma = \sigma$.

However, we note that there are also undesired projections on eigenspaces with $\Delta \gtrsim 2$. This is likely due to limitations in the numerical precision or to the fact that the eigenspaces are not strictly orthogonal. 
We also note that the above comparison is done in the subspace spanned by the first few eigenvectors of lTRG, in order to avoid the UV details from lattice implementation and the ``junk'' part~\cite{yang2017loopTNR} that corresponds to the higher indices of the tensor.

We further compare the higher eigenvectors with the conformal descendants obtained by finite-differencing the coarse-grained tensors with different operator insertion positions. The scheme is elaborated in Table~\ref{tab:OperatorInsertion}. The procedure consists of inserting operators at different positions in the original lattice and then coarse-graining the resultant tensors. Finally, finite differences are taken between the coarse-grained tensors with operator insertions to obtain the lattice version of conformal descendants.

\begin{table}[h]
    \centering
    \begin{tabular}{|c|c|c|}
    \hline
        \parbox[t]{2.5cm}{Coarse-grained tensor} & $\Delta$ & \parbox[t]{4cm}{Inserted operator at lattice level} \\
    \hline
        $\mathbb{1}$    & 0    & $\mathbb{1}$ \\
        $\sigma$    & 1/8   & $\sigma_{0,0}$ \\
        $\sigma\times\sigma=\mathbb{1}+\varepsilon$    & 0,1   & $\sigma_{l,0}\,\sigma_{-l,0}$ \\
        $\sigma\times\sigma\times\sigma=\sigma$     & 1/8   & $\sigma_{l,0}\,\sigma_{-\frac12l,\frac{\sqrt{3}}{2}l}\,\sigma_{-\frac12l,-\frac{\sqrt{3}}{2}l}$ \\
    \hline
        $\partial_x \sigma$ & 1+1/8 & $(\sigma_{d,0}-\sigma_{-d,0})$ \\
        $\partial_y \sigma$ & 1+1/8 & $(\sigma_{0,d}-\sigma_{0,-d})$ \\
        $\partial_x \varepsilon$   & 2   & $(\sigma_{d,l}\,\sigma_{d,-l}-\sigma_{-d,l}\,\sigma_{-d,-l})$ \\
        $\partial_y \varepsilon$   & 2   & $(\sigma_{l,d}\,\sigma_{-l,d}-\sigma_{l,-d}\,\sigma_{-l,-d})$ \\
        $T_{\times}$ & 2 & $(\sigma_{l,l}\,\sigma_{-l,-l}-\sigma_{l,-l}\,\sigma_{-l,l})$ \\
        $T_{+}$ & 2 & $(\sigma_{l,0}\,\sigma_{-l,0}-\sigma_{0,l}\,\sigma_{0,-l})$ \\
        $\partial^2_x \sigma$   & 2+1/8 & $(\sigma_{d,0}-\sigma_{d',0}-\sigma_{-d',0}+\sigma_{-d,0})$ \\
        $\partial^2_y \sigma$   & 2+1/8 & $(\sigma_{0,d}-\sigma_{0,d'}-\sigma_{0,-d'}+\sigma_{0,-d})$ \\
        $\partial_x \partial_y \sigma$ & 2+1/8 & $(\sigma_{d,d}-\sigma_{-d,d}-\sigma_{d,-d}+\sigma_{-d,-d})s$ \\
    \hline
    \end{tabular}
    \caption{Coordinates and patterns of operator insertion for \figref{fig:NptFunction} and \figref{fig:FiniteDifference}. The stress energy tensor $T$ has spin 2, which is the descendent of $\mathbb{1}$, which is decomposed into its quadratic momentum components $T_+$ and $T_{\cross}$.}
    \label{tab:OperatorInsertion}
\end{table}

\begin{figure}[h]
    \centering
    \includegraphics[width=1\hsize]{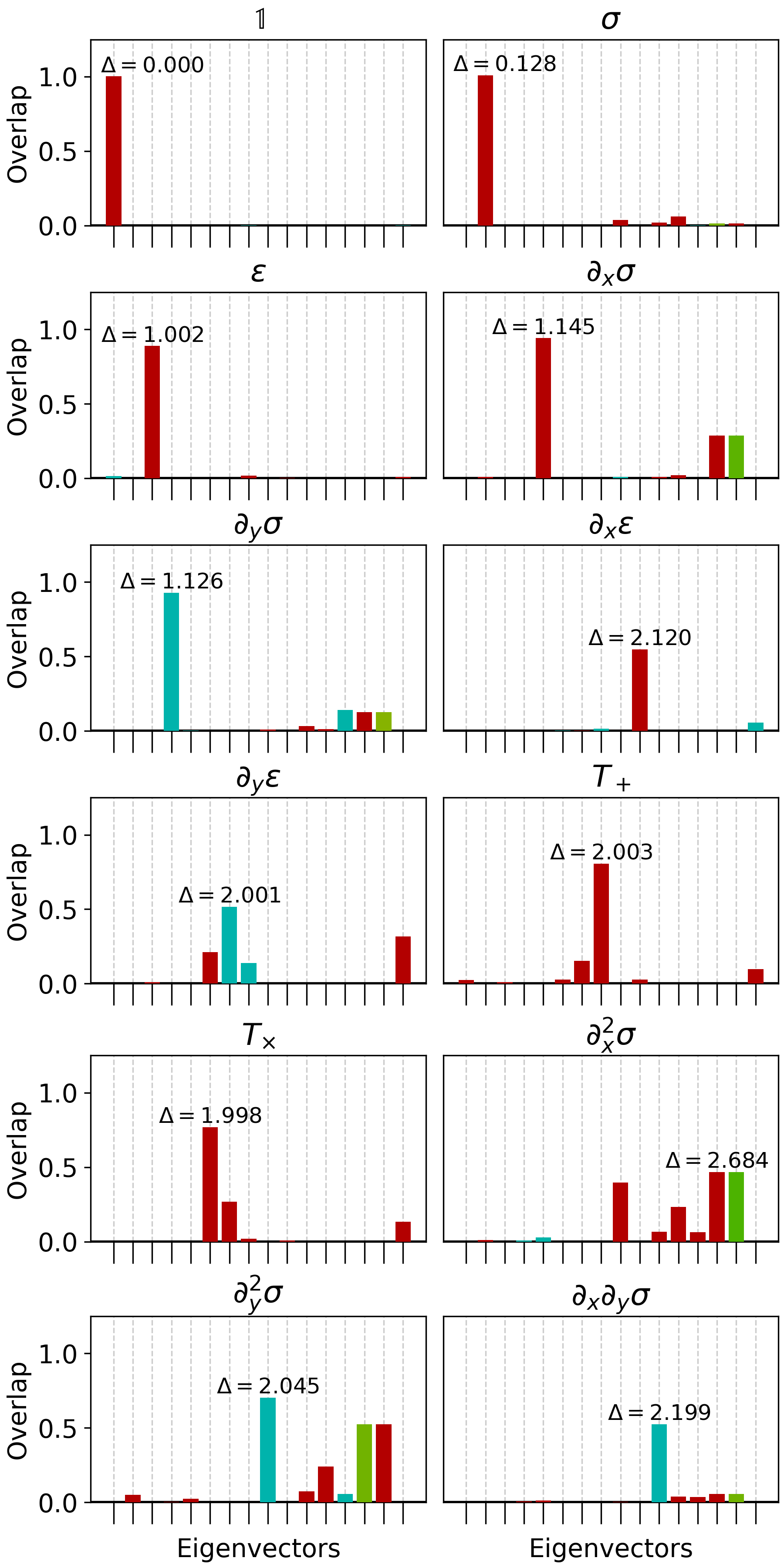}
    \caption{(color online) Finite difference of coarse-grained tensors onto eigenvectors of linearized TRG. The color indicates the argument $\theta$ of the overlapping $\braket{T_\text{lattice}}{T_\text{lTRG}}=re^{i\theta}$: red indicates positive values $\theta=0$ and blue indicates negative values $\theta=\pi$. Here, the same parameters are used as in \figref{fig:NptFunction}.}
    \label{fig:FiniteDifference}
\end{figure}

Figure~\ref{fig:FiniteDifference} displays the projection of the conformal descendants realized in the lattice onto the eigenvectors of lTRG. We observe that $\partial_x \sigma$ and $\partial_y \sigma$ belong to a nearly degenerate 2-dimensional subspace with a conformal dimension of approximately 1.125. Similarly, the other descendants, such as $T_{\times}$, $T_{+}$, $\partial_x \varepsilon$, and $\partial_y \varepsilon$ have their maximal overlap near conformal dimension 2. Meanwhile, the descendants $\partial_x^2 \sigma$, $\partial_x \partial_y \sigma$, and $\partial_y^2 \sigma$ appear near the subspace corresponding to $\Delta=2.125$. 

 However, we note that the order of the eigenvalues of the eigenvectors with the most significant overlap between $\partial_x \varepsilon$ and $\partial_x^2 \sigma$ $\partial_y^2 \sigma$ are not correct.  
 The most significant overlap of $\partial_x \varepsilon$ is at 2.120, which shows a great discrepancy with the overlap of $\partial_y \varepsilon$ between 1.998 and 2.003. Moreover, the most significant overlaps of the next two operators, $\partial_x^2 \sigma$ and $\partial_y^2 \sigma$, are at the scaling dimension 2.045, which is smaller than 2.120. So the order of the scaling dimensions of some operators is wrong, and operators at different degeneracy subspaces are being mixed up due to numerical error.
 
We also note that the pair $T_{\cross}$, $\partial_y \varepsilon$, and the pair $\partial_x^2 \sigma$, $\partial_y^2 \sigma$ are not very distinguishable at the first few eigenvectors. This indicates that the information to distinguish those states is lost during the coarse-graining. This results in the unexpected behavior of $\partial_y^2 \sigma$ in Fig.~\ref{fig:EncodePosition}, which will be discussed in the next subsection.

\subsection{Expansion of One-point Operators into the Conformal Family}
\label{sec:Results_encoding_position}

\begin{figure}[h]
    \centering
    \includegraphics[width=.9\hsize]{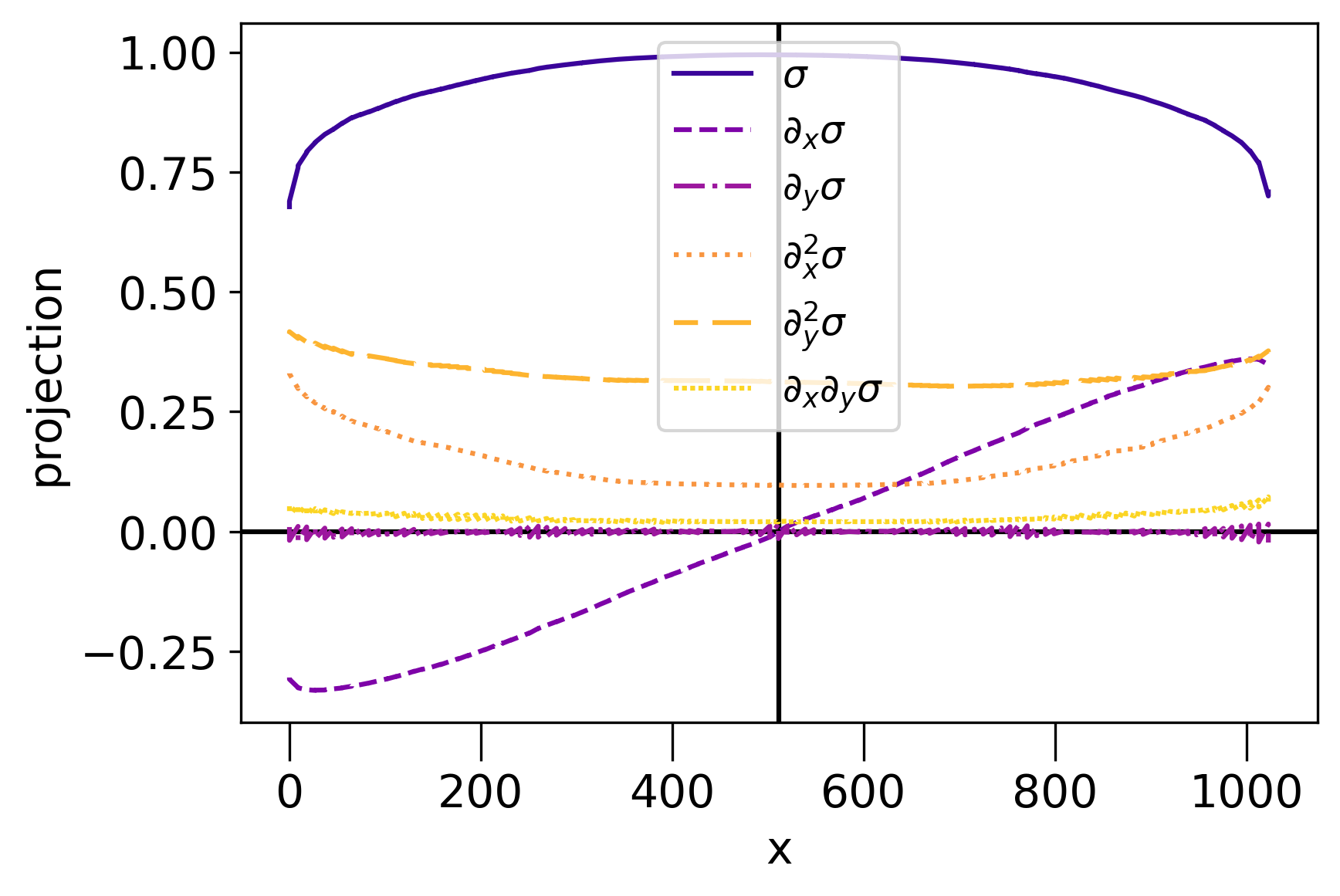}
    \caption{(color online) Projection of coarse-grained tensor with different operator insertion positions onto the conformal state obtained from linearized Tensor Renormalization Group. $L=1024$ is the lattice size. Same parameter as \figref{fig:NptFunction}}
    \label{fig:EncodePosition}
\end{figure}

\smallskip\noindent\textbf{Expansion on flat space}. 
Another way to show that tensors such as $\partial_x \sigma$ and $\partial_x \partial_x \sigma$ obtained from lTRG eigenvectors in the previous subsection have physical meaning just as their name suggests, i.e., one can check if the one-point tensor $T_{\sigma_{x,0}}$ can be Taylor expanded into the conformal family as follows,
\begin{equation}
    \label{eq:TaylorExpansion}
\begin{aligned}
    &\sigma(x,y)=\sigma(0,0)+x \partial_x \sigma(0,0)+y \partial_y \sigma(0,0) \\
    &\,\,+ \frac{1}{2}(x^2 \partial_x^2 \sigma(0,0) + xy \partial_x \partial_y \sigma(0,0) + y^2 \partial_y^2 \sigma(0,0))+...
\end{aligned}
\end{equation}
We thus compute operators on both sides and compare whether the above expansion holds. Figure~\ref{fig:EncodePosition} shows the decomposition of $T_{\sigma_{x,0}}$ onto the conformal tensors $\sigma$, $\partial\sigma$, $\partial^2\sigma$. 
We see that, as the operator $\sigma(x)$ moves along the x-axis, its projection on $T_{\partial_x \sigma}$ changes linearly and its projection on $T_{\partial_x^2 \sigma}$ changes quadratically. Also, its projection on $T_{\sigma}$ remains almost constant, except near the edge. This confirms the Taylor expansion shown in \eqref{eq:TaylorExpansion}. 
However, the unexpected behavior of projections in $T_{\partial_y^2 \sigma}$ might be a result of a numerical error. From \figref{fig:FiniteDifference}, one can see that the lattice tensors of $\partial_x^2 \sigma$ and $\partial_y^2 \sigma$ are not quite distinguishable from the projection onto the first few eigenvectors of lTRG.  Therefore, one can conclude that the coarse-graining process erroneously removes the information that is crucial to distinguish the two operators. This results in the unexpected value $\partial_y^2 \sigma$ in \figref{fig:FiniteDifference}.

The suppression of $T_{\sigma}$ near the edges of the coarse-grained block $x\rightarrow 0$ and $x\rightarrow 1023$ may be the result that GILT artificially suppresses defects near the edges and corners of the coarse-grained block, which will be discussed in \secref{sec:Discussion_RemoveEdgeModes}.

We do not directly compare $T_{\sigma_{x,0}}$ with coarse-grained tensors $T_{\sigma}$, $T_{\partial\sigma}$... It is because, as suggested by \cite{yang2017loopTNR}, the higher index components of the coarse-grained tensors, or equivalently, the subspace spanned by less significant eigenvectors, suffer substantially from the unphysical `noise' of bond dimension truncation and numerical errors. So we only did the comparison in the subspace spanned by the most significant eigenvectors.
 
\smallskip\noindent\textbf{Expansion on a cylinder}.
In parallel with the eigenvalues of lTRG, it is well-known that scaling dimensions can also be extracted from the eigenvalues of the transfer matrix. In particular, we consider the periodic boundary condition where the transfer matrix is a chain of $N=2$ coarse-grained tensors:
\begin{equation}
\label{eq:transfer_matrix_N2}
    \vcenter{\hbox{\includegraphics[page=56]{TRGCFTGraphics.pdf}}}
\end{equation}
At the radial quantization, one would expect the following expansion,
\begin{equation}
\label{eq:cyl_expansion}
    \sigma(r_0,\theta)\sim \sigma+e^{i\theta} \partial\sigma+e^{-i\theta}\bar\partial\sigma +... ,
\end{equation}
where $\sim$ indicates omitting the coefficient at each term. 

\begin{figure}[h]
    \centering
    \includegraphics[width=.9\hsize]{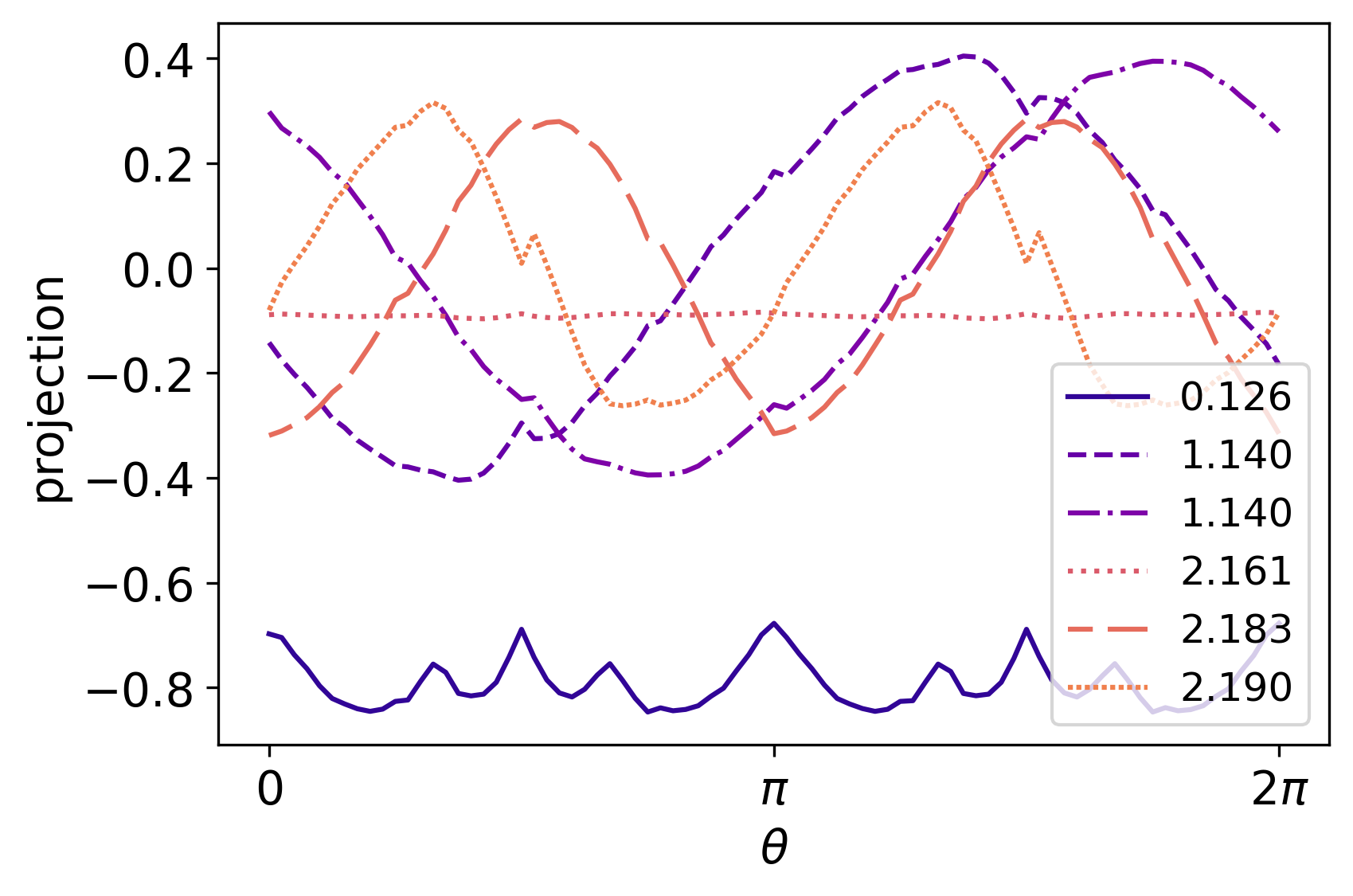}
    \caption{(color online) Projection $\bra{O_i}\tr (T T)_{\sigma_{(r_0,\theta)}}\ket{0}$ of a coarse-grained tensor with different operator insertion positions onto the conformal state obtained from transfer matrix method. $\bra{O_i}$'s are the eigenvectors of the transfer matrix on a cylinder \eqref{eq:transfer_matrix_N2}, $\ket{0}$ is the ground state of the transfer matrix, and $\tr (T T)_{\sigma_{(r_0,\theta)}}$ is a cyclic chain of two coarse-grained tensors, with a defect on one of the tensor, depending on the position of $\theta$. The eigenvectors $\mathcal{O}_i$'s are labeled by their estimated scaling dimensions. }
    \label{fig:EncodePosition_cyl}
\end{figure}

Figure~\ref{fig:EncodePosition_cyl} compares the eigenvectors of the transfer matrix \eqref{eq:transfer_matrix_N2} and the lattice state associated with the placement of a defect at $(r_0,\theta)$. The lattice state is obtained by applying a chain of $N=2$ coarse-grained tensors $\tr (T_{\sigma_{(r_0,\theta)}} T_0)$ (or $\tr (T_0 T_{\sigma_{(r_0,\theta)}})$) to the ground state $\ket{0}$ of the transfer matrix. The size of the coarse-grained block is $L/2$. At $\theta<\pi$, the defect is in the first tensor. At $\theta\ge\pi$, the defect is in the second tensor.

In \figref{fig:EncodePosition_cyl}, one can see that the eigenvector of the scaling dimension $\Delta\approx0.125$, which corresponds to $\sigma$, is approximately constant in expansion \eqref{eq:cyl_expansion}; the eigenvectors of the scaling dimension $\Delta\approx1.125$, which correspond to $\partial\sigma$ and $\bar\partial\sigma$, have a sine and cosine coefficients in the expansion; similarly, eigenvectors associated with $\Delta\approx2.125$ have coefficients $1$, $\sin(2\theta)$ and $\cos(2\theta)$, respectively. This observation confirms that, as in lTRG, the eigenvectors of the transfer matrix can also be seen as conformal states.

\section{Further discussions}
\label{sec:Discussion}

In this section, we evaluate the capabilities and limitations of our adopted coarse-graining scheme for tensor networks on a lattice with point-like defects, consisting of HOTRG coarse-graining, GILT local entanglement filtering, and MCF gauge fixing as outlined in \secref{sec:CoarseGrainingScheme}.

Initially, we focused on improving our numerical method without the presence of defects, as having a good vacuum tensor could facilitate further computations with defects. Following the work of~\cite{lyu2021lTRG}, we have analyzed the change in the tensor components (see the result presented earlier in Fig.~\ref{fig:Tdiff}) and the scaling dimension spectrum (see Table~\ref{tab:scdim}). Here, our goal is to confirm the crucial role of GILT in achieving stable RG flow and show that the MCF gauge fixing method further enhances stability by showing the per-component tensor differences, the scaling dimension spectrum, and the central charge (extracted via the transfer matrix method) throughout the RG flow in \secref{sec:Discussion_TdiffFlow}. 

We also demonstrate how GILT effectively suppresses the accumulation of the CDL tensor by comparing the transfer matrix spectrum between even and odd RG steps in \secref{sec:Discussion_RemoveCDL}. (Note that in our HOTRG procedure, we define each step of the RG as coarse-graining two tensors either horizontally to vertically, reducing the effective number of sites by half.) However, in \secref{sec:Discussion_ErrorFlow}, we show that even with the GILT truncation implemented, the error in the scaling dimensions from the transfer matrix spectrum and the lTRG spectrum continues to accumulate along the coarse grain flow. This eventually results in the system flowing away from the conformal fixed point.

With a good vacuum tensor at hand, a natural next step is to study the two-point function. However, our initial ``on-paper'' discussion in \secref{sec:Discussion_GILTProblemDefect} suggests that GILT might introduce a significant error when calculating correlation functions with more than one defect point, compared to the vanilla HOTRG. Surprisingly, our actual numerical calculations in \secref{sec:Discussion_Compare2PTGILT} show that GILT can achieve results in two-point functions as well as the vanilla HOTRG, as long as the two points are randomly sampled within the coarse graining region, and outliers are properly treated. However, if the points are located at the corner or edge of the coarse-graining region, GILT introduces a significantly larger error.

Further evidence shows that corners and edges are special places that can be seen in the two-point function on a torus, as discussed in Sec.~\ref{sec:Discussion_RemoveEdgeModes}. Only without GILT and when the two operators are located at the corners of the coarse-grained region can the two distantly separated defects give a strong correlation value when they are reunited after the region is wrapped around the torus, under the periodic boundary condition.

The above clue strongly suggests the smearing of point-like operators due to coarse graining, which is a consequence of removing short-distance details in traditional RG techniques. For a pair of smeared operators, when close to each other, they yield a much lower and smoother peak compared to their unsmeared counterparts. In the vanilla HOTRG, the modes at the edges and the corners experience less smearing, because of the algorithm's preference for retaining more information about the CDL tensors. This explains why it produces a significant correlation function when two corner defects reunite after the region is wrapped around the torus. In contrast, GILT introduces some suppression of the boundary modes, which explains why it yields a much smaller correlation function when the two points are located on the boundary of the coarse-graining region.

Suppose that the smearing radius is correlated with the size of the coarse-graining block; when the operators are randomly sampled within the region, it is less likely for them to be within each other's smearing radius. This results in a more reliable calculation of the correlation function.

To verify our conjecture, in \secref{sec:Discussion_SmearingExplained}, we compare the correlation function between points with different degrees of smearing. The more coarse grain two defect points have experienced before they fuse into one tensor, the more smearing we believe they have suffered. We find that the correlation for points at the same coarse-graining level before fusion aligns well with the smeared two-point function between two Gaussian profiles. The outlier in our data confirms our conjecture that GILT and non-GILT treat boundary modes differently.

We further investigated the relationship between the smearing radius and the size of the coarse-graining blocks. Unfortunately, we find that the smearing radius grows faster than the size of the coarse-graining blocks. This implies that as the coarse-graining process continues, the smearing radius eventually overtakes the block size, leading to an increasing loss of information in the correlation functions.

In summary, we conclude that it is possible to evaluate $N$-point functions using HOTRG+GILT. To avoid the issue of operator smearing, it is necessary to sample the points randomly in the bulk, avoid the edges and corners of the coarse-grained regions, and filter out the outliers, e.g., using Huber loss. GILT introduces some artificial suppression for operators at the boundary, but this minor flaw can be ameliorated as discussed above and is compensated by the method's ability to stabilize the RG flow and facilitate the extraction of the fixed-point tensor.

\subsection{Improved Stability of RG Flow with GILT and MCF}
\label{sec:Discussion_TdiffFlow}

\begin{figure}[h]
    \centering
    \includegraphics[width=\linewidth]{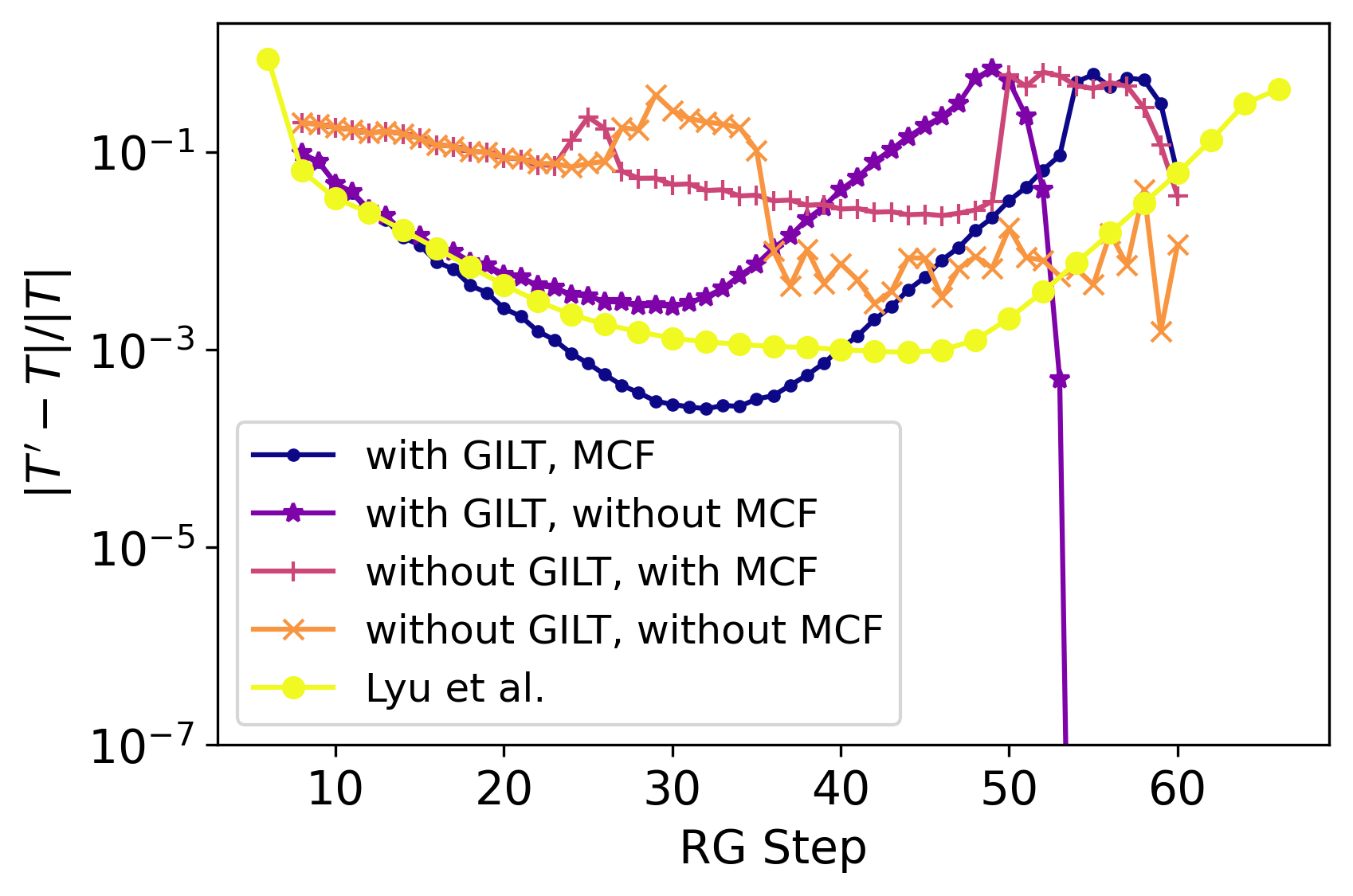}
    
    \caption{(color online) Comparison of HOTRG with and without GILT, with and without MCF by the difference per component of the coarse-grained tensors in successive steps of RG. The $U(\chi)$ gauge-fixing procedure (see \appendixref{sec:Fix_U1_alg}) is always performed. The yellow (lightest in greyscale) line with large circle markers is the result of running the code of~\cite{lyu2021lTRG} at $\chi=24$, adjusting the number of RG steps and tensor normalization to our definition. The bond dimension we used here is $\chi=24$,  and some GILT parameters are: $\epsilon_{\text{GILT}}=8\times 10^{-7}$ and $\text{nIter}_{\text{GILT}}=1$.}
    \label{fig:CompareGilt}
\end{figure}

\begin{figure*}[hbt]
    \centering
    \includegraphics[width=\linewidth]{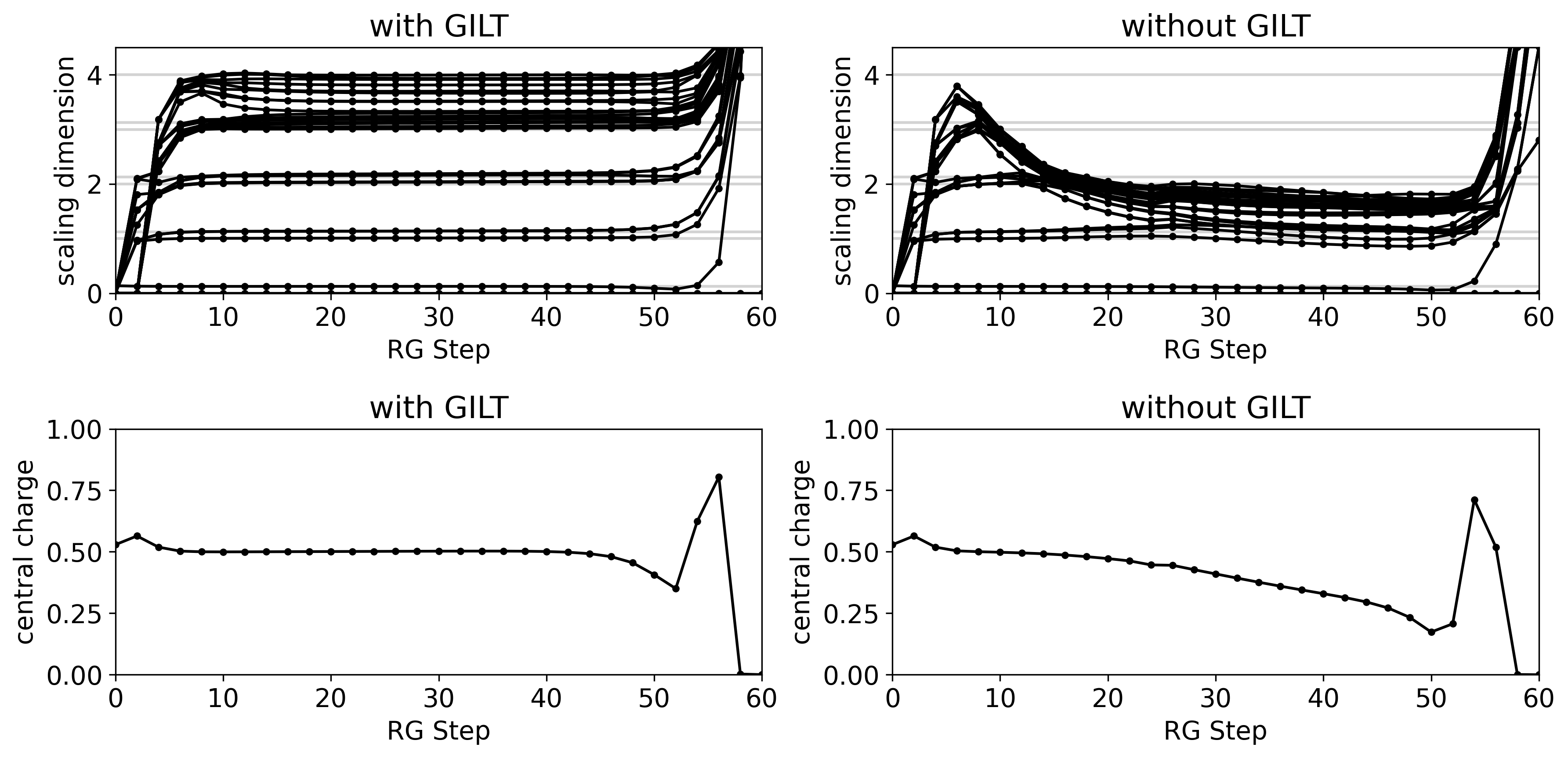}
    
    \caption{Comparison of scaling dimensions and central charge obtained by diagonalizing the transfer matrix, as well as their difference compared to analytic results at the conformal point, via HOTRG with and without GILT. The parameters are the same as in \figref{fig:CompareGilt}.}
    \label{fig:CompareGiltScdim}
\end{figure*}

The use of GILT, as discussed in Ref.~\cite{hauru2018GILT} and briefly reviewed in Sec.~\ref{sec:GILT}, is critical to ensure the stability of the RG flow. As shown in \figref{fig:CompareGilt}, GILT enables the coarse-grained tensor to converge to a fixed point tensor, with the change in coarse-grained tensors decreasing exponentially after each RG iteration. On the contrary, without GILT, the change in coarse-grained tensors remains constant, indicating either continued system evolution during the RG steps or the failure of the gauge-fixing procedure due to degeneracy in spectrum decomposition (see \appendixref{sec:fix-rotation}). The use of MCF furthermore increases stability. As shown in \figref{fig:CompareGilt}, with GILT and without MCF, the stability of our RG flow is comparable to Ref.~\cite{lyu2021lTRG}. With the introduction of MCF, the stability of our RG flow has improved further.

Figure~\ref{fig:CompareGiltScdim} illustrates the scaling dimensions and central charge extracted from the spectrum of the transfer matrix of the coarse-grained tensors. The method is described in \appendixref{sec:Appendix_scdim_TM}. With GILT, the scaling dimensions and the central charge remain consistent throughout most of the RG flow. In contrast, without GILT, information about the scaling dimension is lost at the beginning of the coarse-graining process, and the central charge is less stable during the flow.

\subsection{Finite Size Effect}
\label{sec:Finite_Scale_Effect}

\begin{figure*}[hbt]
    \centering
    \includegraphics[width=\linewidth]{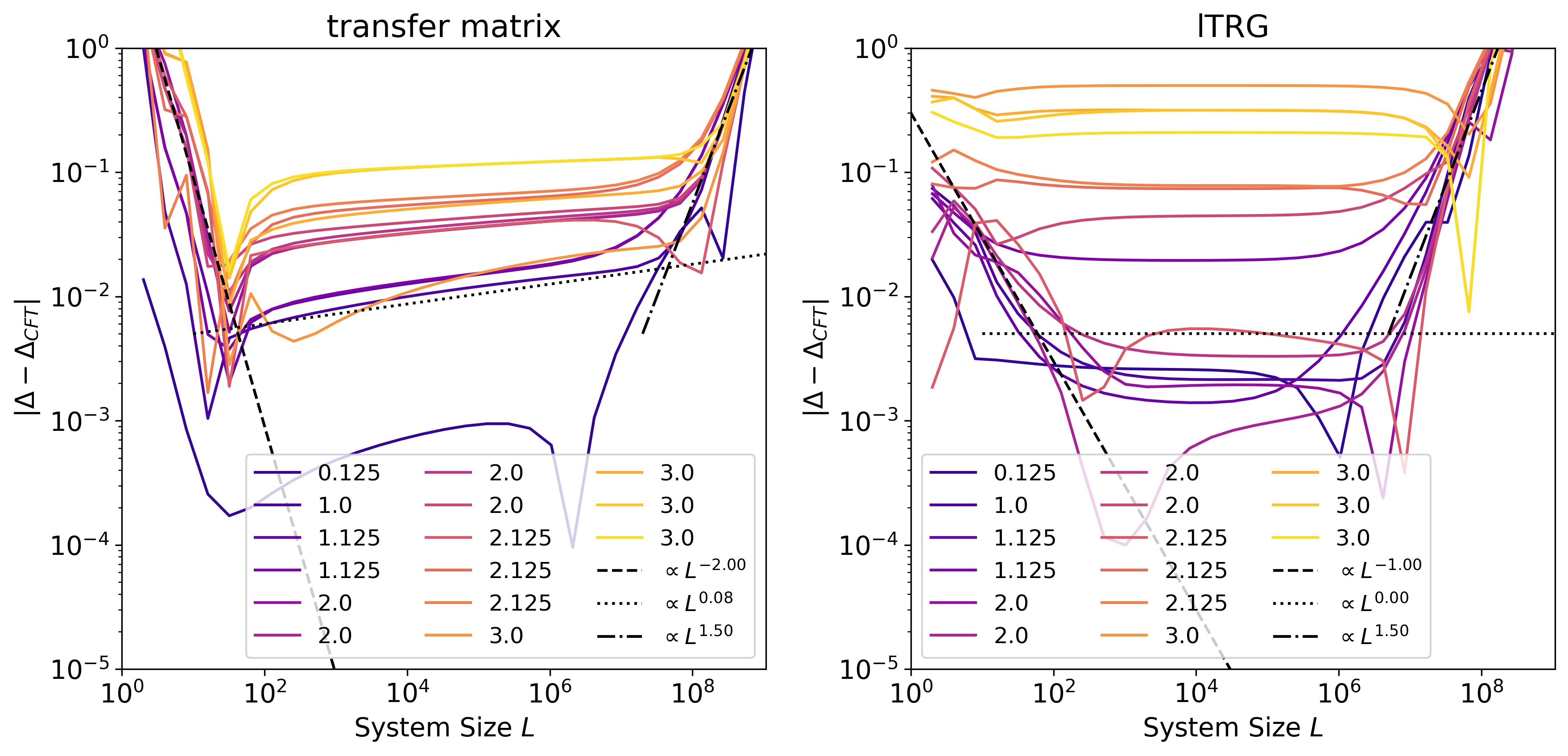}
    
    \caption{(color online) Shift of the scaling dimension in a finite-size lattice. The tensor is obtained using HOTRG, with GILT and MCF, with bond dimension $\chi=24$, $\epsilon_{\text{GILT}}=8\times 10^{-7}$, and $\text{nIter}_{\text{GILT}}=1$. The scaling dimension is obtained by diagonalizing the transfer matrix and the lTRG.}
    \label{fig:FiniteScaleScd}
\end{figure*}

At the critical point, but with a finite-size lattice, the extracted scaling dimension is subject to both finite-size and finite-bond-dimension effects, as depicted in Fig.~\ref{fig:FiniteScaleScd}. At the beginning of the RG flow, the shift in the scaling dimension $\Delta-\Delta_\text{CFT}$, as estimated from the transfer matrix, scales proportionally to $L^{-2}$, where $L$ is the size of the system. This behavior aligns with the expectations for the finite-size effect, as discussed in Ref.~\cite{ueda2023finite}.  However, the shift estimated from the lTRG method is $L^{-1}$.

During the RG flow, when the system reaches a relatively stable state in the ``valley'', the shifts obtained from both the transfer matrix and the lTRG method exhibit a scaling behavior of $L^{0.08}$ and $L^{0}$, respectively. As the system deviates from the RG fixed point, both methods produce shifts that scale as $L^{1.5}$.

In contrast to the findings in Ref.~\cite{ueda2023finite}, our observations do not reveal the emergence of perturbations such as $\epsilon$ (scaling as $L$) and $\sigma$ (scaling as $L^{3.75}$) at larger length scales due to finite-$\chi$ effects. This discrepancy may be attributed to the difference that Ref.~\cite{ueda2023finite} employed the Loop-TNR method, whereas we utilized the HOTRG+GILT scheme.

\subsection{Removal of CDL Tensors}
\label{sec:Discussion_RemoveCDL}

\begin{figure*}[hbt]
    \centering
    \includegraphics[width=\linewidth]{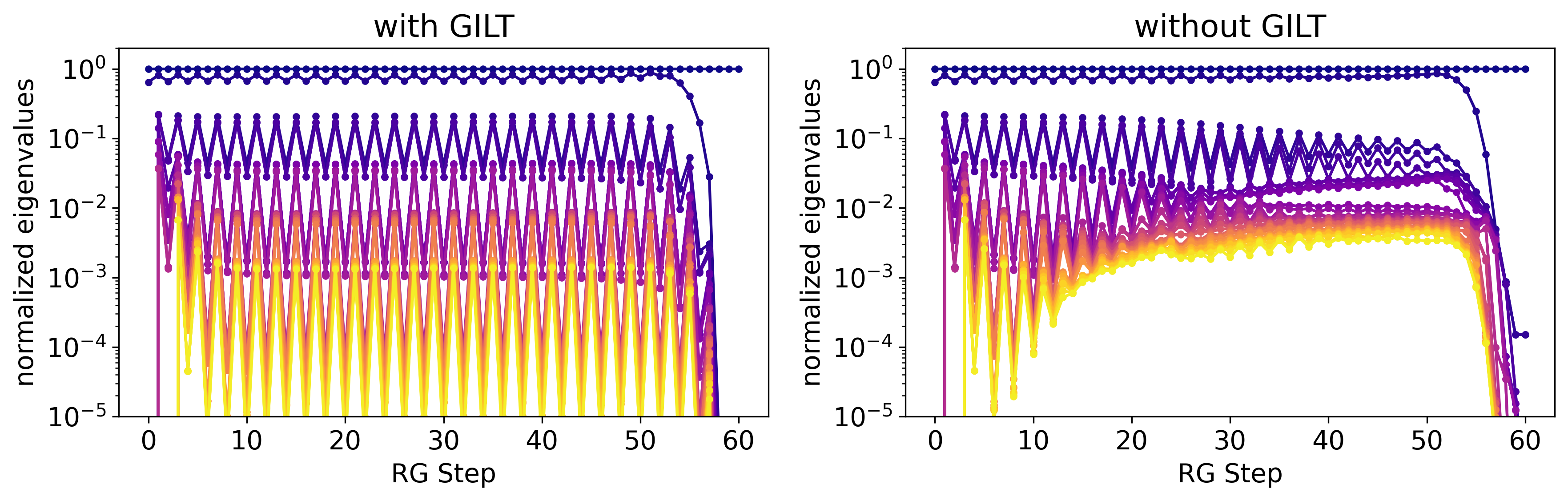}
    \caption{(color online) The eigenvalues $s^{(l)}_i$ of the transfer matrix of coarse-grained tensors with and without GILT, normalized by the largest eigenvalue $s^{(l)}_0$. At even RG steps, the coarse-grained region is a square, and at odd RG steps, the coarse-grained region is a 2:1 rectangle. This results in an oscillation of the eigenvalues of the transfer matrix, as shown in the graph. Without GILT, the oscillation decays after a few RG steps, indicating that the tensor network behaves like CDL tensors. The parameters used here are the same as those in \figref{fig:CompareGilt}.}
    \label{fig:CompareGILTEigs}
\end{figure*}

As suggested in Ref.~\cite{hauru2018GILT}, the accumulating error during coarse graining without GILT may be due to the accumulation of CDL tensors during the RG process (i.e., after vertical and horizontal graining steps, respectively), which consume all the bond-dimension resources at the end. To investigate this, we compare the spectrum $s^{(l)}_i$ of the transfer matrix between even and odd RG steps, normalized by the largest eigenvalue $s^{(l)}_0$. At odd RG steps, the coarse-grained region of each tensor is a 2:1 rectangle compared to the square at even RG steps. Therefore, we expect $(s^{(odd)}_i)^2\approx s^{(even)}_i$ near the fixed point. However, for the fixed point of the CDL tensor network, we expect $s^{(odd)}_i\approx s^{(even)}_i$. Figure~\ref{fig:CompareGILTEigs} shows the oscillation of the eigenvalues of the transfer matrix between odd and even RG steps. Without GILT, the oscillation decays during the coarse-graining process, indicating that the tensor network behaves more like CDL tensors after coarse-graining.
This result confirms the capability of GILT to remove CDL tensors.

\subsection{Accumulation of Error from Truncations}
\label{sec:Discussion_ErrorFlow}

\begin{figure*}[hbt]
    \centering
    \includegraphics[width=1\hsize]{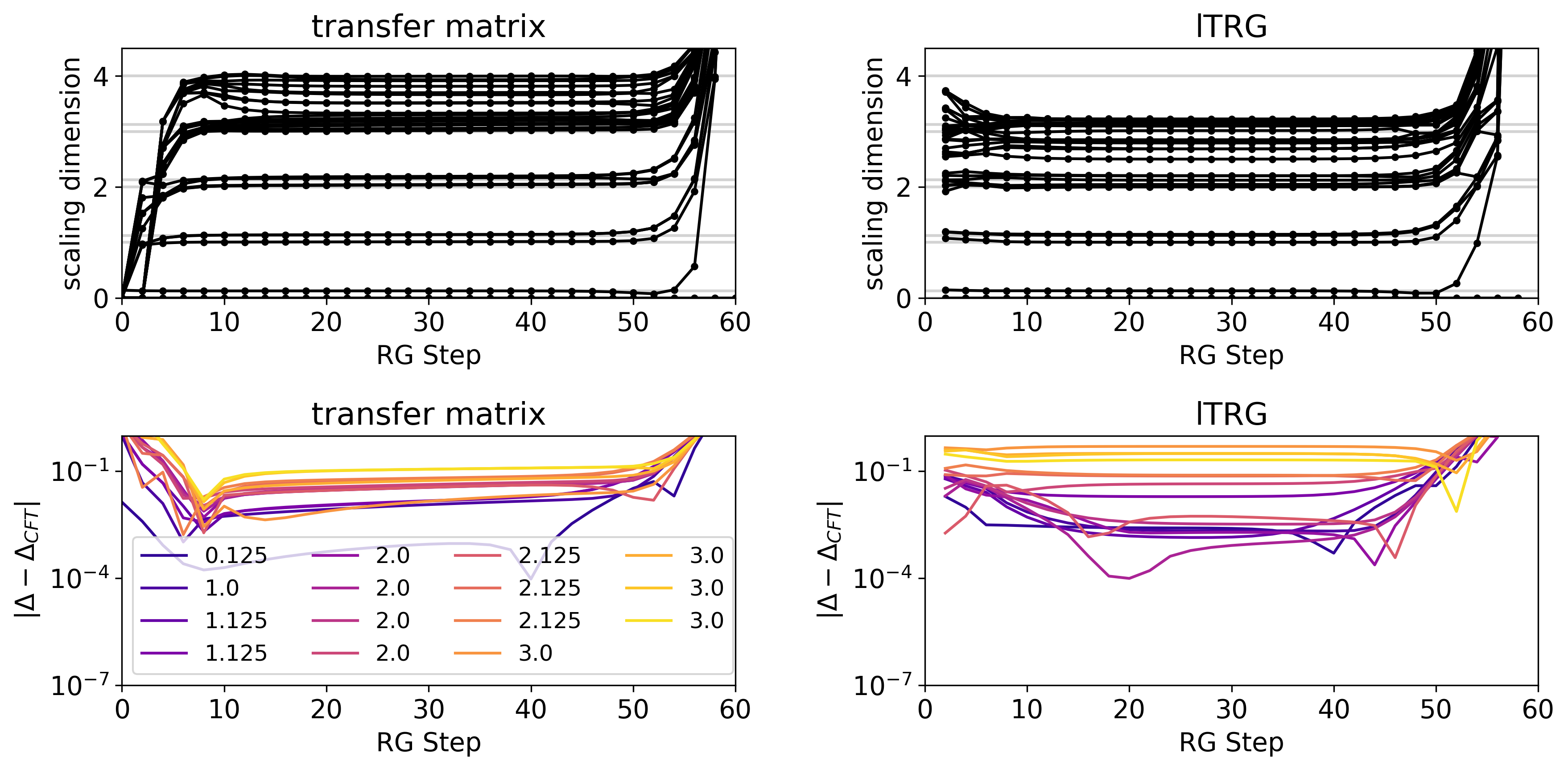}
    \caption{(color online) The flow of scaling dimensions obtained from the transfer matrix and lTRG methods, and their absolute error compared to the Ising Conformal Field Theory $\mathcal{M}(4/3)$. Grey straight lines (in the background) indicate the analytic results. The bond dimension is $\chi=24$, and GILT is applied with $\epsilon_{\text{GILT}}=8\times 10^{-7}$ and $\text{nIter}_{\text{GILT}}=1$. The critical temperature $T_c=0.44068381309509275$ is numerically determined with respect to the current bond dimension.}
    \label{fig:flow_scd}
\end{figure*}

The scaling dimensions error obtained from the transfer matrix and from the lTRG method is summarized in \figref{fig:flow_scd}. The error in the transfer matrix spectrum increases monotonically because, in principle, one can extract the precise scaling dimension from the initial tensor before coarse-graining if the chain length is infinitely long. Thus, when the chain length is long enough to suppress finite-size effects, the only source of error is the truncation error after each RG step, which accumulates throughout the RG flow. This accumulation starts far before the system reaches its proximity to the conformal fixed point (around step 30). 

However, the error in the linearized Tensor Renormalization Group generally improves as the system approaches the conformal fixed point. This improvement is due to the scaling invariance on which the lTRG equation is based, which emerges at the continuous limit. Thus, as the system becomes more conformal, the lTRG method becomes more accurate, leading to a reduced error in the calculated scaling dimensions.

In \figref{fig:Tdiff}, one can see that at approximately the critical temperature, the coarse-grained tensor will still flow away from the fixed point tensor. This can be due to floating-point errors or to the imprecision of the critical temperature we found. Note that due to the finite dimension of the bonds, we do not just use the exact $T_c$ value but optimize the temperature using the method described in Ref.~\cite{lyu2021lTRG} to find the corresponding effective $T_c$. We also want to emphasize that the truncation from HOTRG and GILT introduces an artificial length scale that breaks conformal symmetry.  This might be the reason for the instability of the critical fixed-point tensor and the shift of the critical temperature.

\subsection{Theoretical Insufficiency of HOTRG and GILT in Coarse-graining the Lattice with Defects}
\label{sec:Discussion_GILTProblemDefect}

Most prior use cases in HOTRG and GILT have focused on translational invariant tensor networks, where the same tensor is placed at all the lattice sites. These studies typically consider the partition function/free energy without any defects/impurity operators inserted. The coarse-graining tensor network, with insertions $w$, $g$, and $h$, can be seen as a tree-like tensor network ansatz that tries to approximate the ground-state wavefunction ($g$, $h$ are omitted) or the partition function:
\begin{equation}
\label{eq:MERA_like_ansatz}
    \vcenter{\hbox{\includegraphics[page=53]{TRGCFTGraphics.pdf}}}\rightarrow \vcenter{\hbox{\includegraphics[page=54]{TRGCFTGraphics.pdf}}}
\end{equation}

In this section, we discuss how the vanilla HOTRG and HOTRG+GILT may respond if we replace certain tensors in the MERA-like coarse-graining scheme with defect tensors, i.e., we change a few of the tensors $T$ in \eqref{eq:MERA_like_ansatz} to defect tensors such as $T_\sigma$, but keep the $w$, $g$, $h$ invariant.
HOTRG can produce good fidelity of one-point functions as long as the bond dimension is sufficiently large for the fidelity of the `vacuum' partition function (or free energy). Note that the HOTRG equation~\eqref{eq:HOTRG_equation} holds if both $T_1$ and $T_2$ are vacuum tensors. Therefore, the environment $E$ of a defect tensor $T_{\mathcal{O}}$, which is the contraction of the tensor network with $T_{\mathcal{O}}$ removed and only contains vacuum tensors, can be faithfully computed using the HOTRG coarse grain method.

What about two-point functions and higher? The environment can still be obtained from coarse-graining the remaining vacuum tensor. The source of error comes from the insertion of $w^\dagger w$ between two coarse-grained defect tensors. Note that only one side of $w^\dagger w \simeq \mathbb{1}$ is connected to a vacuum tensor, sufficient for the HOTRG equation to hold.

The GILT equation \eqref{eq:GILT} holds only when the tensors $ABCD$ in the subnetwork \eqref{eq:GILT} are all vacuum tensors. One may ask whether the Hilbert space that GILT truncates is the subspace corresponding to the defect tensors. This conjecture is supported by the results shown in Fig.~\ref{fig:CorrCompareBlockSize} showing that HOTRG + GILT gives a lower estimation of the two-point correlation function.

 In calculating the two-point function, the HOTRG equation only fails when the two coarse-grained defects meet each other in the coarse-graining. On the contrary, the GILT equation introduces some errors at each coarse-grained level. This suggests that GILT introduces more errors in two-point functions. This is consistent with what we have observed. Nevertheless, we will demonstrate that it is still possible to make HOTRG+GILT yield as good results as the vanilla HOTRG for correlation functions.

\subsection{Comparison of Two-point Correlation Functions with and without GILT}
\label{sec:Discussion_Compare2PTGILT}

\begin{figure}[h]
    \centering
    \includegraphics[width=\hsize]{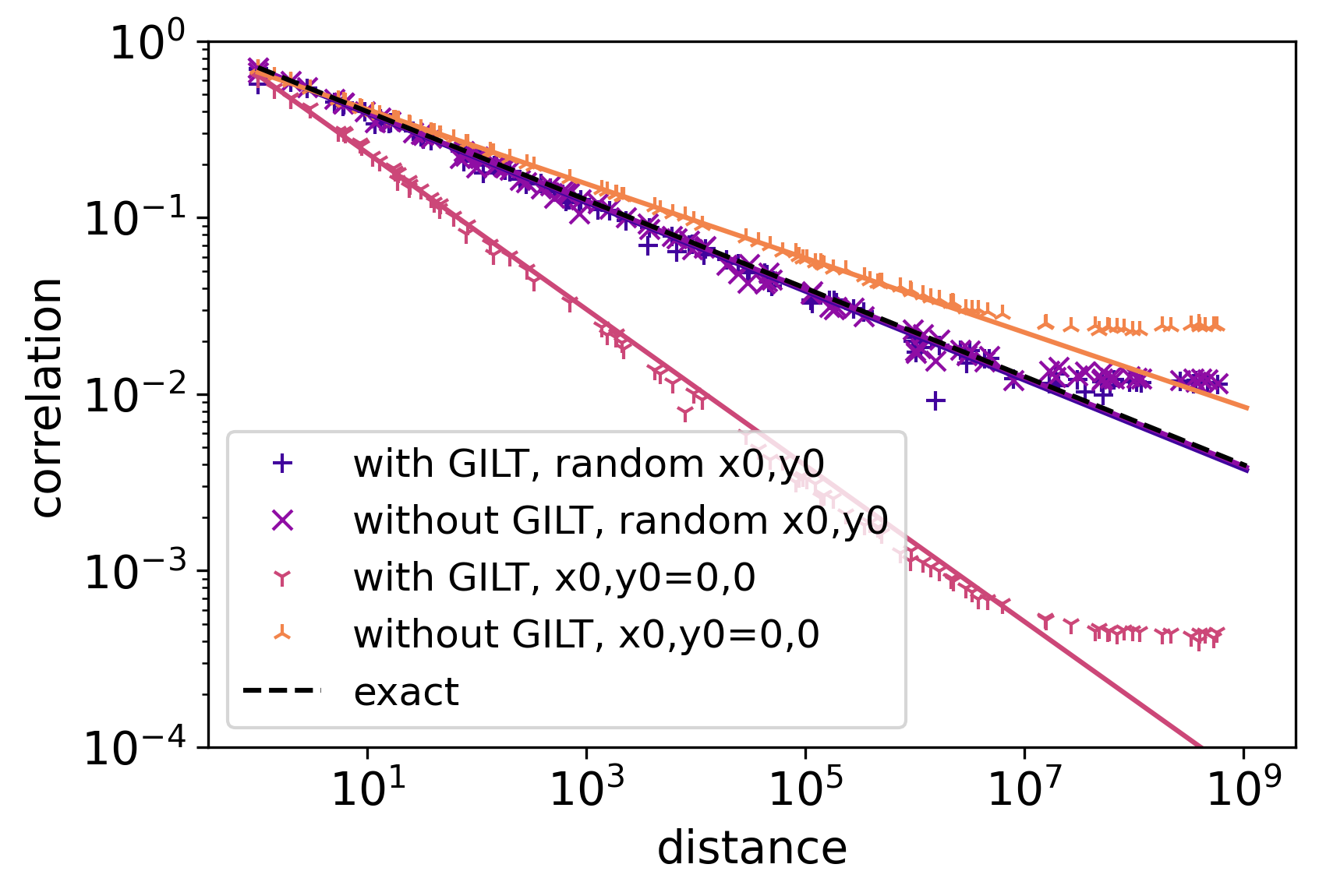}
    \caption{(color online) The two-point correlation function $\expval{\sigma(x_0,y_0)\sigma(x_1,y_1)}$ can be calculated using different strategies. One option is whether to use GILT, while another is whether to take the correlation function averaged over different starting points $(x_0,y_0)$ or to start from the corner $(0,0)$ of the coarse-grained block $0\leq x < L_x, 0 \leq y < L_y$. When using GILT and a fixed starting point, the correlation function gives an incorrect scaling factor. On the other hand, without GILT and with a fixed starting point, the correlation is slightly overestimated. However, by taking the average starting points $(x_0,y_0)$ in the block, both cases with or without GILT agree well with the analytic result.}
    \label{fig:CorrCompareGILT}
\end{figure}

Figure~\ref{fig:CorrCompareGILT} shows the two-point correlation function $\expval{\sigma(x_0,y_0)\sigma(x_1,y_1)}$ calculated by different methods. 
When GILT is used, the correlator gives the wrong scaling dimension when one of the defects is localized at a corner of the coarse-grained block. The reason seems to be that the corner modes are artificially suppressed by GILT, as discussed in \secref{sec:Discussion_RemoveEdgeModes}.

The overall best results are given when GILT is used, but the correlation function is averaged over the positions of the defects. GILT gives a more reliable vacuum tensor, and the averaging helps to avoid corners of coarse-grained blocks at various levels. We used this strategy for the earlier part of our results.

\subsection{Smearing of Point Defects}
\label{sec:Discussion_SmearingExplained}

\begin{figure}
    \centering
    \includegraphics[width=.65\hsize]{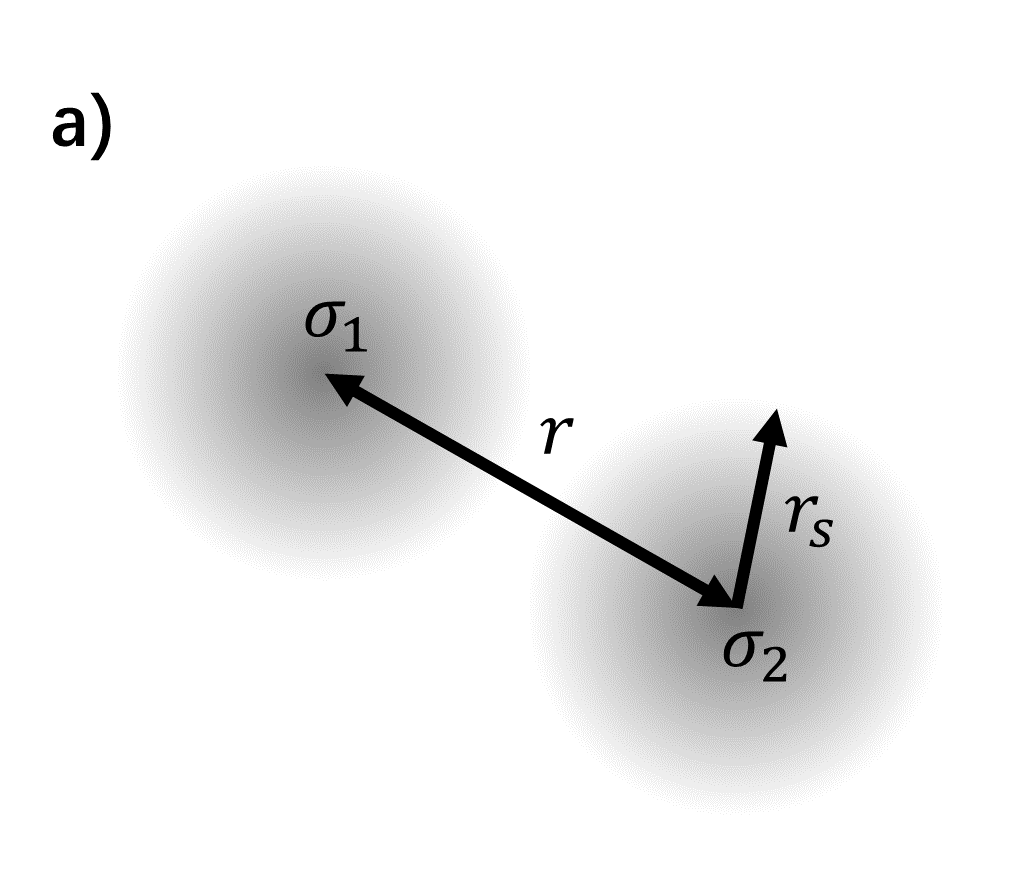}
    \includegraphics[width=\hsize]{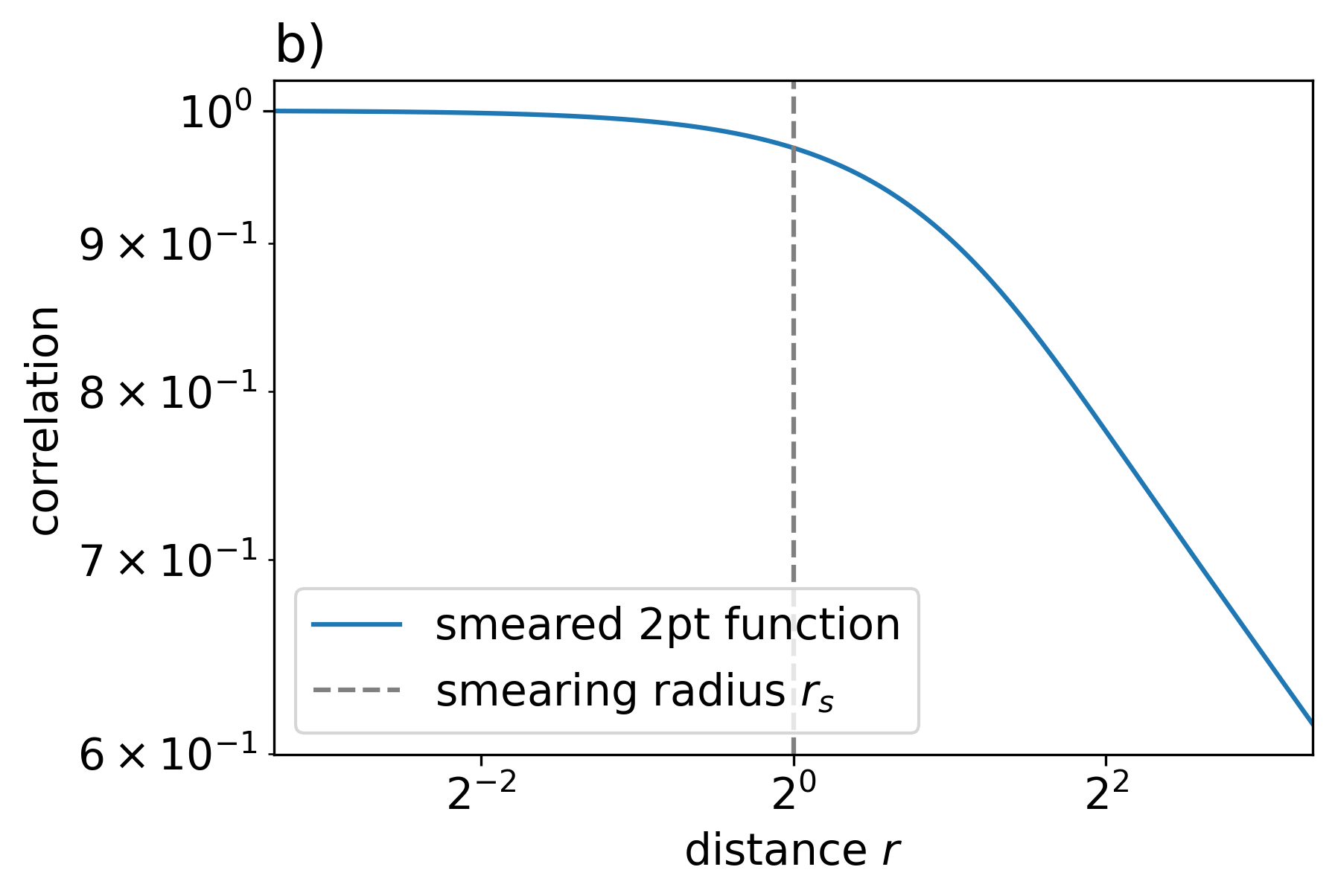}
    \caption{(color online) a) In a smeared two-point function, the two-point-like defects are replaced by a Gaussian profile with a smearing radius $r_s$. b) The shape of a smeared two-point function \eqref{eq:SmearingGaussian} in a log-log plot. The vertical grey line indicates the smearing radius $r_s$. }
    \label{fig:Smeared2PtFunctionIllustration}
\end{figure}

In the coarse-graining process, a point-like defect is expected to be ``smeared'' into a ``particle cloud''. When the two smeared defects move toward each other, one might expect a regularized two-point correlation function, with a smoother, lower peak compared to the correlation function between two ``bare'' defects. \eqref{eq:SmearingGaussian} below shows a type of regularized two-point function, where the point-like defect is replaced by a Gaussian profile \eqref{eq:GaussianKernel}, as illustrated in \figref{fig:Smeared2PtFunctionIllustration}. 
In the log-log graph, \eqref{eq:SmearingGaussian} features a flat line, which indicates the smeared tip of the correlation function, followed by a negative slope, which indicates the unsmeared part of the correlation function, which follows the power law \eqref{eq:PowerLaw},
\begin{equation}
\label{eq:SmearingGaussian}
\begin{aligned}
    f_\text{smeared}(x,y)&=\int\int\int\int dx_1 dy_1 dx_2 dy_2\\
    G(x-x_1,y-y_1)& G(x-x_2,y-y_2) P(x_1-x_2,y_1-y_2) ,
\end{aligned}
\end{equation}
\begin{equation}
    \label{eq:GaussianKernel}
    G(x,y)=\frac{1}{2\pi r_s^2}e^{-\frac12\frac{x^2+y^2}{r_s^2}},
\end{equation}
\begin{equation}
    \label{eq:PowerLaw}
    P(x,y)=\frac{1}{(x^2+y^2)^{\Delta}}.
\end{equation}

\begin{figure}[h]
    \centering
    \includegraphics[width=\hsize]{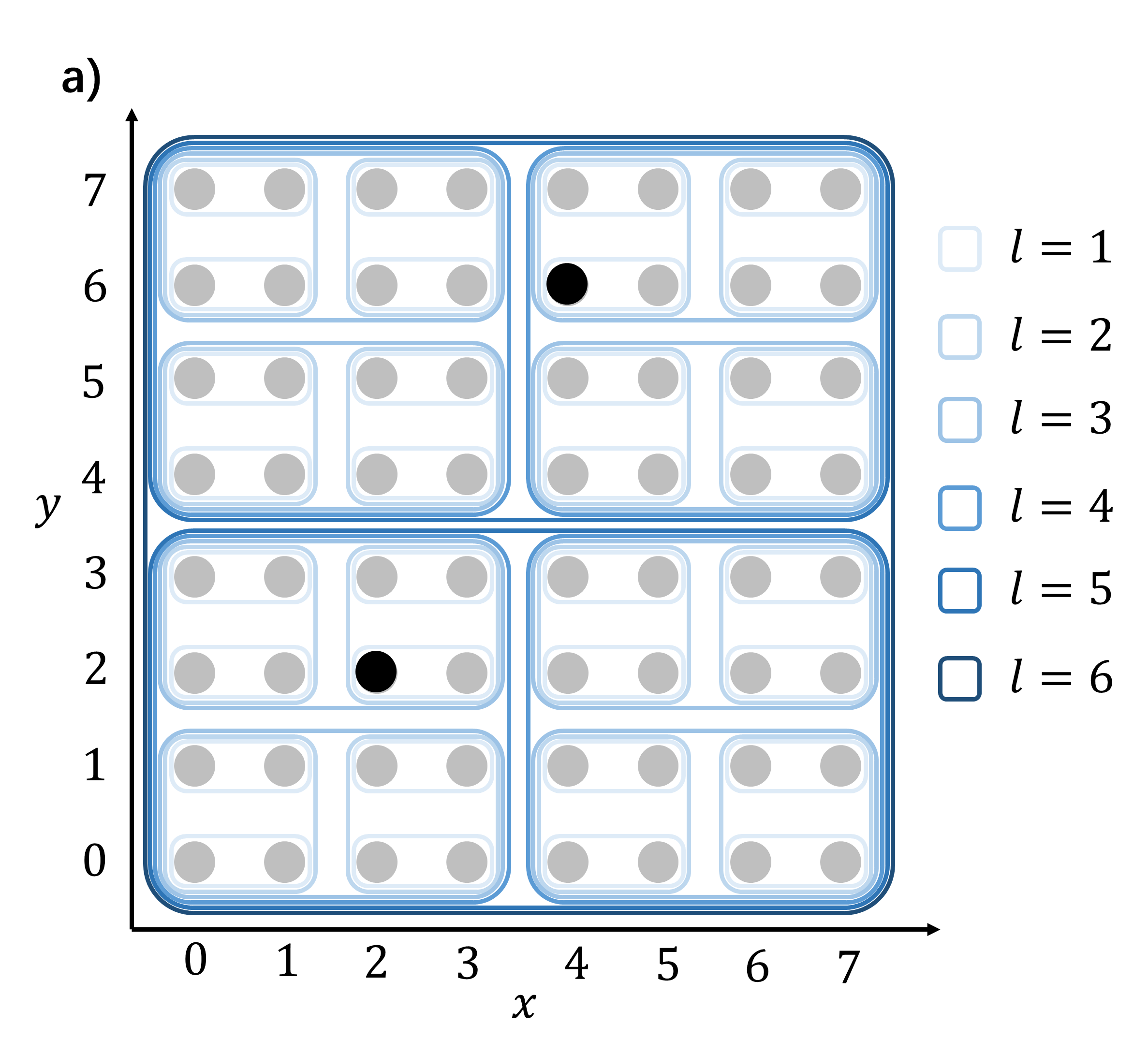}
    \includegraphics[width=\hsize]{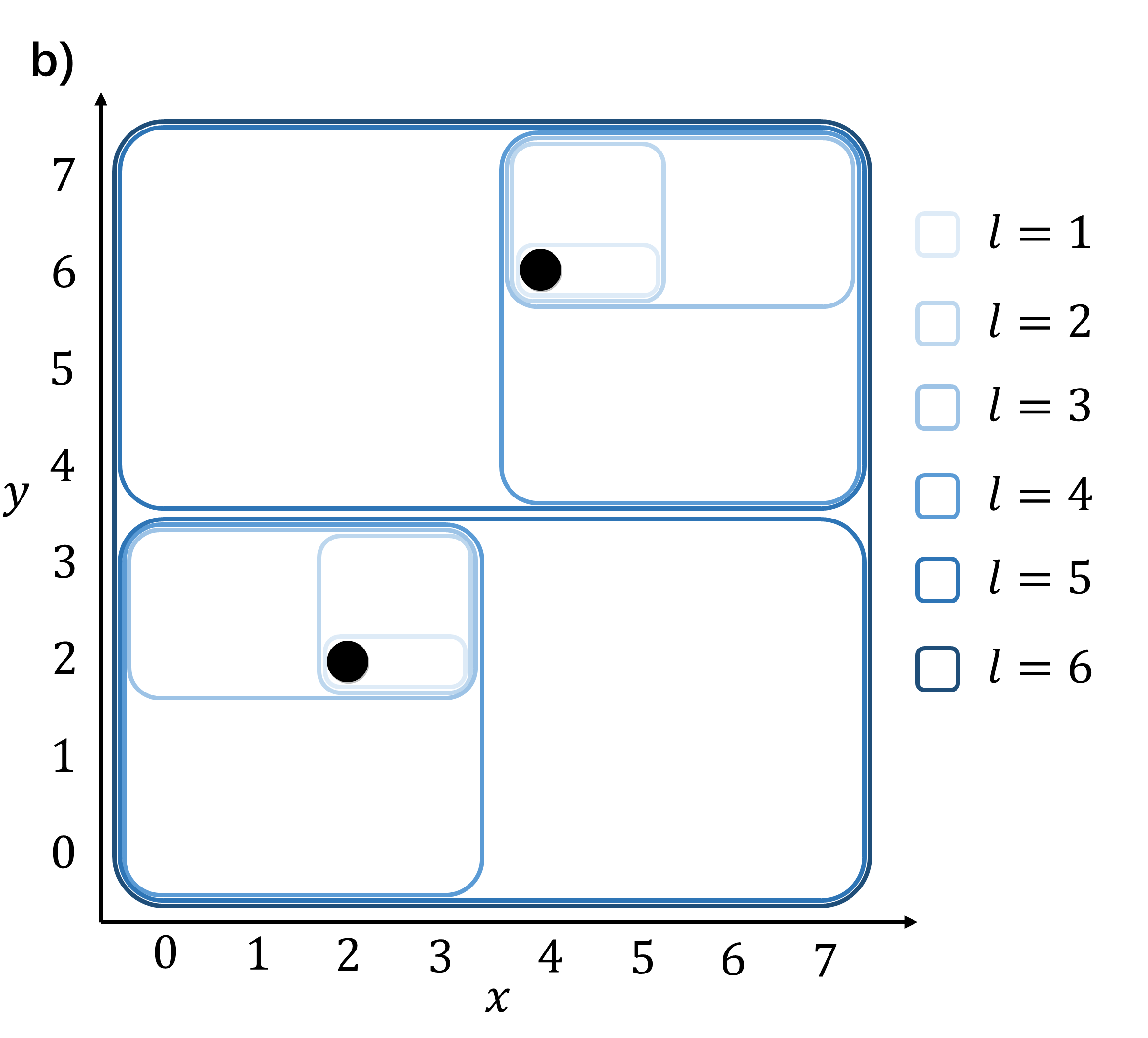}
    \caption{(color online) a) The sites on a $8\times 8$ lattice are grouped into coarse-graining blocks at various levels $l$. The coordinate system originated at the lower left corner of the biggest block $l=6$. The above illustration shows the coarse-graining blocks of a lattice that contains two defects at $(2,2)$ and $(4,6)$. b) At level $1\leq l \leq 5$, there are two defect tensors at each level,  each contained at a separate block. At level $l=6$, the two defect tensors are fused into one. }
    \label{fig:CGBlocks}
\end{figure}

In HOTRG, the fusion of two defects occurs when two coarse-grained tensors, each with one defect in its region, are coarse-grained into a larger tensor. As illustrated in \figref{fig:CGBlocks}, the two point defects $(2,2)$ and $(4,6)$ meets in a $8\times8$ block at level $l=6$. Naturally, one would expect that the amount of smearing that the two defects suffer is related to the block size of the largest coarse-grained tensor before the two defects meet.

\begin{figure}[h]
    \centering
    \includegraphics[width=\hsize]{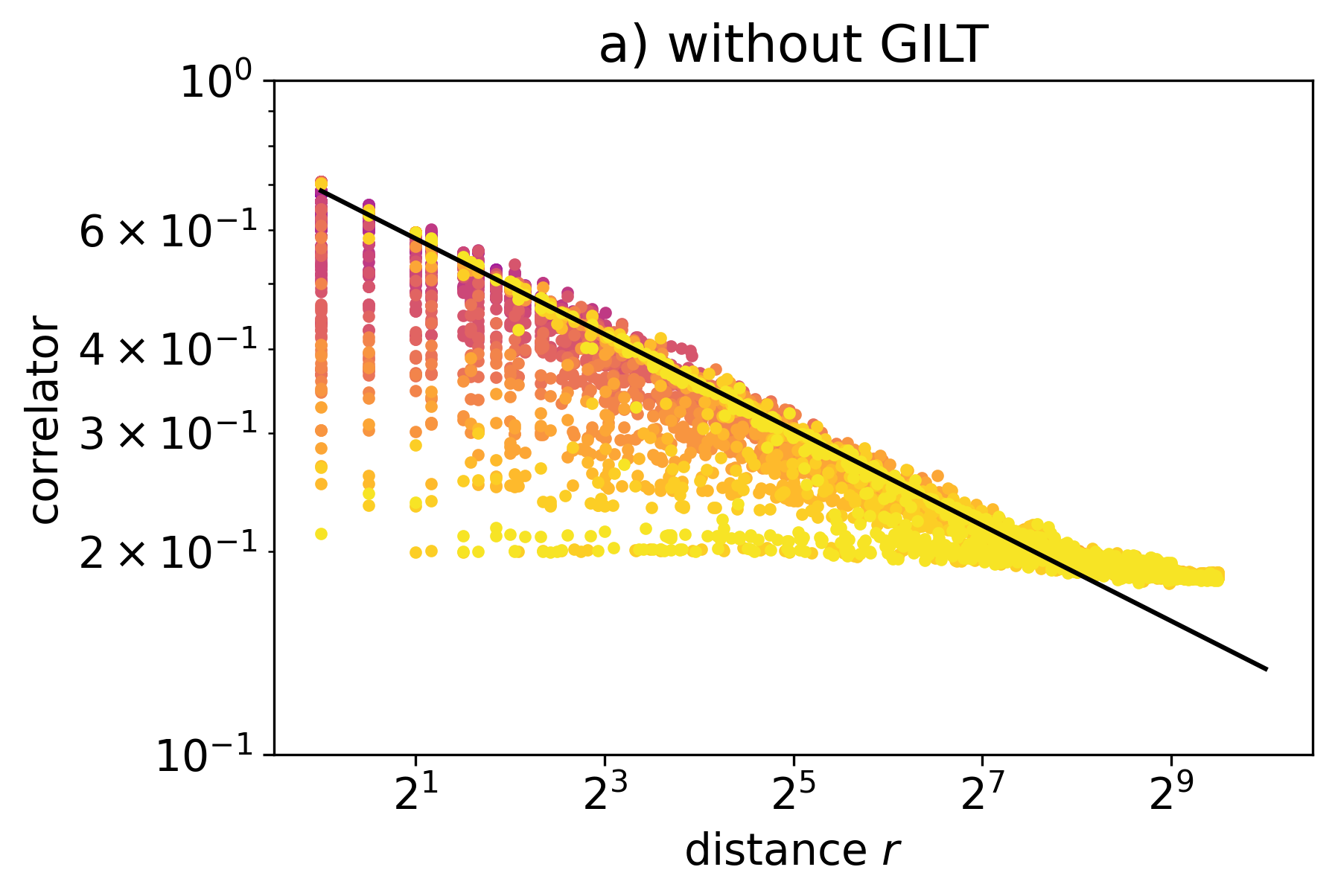}
    \includegraphics[width=\hsize]{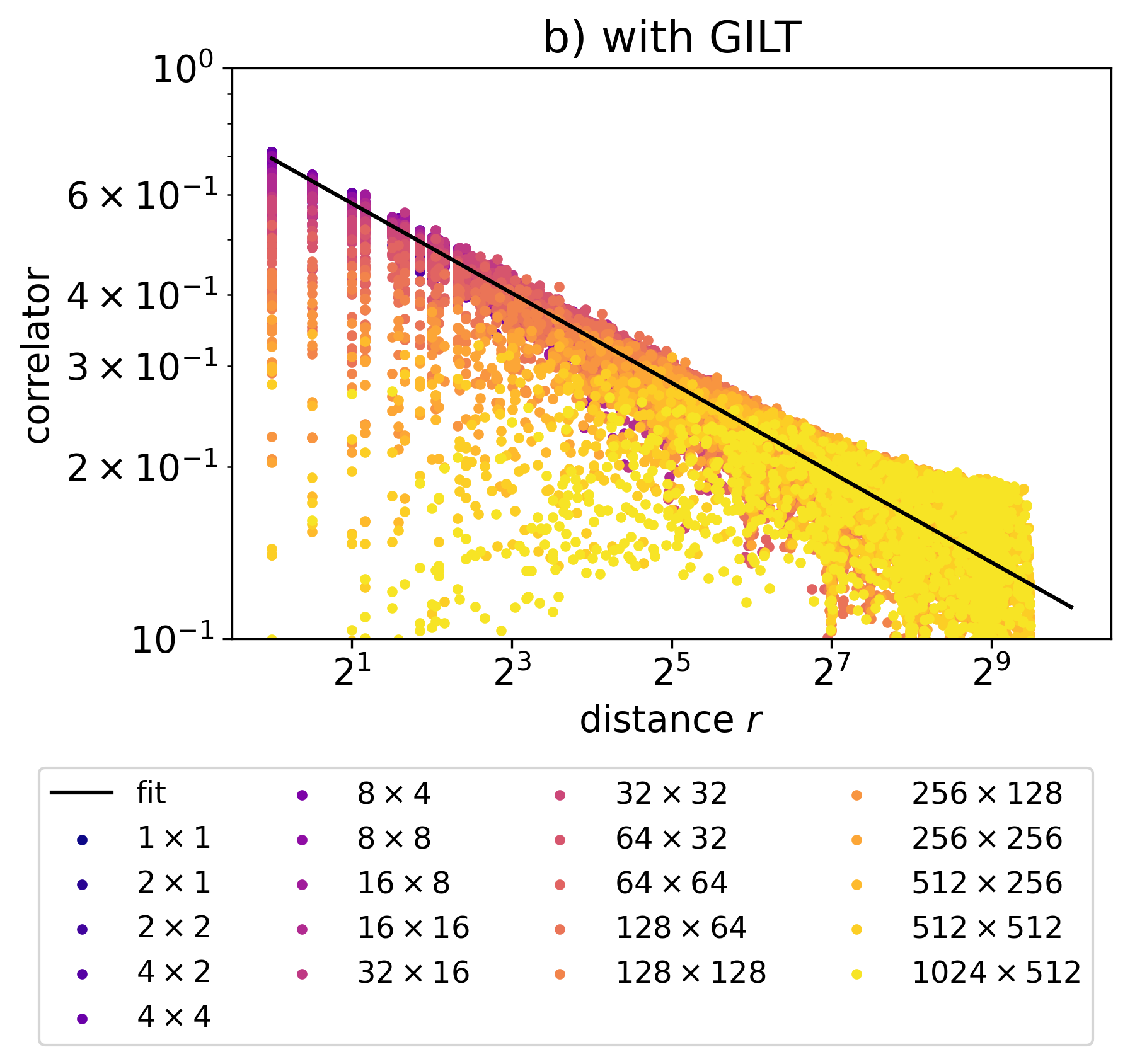}
    \caption{(color online) The two-point correlation function evaluated at different coarse-graining blocks. The color of the datapoints indicates the size of the biggest block before the two defects are fused into one block. The black line is fitted at the log2-log2 scale by a one-sided Huber loss with $\epsilon=0.1$ for $r<256$ with scaling dimension $\Delta_\text{No GILt}=0.118$ and $\Delta_\text{GILT}=0.131$. The points are sampled in a fine-tuned probability distribution that prioritizes the outliers below the main trend. For the trend of different block sizes, see \figref{fig:CorrCompareBlockSizeSeparate}-\figref{fig:CorrLowerBoundGILT}}
    \label{fig:CorrCompareBlockSize}
\end{figure}

\begin{figure*}[hbtp]
    \centering
    \includegraphics[width=\hsize]{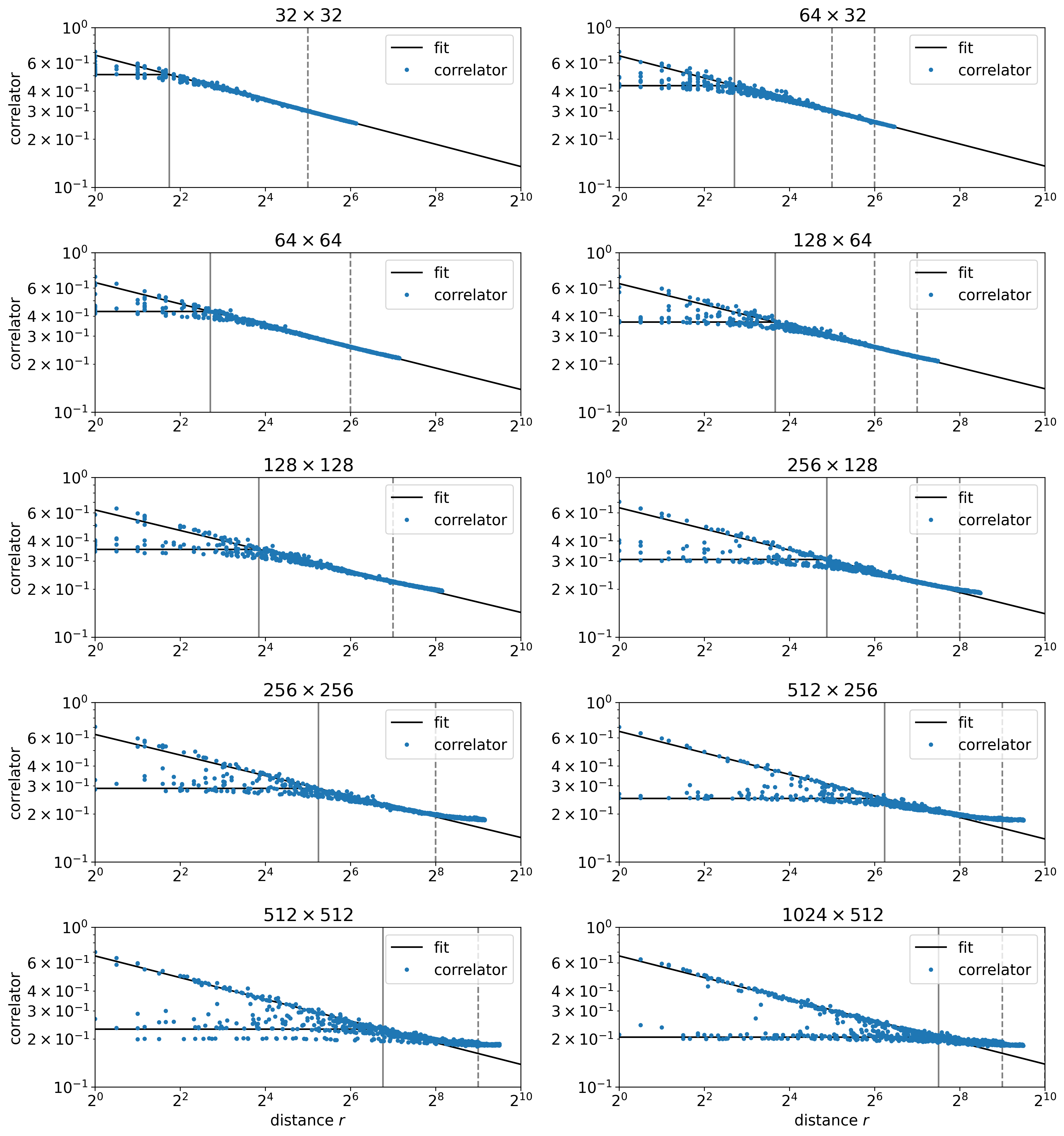}
    \caption{(color online) The two-point correlation function evaluated at different coarse-graining blocks. GILT is not used. This figure represents the same data as in \figref{fig:CorrCompareBlockSize}, but separated in different frames by different block sizes. Each subfigure represents a certain size $l_x\times l_y$ of the largest block before the two defects are fused into one block. The two black lines are the fitted main trend and the lower bound, respectively. The vertical gray dashed line indicates the block size $l_x, l_y$. The solid vertical gray line indicates the smearing radius $r_s$.}
    \label{fig:CorrCompareBlockSizeSeparate}
\end{figure*}

\begin{figure*}[hbtp]
    \centering
    \includegraphics[width=\hsize]{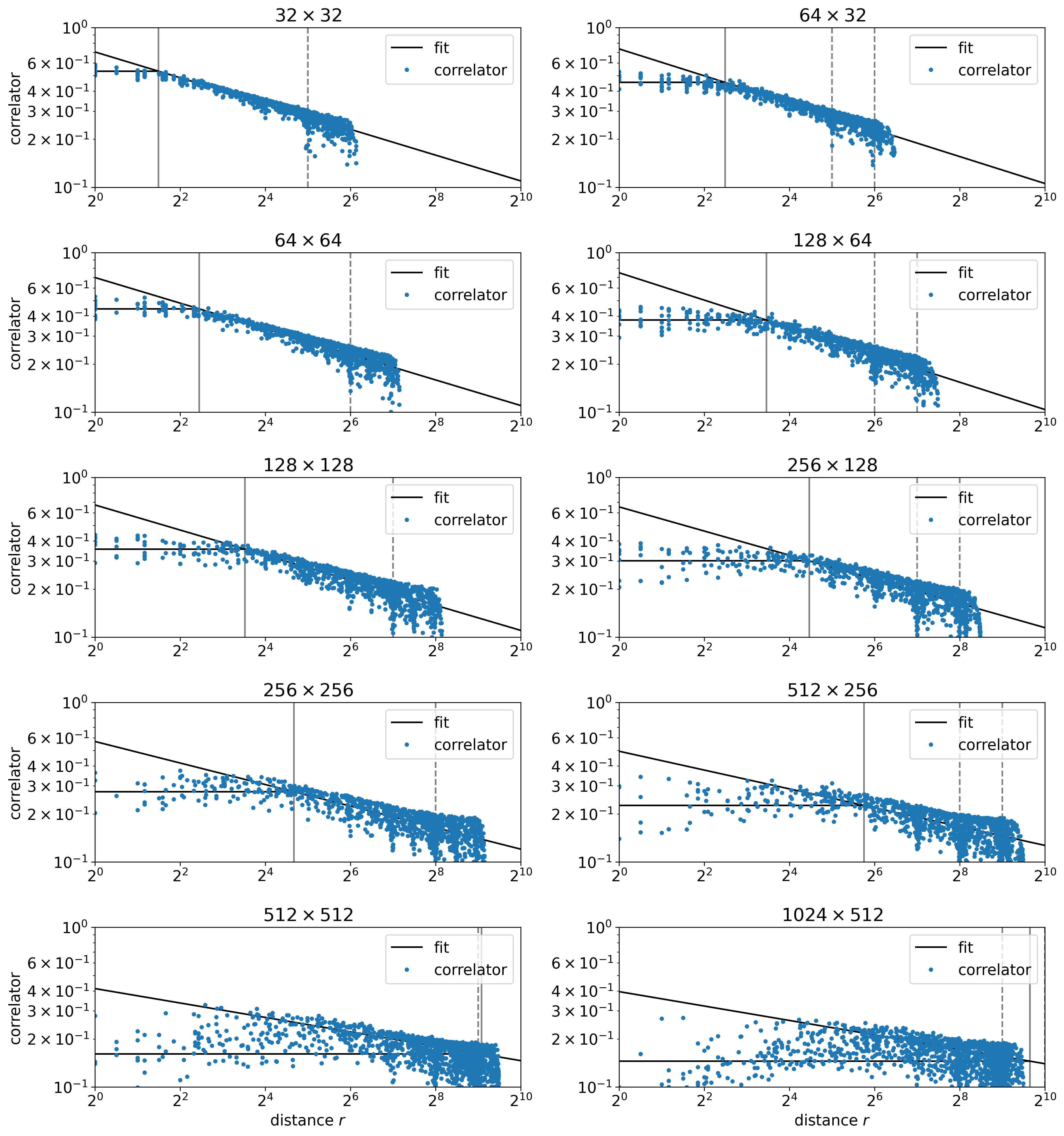}
    \caption{(color online) The two-point correlation function evaluated at different coarse-graining blocks. GILT is used. This figure represents the same data as \figref{fig:CorrCompareBlockSize},  but separated in different frames by different block sizes. Each subfigure represents a certain size $l_x\times l_y$ of the largest block before fusing the two defects into one block. The two black lines are the fitted main trend and the lower bound, respectively. The vertical gray dashed line indicates the block size $l_x, l_y$. The solid vertical gray line indicates the smearing radius $r_s$. }
    \label{fig:CorrCompareBlockSizeSeparateGILT}
\end{figure*}

Figure \ref{fig:CorrCompareBlockSize} illustrates how the size of the coarse-grained block, $l_x\times l_y$, just before two defects fuse, regulates the short-distance behavior of the two-point correlators. The color indicates the size of the biggest block before the two defects are fused into one block. The black line is fitted at the log2-log2 scale by a one-sided Huber loss with $\epsilon=0.1$ for $r<256$ with respective fitted scaling dimensions $\Delta_\text{No GILt}=0.118$ and $\Delta_\text{GILT}=0.131$. The points are sampled with a \textit{fine-tuned} probability distribution that prioritizes the outliers below the main trend to illustrate the smearing effect. 

As mentioned earlier, the size of the coarse-grained block $l_x \times l_y$  implies the amount of regularization that the two-point correlator suffers. For correlators with the same separation $r$, the larger the coarse-grained block, the lower the regularized correlator. Also note that the amount of smearing is determined not only by the size of the block but also by the relative position of a defect inside a block, which has not been examined carefully yet.

Figures~\ref{fig:CorrCompareBlockSizeSeparate} and~\ref{fig:CorrCompareBlockSizeSeparateGILT} further demonstrate the trend of the smeared correlation function for a given coarse grain size $l_x\times l_y$. There are two main trends in the data. On the log-log plot, the first one is a straight line with a negative slope, implying the unsmeared exact correlation function. The second trend is a flat line, indicating the smeared two-point function. Together, the two trends confirm the smearing picture described in \eqref{eq:SmearingGaussian}. Note that there are also data scattered between the two trends, as we have already seen in \figref{fig:CorrCompareBlockSize}.

The first trend resembles the exact, unsmeared correlation function. In \figref{fig:CorrCompareBlockSize}, the first trend of data with various block sizes almost overlaps with each other, showing consistency between different levels of coarse graining. It might indicate that certain pairs of defects do not suffer from smearing or that the smearing radius is much smaller than the separation $r$. 

The second trend resembles the smeared correlation function, which is a flat line that clipped the tip of the first trend at short distances. The flat line indicates that the correlation function is almost invariant when the separation of the smearing operator changes. It might be explained by the picture that the separation is less than the smearing radius $r_s$. The smearing radius $r_s$ is estimated by the intersection of the first and the second trend.

\begin{figure}[h]
    \centering
    \includegraphics[width=.7\hsize]{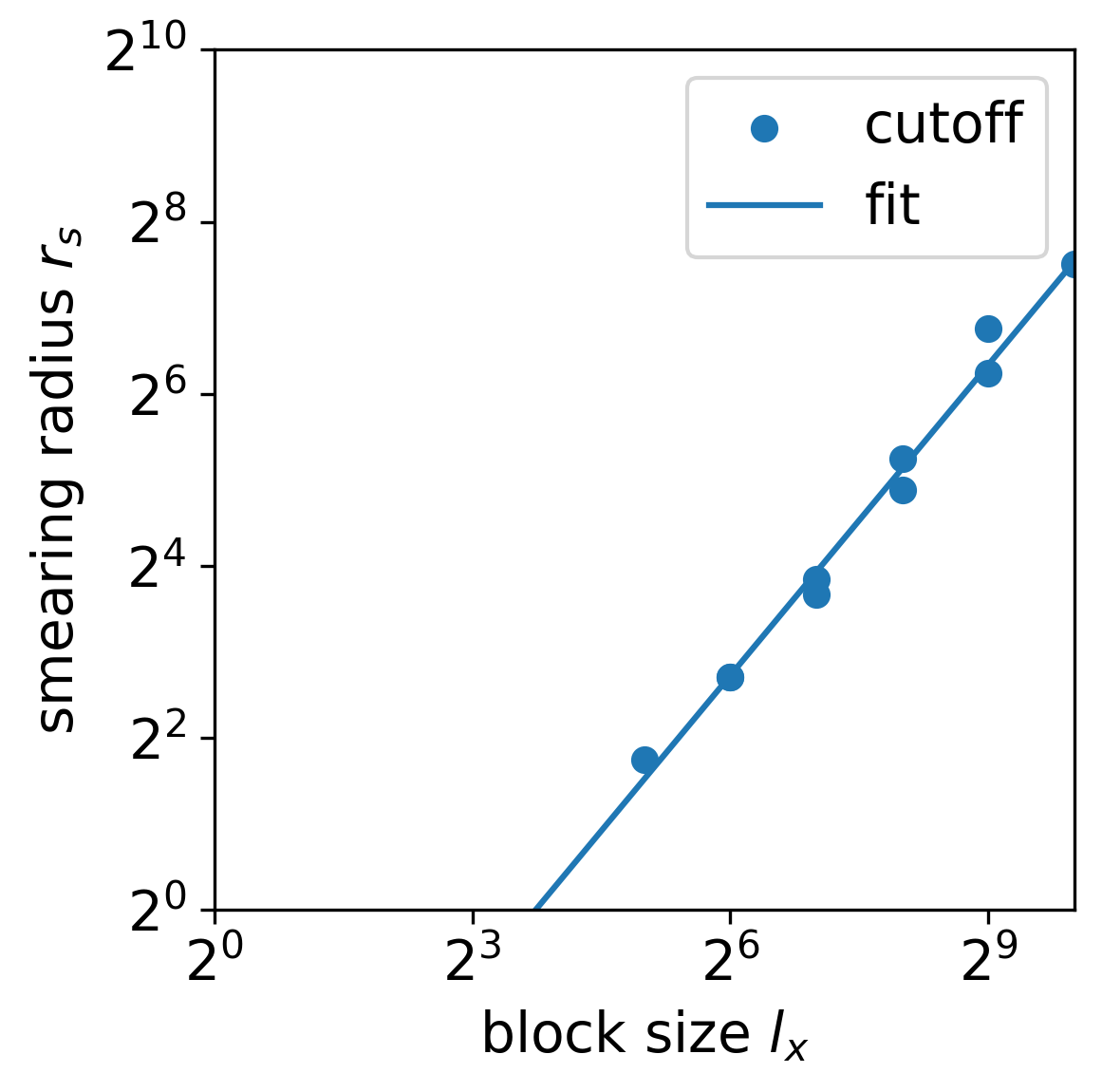}
    \caption{(color online) The smearing radius $r_s$ in \figref{fig:CorrCompareBlockSizeSeparate} and the size of the coarse-graining block. GILT is not used here.}
    \label{fig:CorrLowerBound}
\end{figure}

\begin{figure}[h]
    \centering
    \includegraphics[width=.7\hsize]{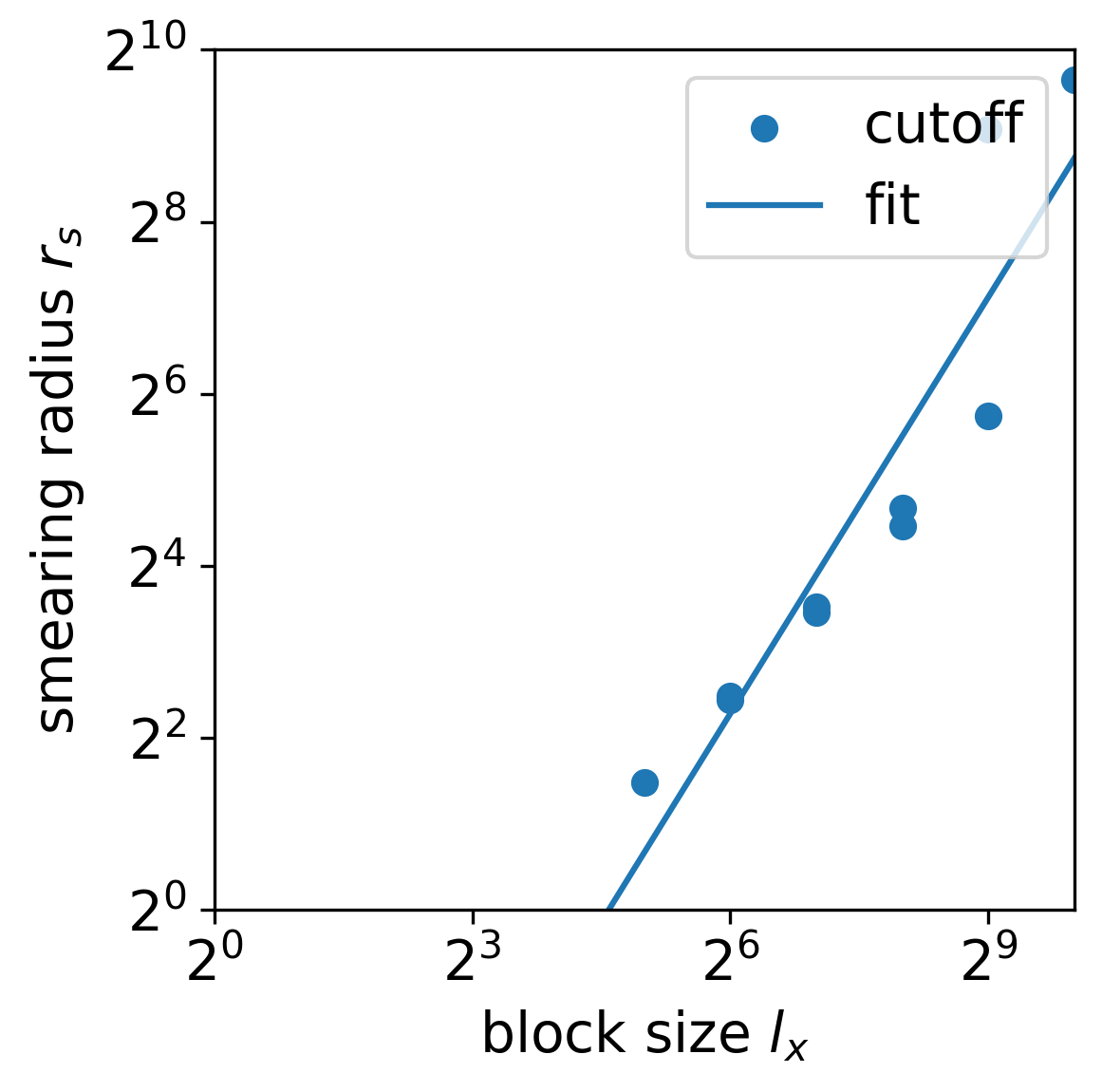}
    \caption{(color online) The smearing radius $r_s$  in \figref{fig:CorrCompareBlockSizeSeparateGILT} and the size of the coarse-graining block. GILT is used here.}
    \label{fig:CorrLowerBoundGILT}
\end{figure}

The smearing radius $r_s$ increases with the increase of the coarse grain block size ($l_x\times l_y$) before two defect tenors merge, as shown in \figref{fig:CorrLowerBound} and \figref{fig:CorrLowerBoundGILT}. However, the smearing radius grows faster than the growth of the block size: $r_s\propto l_x^{\delta_\text{LS}}$, with $\delta_\text{LS, No GILT}=1.20$ and $\delta_\text{LS, GILT}=1.61$. This indicates that as the coarse-graining iteration proceeds, the smearing radius will catch up with the lattice size and may result in errors in further coarse-graining.

Now, we compare the cases with and without GILT. When GILT is applied, most of the short-distance correlation points on the first (unsmeared) trend are moved to the second (smeared) trend. This means that GILT successfully removes most of the short-distance details of the system, compared to the case without GILT. However, at the long-distance end, the data without GILT follow more precisely to the fitted line, while the data with GILT have a more significant error with a much lower value. This means that GILT also removes some of the information on the long-range behavior of the defects.

The data analysis in this subsection confirms that the defect is smeared after coarse graining. Without GILT, defects in certain places might not be smeared as much. With GILT, most defects are smeared to approximately the same degree, but GILT also substantially removes some of the long-range information about the defects, resulting in an error that lowers the calculated two-point function value.

We also find the positioning of the defects plays a crucial role in determining whether the defects are being missed from smearing in the non-GILT case, and the defects are being over-suppressed in the GILT case. There is a sign that the corner of the largest unfused block is the aforementioned special place, but further numerical evidence is required to support this conjecture.

Finally, we provide an explanation for why naive averaging in different places when sampling the two-point function in \figref{fig:Corr} yields the best result for the scaling dimension when GILT is used. As shown in \figref{fig:CorrCompareBlockSizeSeparateGILT}, to obtain the unsmeared two-point function, we require $r > r_s$. When points are randomly sampled, it is most likely that the size of the largest block $l_x \times l_y$ before merging is comparable to the distance between the two points $r$ and $l_x > r_s$. The only situation where $l_x \gg r$ is when $r$ is located at the corner or edge of the largest unfused block, which is a very unlikely scenario. Thus, in \figref{fig:CorrCompareBlockSizeSeparateGILT}, we intentionally increase the probability that $r$ is inside the smearing radius, so we can observe the reduction in correlation function due to smearing, and the fitted scaling dimension has a larger error.

However, in \figref{fig:Corr}, when the points are randomly sampled, most of them satisfy $r > r_s$, which guarantees that smearing occurs only on a shorter scale compared to $r$, thus ensuring the correctness of the two-point function. Therefore, the naive averaging over different places yields the best scaling dimension result in this case.

\subsection{Removal of Localized Edge Modes}
\label{sec:Discussion_RemoveEdgeModes}

When GILT is not applied, the dominance of the CDL tensor suggests that the truncated Hilbert space of the coarse-grained tensor mainly captures the exciting modes near the boundaries and corners of the coarse-grained block. This can be shown from the observation that there is no smearing at the corner of the coarse-grained block when GILT is turned off.

\begin{figure}[h]
    \centering
    \includegraphics[width=\hsize]{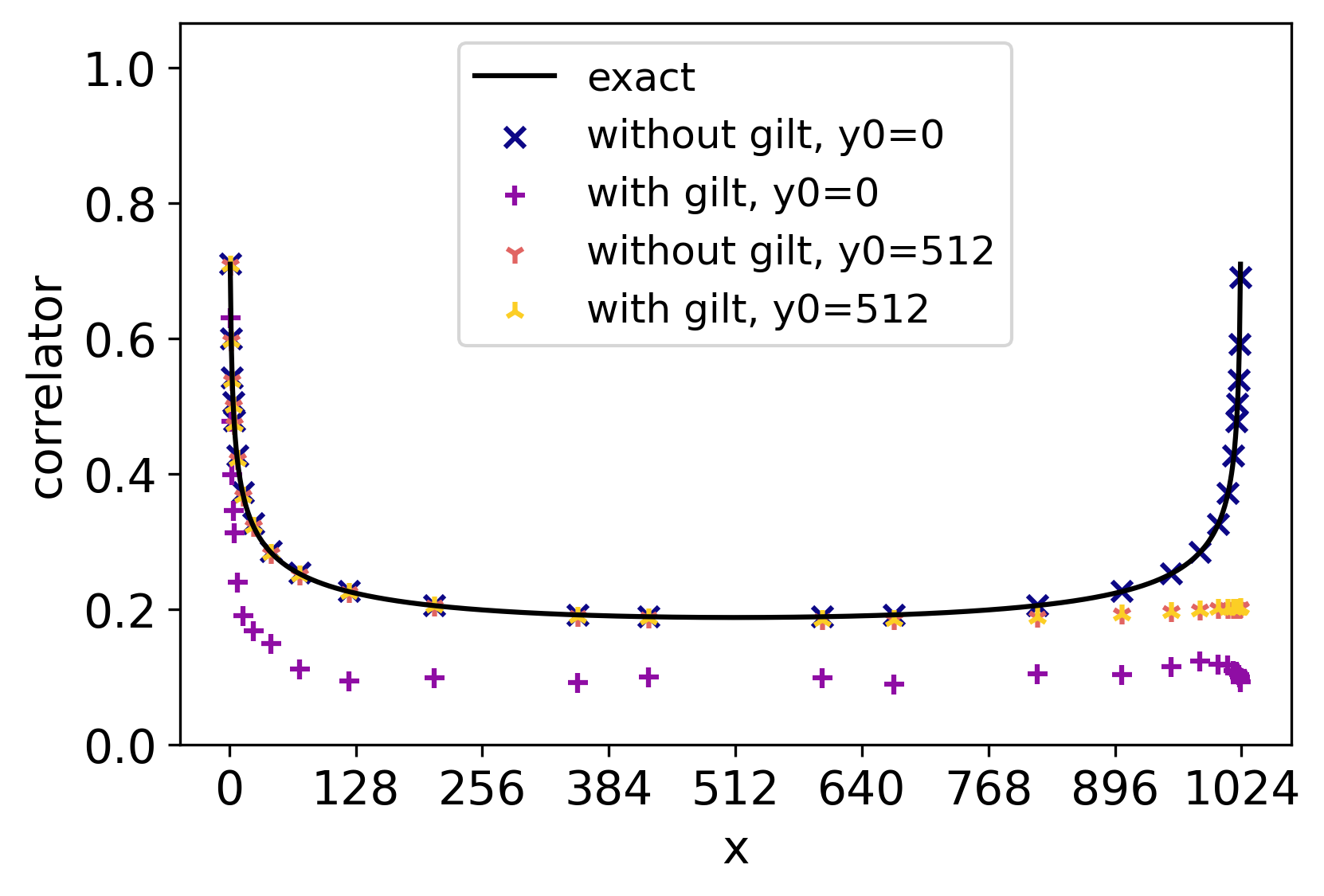}
    \caption{(color online) The two point correlation function $\expval{\sigma(0,y_0)\sigma(x,y_0)}$ on a torus $x\sim x+L_x, y\sim y+L_y$ of size $L_x\times L_y=1024\times 1024$. Different numerical schemes are compared: One option is whether to use GILT, while another is whether the start point is at the corner ($x=0,y=0$) of the coarse-grained block $0\leq x < L_x, 0 \leq y < L_y$, or near the center ($x=0,y\approx L_y/2$) of the block edge.}
    \label{fig:TorusGilt}
\end{figure}

\begin{figure}
    \centering
    \includegraphics[width=\hsize]{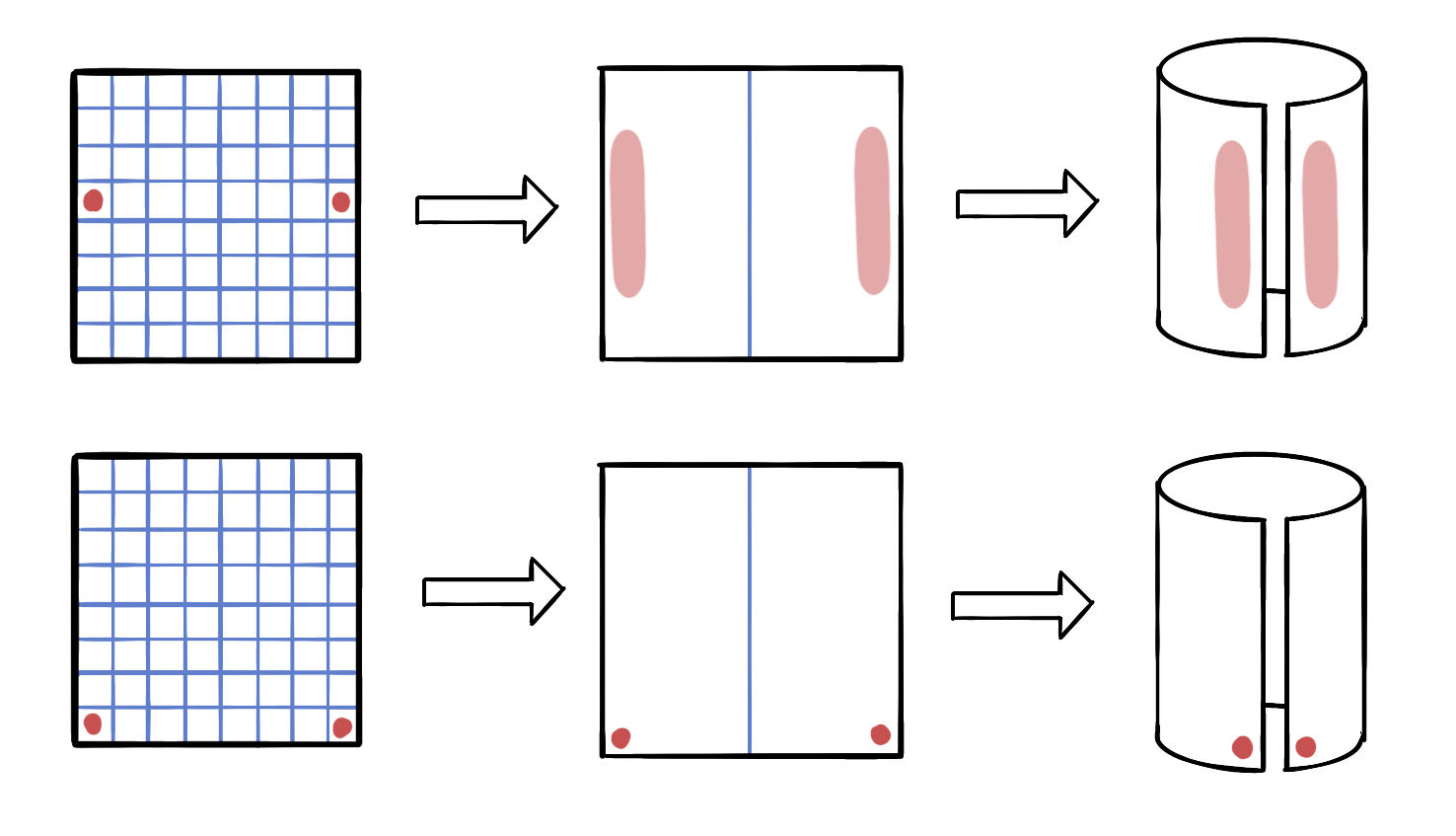}
    \caption{(color online) On a torus geometry, the two operators will experience coarse graining before being fused together. a) When the operator is at the center of the edge, it will be ``smeared'' into a ``particle cloud'', resulting in a lower correlation function. b) When the operator is at the corner, and GILT is not applied, smearing will not happen, resulting in a sharp peak at the end of the two-point function in \figref{fig:TorusGilt}.}
    \label{fig:TorusIllust}
\end{figure}

\begin{figure*}[hbtp]
    \centering
    \includegraphics[width=\hsize]{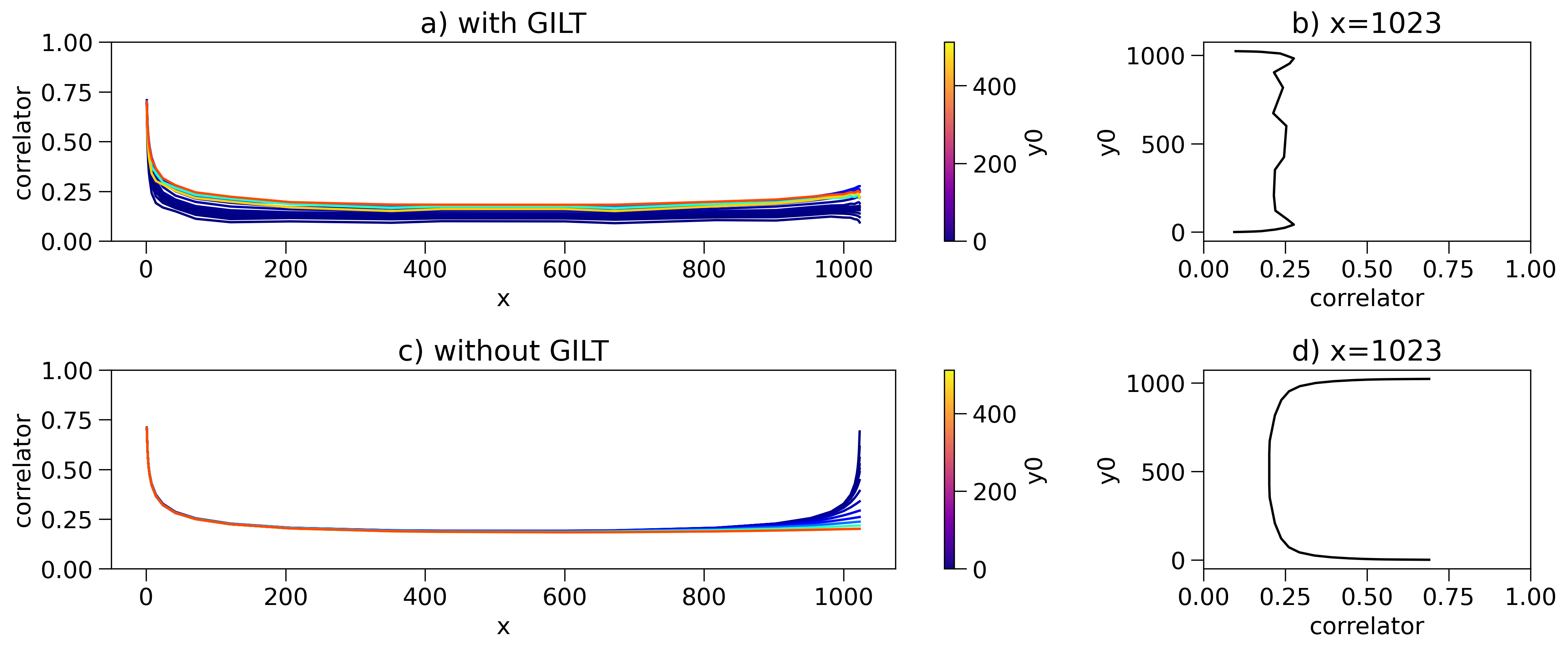}
    \caption{(color online) The two-point correlation function $\expval{\sigma(0,y_0)\sigma(x,y_0)}$ on a torus $x\sim x+L_x, y\sim y+L_y$ of size $L_x\times L_y=1024\times 1024$. In a) and c) the color denotes the values of $y_0$, where red indicates $y_0=L_y/2$ and blue indicates $y_0=0$. In b) and d), $x=L_x-1$. }
    \label{fig:TorusYDependency}
\end{figure*}

In particular, we compare the torus correlation function at $x\rightarrow 0$ and $x\rightarrow L_x$. As shown in \figref{fig:TorusIllust}, the correlator at $x\rightarrow L_x$ is evaluated by contracting opposite edges of a $L_x \times L_y$ block. So, the largest coarse-grained block containing a single defect is of size $L_x/2$. One would expect a very large smearing effect, resulting in a vanishing peak at $x\rightarrow L_x$. However, \figref{fig:TorusGilt} shows that the peak at $x\rightarrow L_x$ is as sharp as the bare correlation function evaluated at smaller scales at $x\rightarrow 0$, in the case where GILT is not used and the two defects are localized at the corner of the block.

It suggests that the defect is not smeared after the coarse-graining iterations. In other words, the truncated Hilbert space contains point-like excitations at the corners of the block. To further verify it, we plot the height of the peak when $y_0$ moves from the corner to the midpoint of the block edge. As shown in \figref{fig:TorusYDependency}c,d, the smearing effect is stronger when the two defects are on the midpoint of the opposite edges. Given the finiteness of the dimension of the truncated Hilbert space, it suggests that the defects are smeared into a one-dimensional ``cloud'' along the edge.

In the case of GILT, one would expect that the edge modes are being suppressed, and there is no peak at $x\rightarrow L_x$,  as confirmed in \figref{fig:TorusYDependency}a,b. Note that there are some inconsistencies in the correlation function when the defects are located in the corners. It suggests that GILT introduces some artificial suppression for corner excitation.

\section{Conclusion}
\label{sec:Conclusion}

In this work, we have integrated the MCF gauge-fixing method~\cite{acuaviva2022MCF}, with the tensor renormalization scheme~\cite{lyu2021lTRG} that combines the local entanglement filtering method GILT \cite{hauru2018GILT} and the HOTRG coarse graining method~\cite{xie2012HOTRG}. The combined method enhances the stability of the RG flow, which is manifested visually in the space of the tensor components. Additionally, we have demonstrated the detection of CDL-like effects from the oscillation of the transfer matrix spectrum between horizontal and vertical coarse-graining steps.

Following \cite{lyu2021lTRG}, we have calculated the scaling dimensions for conformal states with $\Delta<3$ in the critical square lattice Ising model using two different methods: the transfer matrix spectrum~\cite{gu2009TEFR_TMscD} and the lTRG approach~\cite{lyu2021lTRG}. Both methods agree well with the analytical results. We have also calculated the OPE coefficient $C_{\sigma\sigma\varepsilon}$ from lTRG and its eigenvectors. 

Important quantities to calculate in TN include correlation functions. We have computed the two-point spin-spin correlation function for the square-lattice Ising model at temperatures near the critical point (at, above, and below). The extracted exponent is in good agreement with the scaling dimension. We have also examined the correlation function on a finite-size torus and obtained results consistent with analytical results. In addition, we have also computed four-point correlation functions, and from those we were able to extract the scaling dimensions and the OPE coefficient, although the quality was not as good as other approaches we have used.

We have extended the use of lTRG beyond the computation of scaling dimensions and compared the eigenvectors of the lTRG fixed point equation with coarse-grained defect tensors, discovering that inserting operators on the lattice allows for the construction of conformal states/operators, which corroborate well with those obtained from lTRG. We have also found that the position of the inserted operator is reflected in its decomposition onto the conformal states numerically.

Our work has explored the effects of coarse-graining (HOTRG) and local entanglement filtering (GILT) on lattice defects/impurities. We have observed that point-like defects become ``smeared'' into ``particle clouds'' after coarse graining and that the degree of smearing varies depending on the position of the defect in the coarse graining region, particularly when it is located at the corner or edge of the region. GILT also exhibits artificial suppression for edge modes; despite this, we were able to mitigate this and demonstrate that a good quality two-point function on the plane or torus can still be retrieved if the smearing effect and the edge mode suppression are properly circumvented.

We now conclude this work by discussing possible future directions. The HOTRG and GILT algorithms we used in this do not account for lattice defects when determining truncation and filtering strategies. It may be possible to modify these coarse-graining algorithms to be sensitive to lattice defects. Therefore, it is interesting to explore whether this modification leads to more robust coarse graining in $n$-point functions and whether it can improve the overall quality of fixed-point tensors.
We also expect that by studying the tensor-space RG flow, oscillation of CDL tensors, and smearing of point-like operators, one can troubleshoot the current limitations of coarse-graining and local entanglement filtering methods and develop more precise and efficient computational methods for tensor-based RG approaches.
Furthermore, we note that our methods can be extended to other systems, including quantum spin models, such as the transverse-field Ising model, and possibly to 3D classical systems. It may also be interesting to study the tensor-based RG flow of extended defect objects, such as defect lines in 2D and defect planes in 3D.
Finally, it might be useful to investigate the reverse. In this paper, we extract the CFT data from a fixed-point tensor. Then, a natural question is where there exist some prescriptions for constructing a fixed-point tensor from the given CFT data, which might pave a new way of conformal bootstrapping via studying the fixed points of the tensor renormalization group in the tensor space.

\medskip\noindent \textit{Note added:} Close to the completion of our initial manuscript, we became aware of a recent work by Atsuhi Ueda and Masaki Oshikawa on ``finite-size and finite bond dimension effects of tensor network renormalization''~\cite{ueda2023finite}. The part that the two works overlap most is how to extract CFT data, including OPE coefficients and scaling dimensions. They use eigenvectors of the transfer matrix. We compare the method with the transfer matrix and our new proposed method with lTRG. The main focus of their paper is on the finite-size effect, and one focus of ours is the eigenvectors of lTRG and coarse-graining defects.
We have further provided a new section (Sec.~\ref{sec:Finite_Scale_Effect}) for some comparison.

\section{Acknowledgements}

We are grateful for many useful and valuable discussions with Ning Bao, Nikko Pomata, Andreas Weichselbaum, and Alexander B. Zamolodchikov. 
We acknowledge support from the Materials Science and Engineering Divisions, Office of Basic Energy Sciences of the U.S. Department of Energy under Contract No. DESC0012704.

\bibliographystyle{unsrt}
\bibliography{references}

\appendix

\section{List of Algorithms in TRG}
\label{sec:Algorithms}

\subsection{Higher-Order Tensor Renormalization Group}
\label{sec:HOTRG_alg}
Here, we recapitulate the essence of the HOTRG coarse-graining procedure, as schematically shown in Algorithm~\ref{alg:HOTRG}. We define one HOTRG step as merging two tensors into one. The first (or odd number of) step involves finding the isometry $w$ to merge two tensors along the horizontal directions, resulting in a rectangular shape of the whole system. The second (or even number of) step involves merging two resultant tensors from the previous step into one, restoring the whole system to a square shape. 
\begin{algorithmnofloat}
    \caption{Higher-Order Tensor Renormalization Group~\cite{xie2012HOTRG}}\label{alg:HOTRG}
    \begin{equation}
        \label{eq:HOTRG_steps}
        \vcenter{\hbox{\includegraphics[page=20]{TRGCFTGraphics.pdf}}}
        =
        \vcenter{\hbox{\includegraphics[page=21]{TRGCFTGraphics.pdf}}}
    \end{equation}
    
    \begin{equation}
        \label{eq:HOTRG_steps2}
        \vcenter{\hbox{\includegraphics[page=22]{TRGCFTGraphics.pdf}}}
        =
        \vcenter{\hbox{\includegraphics[page=23]{TRGCFTGraphics.pdf}}}
    \end{equation}
    
    \textbf{Inputs:} A translational invariant tensor network on a lattice. The tensor at each site is $T$.
    
    \textbf{Outputs:} The coarse-grained tensor $T'$ by grouping every two adjacent tensors in the chosen spatial dimension, and an isometry $w$ that truncates the bond dimension of the coarse-grained tensor from $\chi\times \chi$ to $\chi$.

    \textbf{Steps:}

    To minimize the truncation error:

    \begin{equation}
        \label{eq:HOTRG_equation1}
        \left\Vert
        \,\vcenter{\hbox{\includegraphics[page=24]{TRGCFTGraphics.pdf}}}
        -
        \vcenter{\hbox{\includegraphics[page=25]{TRGCFTGraphics.pdf}}}\,
        \right\Vert^2,
    \end{equation}
the isometry $w$ can be estimated from the singular value decomposition of the two-site reduced density matrix $M$:
\begin{equation}
    \label{eq:HOTRG_svd_environment}
        \vcenter{\hbox{\includegraphics[page=26]{TRGCFTGraphics.pdf}}}
        =
        \vcenter{\hbox{\includegraphics[page=27]{TRGCFTGraphics.pdf}}}
        \svdeq
        \vcenter{\hbox{\includegraphics[page=28]{TRGCFTGraphics.pdf}}}
    \end{equation}
    On a 2D square lattice, one does the vertical and horizontal coarse-graining step alternatively, as shown in \eqref{eq:HOTRG_steps} and \eqref{eq:HOTRG_steps2}.

\end{algorithmnofloat}

We note that the memory complexity of a HOTRG step is $\mathcal{O}(\chi^5)$, and the time complexity is $\mathcal{O}(\chi^7)$ \cite{xie2012HOTRG}.

\subsection{Graph Independent Local Truncation}
\label{sec:GILT_alg}
Next, we describe the GILT procedure~\cite{hauru2018GILT} that aims to remove CDL-like tensors. It is shown in Algorithm~\ref{alg:GILT}. It involves a local truncation $R$ that interrupts loop-like correlations.

\begin{algorithmnofloat}
    \caption{Graph Independent Local Truncation~\cite{hauru2018GILT}}\label{alg:GILT}
    \begin{equation}
        \vcenter{\hbox{\includegraphics[page=29]{TRGCFTGraphics.pdf}}}
        \approx
        \vcenter{\hbox{\includegraphics[page=30]{TRGCFTGraphics.pdf}}}
        \svdeq
        \vcenter{\hbox{\includegraphics[page=31]{TRGCFTGraphics.pdf}}}
        \label{eq:GILT1}
    \end{equation}
    
    \textbf{Inputs:} a subgraph of a tensor network and an edge in the subgraph to truncate.
    
    \textbf{Outputs:} a truncation matrix $R$ to insert in the chosen edge, or equivalently, a pair of linear transformations $g_1$ and $g_2$, $R=g_1 g_2$ to apply to the legs of the adjacent tensors, which partially remove local entanglements passing through the edge.

    \textbf{Steps:}
    
    \begin{equation}
        \vcenter{\hbox{\includegraphics[page=32]{TRGCFTGraphics.pdf}}}
        =
        \vcenter{\hbox{\includegraphics[page=33]{TRGCFTGraphics.pdf}}}
        \svdeq
        \vcenter{\hbox{\includegraphics[page=34]{TRGCFTGraphics.pdf}}}
        \label{eq:GILT_Env}
    \end{equation}
    The environment $E$ of an edge is defined by removing that edge from the tensor network. $E$ is interpreted as an isometry from the two ends of broken edges to the external edges.
\begin{equation}
        \vcenter{\hbox{\includegraphics[page=35]{TRGCFTGraphics.pdf}}}
        =
        \vcenter{\hbox{\includegraphics[page=36]{TRGCFTGraphics.pdf}}}
        \label{eq:GILT_svd}
    \end{equation}
    By spectral decomposing $E$, one can see that the subspace of variations of $R\in \text{Mat}_{\chi\times \chi}$ which does not significantly affect the tensor output, is spanned by the singular vectors corresponding to smaller singular values. 
    
    In particular, we choose $R$ in the following way:
    \begin{equation}
        \vcenter{\hbox{\includegraphics[page=37]{TRGCFTGraphics.pdf}}}
        =
        \vcenter{\hbox{\includegraphics[page=38]{TRGCFTGraphics.pdf}}}
    \end{equation}
\begin{equation}
        t'_i=t_i \frac{\hat S_i^2}{\hat S_i^2+\epsilon^2}, 
        \hat S_i=\frac{S_i}{\max(S)},
        \vcenter{\hbox{\includegraphics[page=39]{TRGCFTGraphics.pdf}}}
        \;
        =
        \vcenter{\hbox{\includegraphics[page=40]{TRGCFTGraphics.pdf}}}
        \label{eq:GILT_truncate_singular_values}
    \end{equation}
     Split $R$ into $g_1$ and $g_2$ by SVD, and absorb them into the adjacent tensors, as in \eqref{eq:GILT}.
     Furthermore, the local entanglement can be truncated by repeating the process on the edge between $g_1$ and $g_2$.
    
\end{algorithmnofloat}

We note that the memory complexity of a GILT step is $\mathcal{O}(\chi^4)$, and the time complexity is $\mathcal{O}(\chi^6)$ \cite{hauru2018GILT}.

\subsection{Minimal Canonical Form}
\label{sec:MCF_alg}
In order to compare tensors in different coarse-graining steps and determine the change in RG flow, it is necessary to convert them into certain canonical forms. One key tool that we use is the recently proposed MCF~\cite{acuaviva2022MCF}. This fixes the redundancy under the group $GL(\chi)^{\otimes m}$ up to a residue $U(\chi)^{\otimes m}$, where $m$ is the bond dimension of each leg. MCF uses an iterative procedure, described in Algorithm~\ref{alg:MCF}.

\begin{algorithmnofloat}
    \caption{Minimal Canonical Form\cite{acuaviva2022MCF}}\label{alg:MCF}
    \begin{equation}
        \vcenter{\hbox{\includegraphics[page=41]{TRGCFTGraphics.pdf}}}
        =
        \vcenter{\hbox{\includegraphics[page=42]{TRGCFTGraphics.pdf}}}
    \end{equation}
\begin{equation}
        \vcenter{\hbox{\includegraphics[page=43]{TRGCFTGraphics.pdf}}}
        =
        \vcenter{\hbox{\includegraphics[page=44]{TRGCFTGraphics.pdf}}}
        \label{eq:MCF_gauge_transform}
    \end{equation}

    \textbf{Inputs:} A translationally invariant tensor network on a lattice. The tensor at each site is $T$. 
    
    \textbf{Outputs:} A linear transformation $(g_1, g_2) \in GL(\chi)^{\otimes m}$, which gauge transforms the tensor into its minimal canonical form: 
\begin{equation}
        T_{min}=\argmin\{||\tilde{T}||_2:\tilde{T}\in \overline{G\cdot T} \},
    \end{equation}
    where $m$ is the spatial dimension of the lattice, $\chi$ is the bond dimension, and $\overline{G\cdot T}$ is the closure of the gauge orbit generated by the gauge transform \eqref{eq:MCF_gauge_transform}.

    \textbf{Steps:}

    The object to minimize is:
\begin{equation}
        ||T||_2=\tr \rho =\tr \ket{T}\bra{T}=
        \vcenter{\hbox{\includegraphics[page=45]{TRGCFTGraphics.pdf}}}
    \end{equation}
Its first-order variation with respect to gauge transformation is:
\begin{equation}
        \partial_{t=0}||(e^{t X_1},e^{t X_2})\cdot T||_2^2=2\sum_{k=1}^m \tr X_k (\rho_{k,1}-\rho_{k,2}^T),
        \label{eq:MCF_grad}
    \end{equation}
where the reduced density matrices $\rho_{k,1}$,$\rho_{k,2}$ are:
    \begin{equation}
    \begin{split}
        &\rho_{1,1}=
        \vcenter{\hbox{\includegraphics[page=46]{TRGCFTGraphics.pdf}}},
        \rho_{1,2}^T=
        \vcenter{\hbox{\includegraphics[page=47]{TRGCFTGraphics.pdf}}},\\
        &\rho_{2,1}=
        \vcenter{\hbox{\includegraphics[page=48]{TRGCFTGraphics.pdf}}},
        \rho_{2,2}^T=
        \vcenter{\hbox{\includegraphics[page=49]{TRGCFTGraphics.pdf}}},
    \end{split}
    \end{equation}

    According to \cite{acuaviva2022MCF}, the global minimum can be reached by gradient descent. We apply the following gauge transform by the gradient in \eqref{eq:MCF_grad} and the step length suggested in \cite{acuaviva2022MCF},
\begin{equation}
        \vcenter{\hbox{\includegraphics[page=50]{TRGCFTGraphics.pdf}}}
        =
        \exp\left[{-\frac{1}{4m}\frac{1}{tr \rho} (\rho_{k,1}-\rho_{k,2}^T)}\right]
    \end{equation}
and repeat the whole process until converges.
    
\end{algorithmnofloat}

\subsection{Preliminary procedure of fixing the $U(\chi)$  gauge}
\label{sec:Fix_U1_alg}

 After fixing the $GL(\chi)^{\otimes m}$ gauge using MCF, there is a remaining $U(\chi)^{\otimes m}$ gauge. 
 Ref.~\cite{lyu2021lTRG} states that one only needs to fix the residue $U(1)^{\otimes m \chi}$ subgroup, which reduces to the sign ambiguity at each index $Z_2^{\otimes m \chi}$ for the case of real tensors.
However, we found that, after fixing the signs using an improved method modified from Ref.~\cite{lyu2021lTRG}, one can further minimize the difference of $T$ from the reference tensor $\tilde T$ by applying a standard local minimization procedure for the two gauging matrices $g_1,g_2$ by a singular value decomposing the corresponding environment tensors iteratively. 
This is because the argument in Ref.~\cite{lyu2021lTRG} relies on the assumption that there is no level crossing between eigenvalues when one obtains $T$ through the HOTRG method~\eqref{eq:HOTRG_svd_environment}. The potential level crossing mixes two of these indices and can be fixed by the $U(\chi)^{\otimes m}$ gauge.

Here we show our procedure to fix the $U(\chi)^{\otimes m}$ gauge in Algorithm~\ref{alg:Fix_U1}. 
    
\begin{algorithmnofloat}
    \caption{Procedure of fixing the $U(\chi)^{\otimes m}$  gauge \cite{lyu2021lTRG}}\label{alg:Fix_U1}

    \textbf{Input:} $T_{ijkl}$, the tensor to be gauge-fixed. $\tilde{T}_{ijkl}$, the reference tensor for the gauge choice.
    
    \textbf{Output:} $T_{ijkl}$, the gauge-fixed tensor
    
    \textbf{Steps:}
    
    \begin{algorithmic}
    \For{repeat a few steps}
        \For{$\tilde\chi \gets 1$ to $\chi$ }
            \State $\rho_{1,i} \gets \sum_{j,k,l\leq \tilde\chi} T_{ijkl} \tilde{T}_{ijkl}$
            \State $d_{1,i} \gets \sgn{\rho_{1,i}}$
            \State $T_{ijkl} \gets T_{ijkl} d_{1,i} d_{1,j}$
            \State $\rho_{2,k} \gets \sum_{i,j,l\leq \tilde\chi} T_{ijkl} \tilde{T}_{ijkl}$
            \State $d_{2,k} \gets \sgn{\rho_{2,k}}$
            \State $T_{ijkl} \gets T_{ijkl} d_{2,k} d_{2,l}$
        \EndFor
        \For{repeat a few steps}
            \State loss$\gets \bra{\tilde T}\ket{(g_1, g_2)\cdot T}$
            \State optimize $g_1$ by svd the environment tensor of $g_1$
            \State optimize $g_2$ by svd the environment tensor of $g_2$
            \State $T \gets (g_1, g_2)\cdot T$
        \EndFor
    \EndFor
    \end{algorithmic}
\end{algorithmnofloat}

We note that in Algorithm~\ref{alg:Fix_U1}, the overlap is evaluated as 

\begin{equation}
    \bra{\tilde T}\ket{(g_1, g_2)\cdot T}=\vcenter{\hbox{\includegraphics[page=55]{TRGCFTGraphics.pdf}}},
\end{equation}

where the environment tensor of $\bar{g_1}$ is the tensor subnetwork after removing $g_1$ from the network, and the standard iteratively optimization method for unitary $g_1$ is: $g_1 \gets (UV^\dagger)^*$, with $U$ and $V$  obtained from SVD the environment tensor: $\bar{g_1}=UsV^\dagger$. The tensor $g_2$ can be updated similarly.

\section{Counting the number of conformal states}
\label{sec:Appendix_CFTStates}

This section discusses how to obtain the number of conformal states at each conformal level in Ising CFT $\mathcal{M}(3/4)$. It is a brief summary of some relevant formulas from Chapters 7 and 8 in Ref.~\cite{BigYellowBook}. For comparison, we also list the number of degeneracy states of the first 12 scaling dimensions of Ising CFT in \tabref{tab:CFT_degeneracy}.

\begin{table}[h]
    \centering
    \begin{tabular}{|c|c|c|c|c|}
    \hline
        $\Delta$ & Total & $\mathbb{1}$ & $\sigma$ & $\varepsilon$ \\
    \hline
        0                   & 1  & 1  &   &   \\
        $\frac{1}{8}$       & 1  &   & 1  &   \\
        1                   & 1  & 0  &   & 1  \\
        $1 + \frac{1}{8}$   & 2  &   & 2  &   \\
        2                   & 4  & 2  &   & 2  \\
        $2 + \frac{1}{8}$   & 3  &   & 3  &   \\
        3                   & 5  & 2  &   & 3  \\
        $3 + \frac{1}{8}$   & 6  &   & 6  &   \\
        4                   & 9  & 5  &   & 4  \\
        $4 + \frac{1}{8}$   & 9  &   & 9  &   \\
        5                   & 13 & 6  &   & 7   \\
        $5 + \frac{1}{8}$   & 14 &   & 14  &    \\
    \hline
        \end{tabular}
    \caption{Number of conformal states at each scaling dimension. }
    \label{tab:CFT_degeneracy}
\end{table}

\begin{table}[ht]
    \centering
    \begin{tabular}{|c|c|c|c|}
    \hline
        level & $h_{1,1}=0$ & $h_{2,1}=1/16$ & $h_{1,2}=1/2$ \\
    \hline
        0     & 1   & 1      & 1     \\
        1     & 0   & 1      & 1     \\
        2     & 1   & 1      & 1     \\
        3     & 1   & 2      & 1     \\
        4     & 2   & 2      & 2     \\
        5     & 2   & 3      & 2     \\
        6     & 3   & 4      & 3     \\
        7     & 3   & 5      & 4     \\
        8     & 5   & 6      & 5     \\
        9     & 5   & 8      & 6     \\
        10    & 7   & 10     & 8     \\
    \hline
    \end{tabular}
    \caption{Number of conformal states for different chiral representations of Ising CFT}
    \label{tab:degeneracy_chiral}
\end{table}

In Ising CFT $\mathcal{M}(3/4)$, there are three primary operators (states): $\mathbb{1}$, $\sigma$, $\varepsilon$. Their conformal dimensions are $\Delta_{\mathbb{1}}=0$, $\Delta_{\sigma}=\frac{1}{8}$, $\Delta_{\varepsilon}=1$. They are the tensor product of the left and right chiral representations of the Virasoro algebra:

\begin{align}
\ket{\mathbb{1}}&=\ket{h=0}\otimes\ket{\bar{h}=0},\\
    \ket{\mathbb{\sigma}}&=\ket{h=1/16}\otimes\ket{\bar{h}=1/16},\\
    \ket{\mathbb{\varepsilon}}&=\ket{h=1/2}\otimes\ket{\bar{h}=1/2}.
\end{align}

The corresponding descendant states can be achieved by applying left and right Virasoro operators on those highest-weight states:

\begin{equation}
    L_{-m_1}L_{-m_2}L_{-m_3}...\ket{h} \otimes \bar{L}_{-\bar{m}_1}\bar{L}_{-\bar{m}_2}\bar{L}_{-\bar{m}_3}...\ket{\bar{h}}.
\end{equation}
The conformal dimension of a state is the sum of the left and right conformal spin:
\begin{equation}
    \Delta=h+m_1+m_2+...+\bar{h}+\bar{m}_1+\bar{m}_2+...
\end{equation}
The operators $L_m$,$\bar{L}_m$ satisfies the Virasoro Algebra:
\begin{equation}
\label{eq:Virasoro}
    [L_m,L_n] = (m-n)L_{m+n} + \frac{c}{12}(m-1)m(m+1)\delta_{m+n,0},
\end{equation}
where 
\begin{equation}
    L_{-m}=L_m^\dagger,
\end{equation}
and the primary state $\ket{h}$ satisfies:
\begin{equation}
    L_0\ket{h} = h, \qquad
    L_{m>0}\ket{h} = 0.
\end{equation}

Given $c$ and $h$, the inner product between states of the form $L_{-m_1}...\ket{h}$ is uniquely defined and can be computed by moving all annihilation operators $L_{m>0}$ to the right and creation operators $L_{m<0}$ to the left using the commutator relationship \eqref{eq:Virasoro}. For example, the inner product between $L_{-2}\ket{h}$ and $L_{-2}\ket{h}$ is:
\begin{equation}
    \bra{h}L_2L_{-2}\ket{h}=\bra{h}4L_0+\frac{c}{2}+L_{-2}L_2\ket{h}=4h+\frac{c}{2}.
\end{equation}

Note that different descendent states of the same primary state might be linearly dependent if they are at the same level. It can be seen from the inner product of different states. In particular, define the Kac matrix:
\begin{equation}
    M^{(l)}_{m,n}=\bra{m}\ket{n},
\end{equation}
where $\ket{m}$, $\ket{n}$ are the descendent states of the primary state $\ket{h}$ and have the level $l$.

For example, the explicit expression of $M^{(2)}$ of given $c$ and $h$:
\begin{equation}
    \begin{pmatrix}
        \frac{c}{2} + 4h & 6h \\
        6h & 4h + 8h^2
    \end{pmatrix},
\end{equation}
which represents the inner products between $L_{-2}\ket{h}$ and $L_{-1}L_{-1}\ket{h}$.

The rank of $M^{(l)}$ determines the number of linearly independent descendants of $\ket{h}$ at the level $l$.
To compute the degeneracy, one might write down the Kac matrix manually or use symbolic packages such as ~\cite{virasoroCode}.

Instead, the general expression for degeneracy can also be read from the character of irreducible Verma module $M_{r,s}$ :

\begin{eqnarray}
    \chi_{h}(q) &:=&\tr_{M_{r,s}}(q^{L_0}) \\ 
    &= &q^{h-c/24} \sum_{n=0}^\infty (\text{degeneracy at level n})*q^n.\nonumber
\end{eqnarray}

The expression of $\chi_{h}$ is found as follows~(see 8.17 in~\cite{BigYellowBook}):

\begin{equation}
    \label{eq:verma_character}
    \chi_{h_{r,s}}(q)=K^{p,p'}_{r,s}(q)-K^{p,p'}_{r,-s}(q),
\end{equation}
where $K$ is given as
\begin{equation}
    K^{(p,p')}_{r,s}(q)=\frac{q^{-1/24}}{\varphi(q)}\sum_{n\in \mathbb{Z}}q^{(2pp'n+pr-p's)^2/4pp'},
\end{equation}
and the $\varphi(q)$ is defined via
\begin{eqnarray}
    \frac{1}{\varphi(q)}&=&\frac{1}{\prod_{n\geq 1}(1-q^n)}\\
    &=&1+q+2q^2+3q^3+5q^4+7q^5+11q^6+... ,\nonumber
\end{eqnarray}
where $p,p'$ labels the minimal model $M(p'/p)$. Moreover, $r,s$ labels the highest weight state $\ket{h_{r,s}}$. 
The central charge $c$ is given as follows,
\begin{equation}
    c=1-6\frac{(p-p')^2}{pp'},
\end{equation}
and the scaling dimension is given by
\begin{equation}
     h_{r,s}=\frac{(pr-p's)^2-(p-p')^2}{4pp'}.
\end{equation}

The degeneracy of the Kac matrix for $c=1/2$ at the first few levels is summarized in \tabref{tab:degeneracy_chiral}.
By collecting the tensor product of the left and right states by the total conformal dimension, we reached \tabref{tab:CFT_degeneracy}.

\section{The OPE Coefficient of Ising Model}
\label{sec:Appendix_CFTOPE}

The three-point function in CFT has a general form,
\begin{equation}
    \expval{\mathcal{O}_1\mathcal{O}_2\mathcal{O}_3}=
    \frac{C_{123}}{r_{12}^{\Delta_1+\Delta_2-\Delta_3}r_{23}^{\Delta_2+\Delta_3-\Delta_1}r_{31}^{\Delta_3+\Delta_1-\Delta_2}},
\end{equation}
where $r_{ij}=\left| r_i - r_j \right|$ is the distance between operators and the scaling behavior $a,b,c$ is determined by the scaling dimensions of the operators.

The OPE coefficient $C_{123}$ is the only thing not fixed by conformal symmetry. In 2D Ising CFT $\mathcal{M}(3/4)$, the OPE coefficients are:
\begin{align}
    C_{\sigma\sigma \mathbb{1}}=C_{\varepsilon\varepsilon \mathbb{1}}=1, \
    C_{\sigma\sigma\varepsilon}=\frac{1}{2}, \
    {C_{\varepsilon\sigma\sigma}}=\frac{1}{2}. 
\end{align}

Note that the first line indicates field operators $\sigma$ and $\varepsilon$ need to be properly rescaled such that:
\begin{equation}
\label{eq:CFTOperatorRenormalization}
\begin{aligned}
    \expval{\sigma(r_1)\sigma(r_2)}&=\frac{1}{\left|r_1-r_2\right|^{\frac{1}{4}}},\\
    \expval{\varepsilon(r_1)\varepsilon(r_2)}&=\frac{1}{\left|r_1-r_2\right|^{2}}.
\end{aligned}
\end{equation}
In terms of state-operator correspondence, the OPE coefficient can also be written as:
\begin{equation}
    \label{eq:CFTOPE}
    \ket{\mathcal{O}_2\mathcal{O}_3}=\frac{C_{231}}{r_{23}^{\Delta_2+\Delta_3-\Delta_1}}\ket{\mathcal{O}_1}+...,
\end{equation}
where $\braket{\mathcal{O}_i}{\mathcal{O}_i}$ is normalized to 1, and ... denotes operators other than $\ket{\mathcal{O}_1}$, such as the descendent states of $\mathcal{O}_1$ like $\partial \ket{\mathcal{O}_1}$ and the other operators $\mathcal{O}_j$ which also have a nonvanishing three-point function with $\mathcal{O}_2\mathcal{O}_3$.

\section{Four-point Function of 2D Ising CFT}
\label{sec:Appendix_CFT4PT}

Here we briefly conclude the calculation of the four-point function of Ising CFT, following the lecture notes in Ref.~\cite{simmons2017TASICFBSNotes}.
For simplicity, we consider only the CFT on a plane.

In a CFT, the $N$-point correlation functions can be completely determined from the ``CFT data'', which are the scaling dimensions $\Delta_{\mathcal{O}_i}$ and the OPE coefficients $C_{\mathcal{O}_i\mathcal{O}_j\mathcal{O}_k}$.
In particular, \eqref{eq:CFTOPE} is used to replace the product of two operators into the sum of single operators and iteratively convert the problem to the problem of calculating $(N-1)$-point functions.

By fusing the first and second pair of operators, the four-spin correlation function in 2D Ising CFT can be written as:
\begin{equation}
    \label{eq:4ptFunction}
    \expval{\sigma(x_1)\sigma(x_2)\sigma(x_3)\sigma(x_4)}
    =
    \sum_{\mathcal{O}=\mathbb{1},\varepsilon}
    \frac{C_{\sigma\sigma\mathcal{O}}^2}{x_{12}^{2\Delta_\sigma}x_{34}^{2\Delta_\sigma}}
    g_{\Delta_\mathcal{O},l_\mathcal{O}}(u,v),
\end{equation}
where $\mathcal{O}=\mathbb{1},\varepsilon$ are the two families (primary and its descendants) of operators that appear in the OPE of $\sigma\cross\sigma$, and $C_{\sigma\sigma\mathcal{O}}$ is the corresponding OPE coefficient. 
In the above, $u$ and $v$ are the cross-ratios of the four points $x_1$,$x_2$,$x_3$,$x_4$, which captures the degree of freedom for the points which is invariant under conformal transformation. They can be further summarized by a single complex variable $z$, which is the position of the fourth point when one transforms the first three points to $0$, $1$, and $\infty$:
\begin{align}
    \label{eq:crossRatio}
    u=\frac{x_{12}^2x_{34}^2}{x_{13}^2x_{24}^2}, v=\frac{x_{23}^2x_{14}^2}{x_{13}^2x_{24}^2},\\
    \label{eq:crossRatioFromzzbar}
    u=z\bar z, v=(1-z)(1-\bar z),
\end{align}
where
$g_{\Delta_\mathcal{O},l_\mathcal{O}}(u,v)$ is the conformal block for the operator family $\mathcal{O}$. $\Delta$ and $l$ are the scaling dimension and the spin, respectively. ($l_\mathbb{1}$, $l_\varepsilon=0$). $g_{\Delta_\mathcal{O},l_\mathcal{O}}(u,v)$ captures the correlation function between the primaries and their descendants in the OPE of the first and second pair of operators on the family $\mathcal{O}$. In 2D, the conformal block can be conveniently calculated using the hypergeometric function ${}_2F_1$:
\begin{align}
    \label{eq:conformalBlock2DScalar}
    g_{\Delta,l}(u,v)&=k_{\Delta+l}(z)k_{\Delta-l}(\bar z) + k_{\Delta-l}(z)k_{\Delta+l}(\bar z),\\
    \label{eq:conformalBlockK}
    k_{\beta}(x)&=x^{\beta/2}{}_2F_1(\beta/2,\beta/2,\beta,x).
\end{align}

\section{Scaling Dimension and Central Charge from Transfer Matrix}
\label{sec:Appendix_scdim_TM}

For a cylinder of shape $W \times H$, when conformally mapped to an annulus of inner radius $r_1=1$, the outer radius is:
\begin{equation}
    r_2=e^{\frac{2\pi H}{W}} r_1.
\end{equation}
One can extract the scaling dimension of each eigenvector of the transfer matrix as follows, where $\lambda_i$ are the eigenvalues of the transfer matrix,
\begin{equation}
    \frac{\lambda_i}{\lambda_0}=\left( \frac{r_2}{r_1} \right)^{-\Delta_i}.
\end{equation}


The central charge can be estimated from the largest eigenvalue of the transfer matrix \cite{gu2009TEFR_TMscD}:
\begin{equation}
    c=12 \frac{\log{\lambda_0}}{\log{r_2/r_1}},
\end{equation}

where the tensor is required to be renormalized such that $\tr \vcenter{\hbox{\includegraphics[page=69,scale=0.5]{TRGCFTGraphics.pdf}}} = \tr \vcenter{\hbox{\includegraphics[page=53,scale=0.5]{TRGCFTGraphics.pdf}}}$ \cite{gu2009TEFR_TMscD} and $\tr T =\vcenter{\hbox{\includegraphics[page=57,scale=0.5]{TRGCFTGraphics.pdf}}}$ indicate contracting the opposite pairs of legs of the tensor. When GILT is not applied, using trace to renormalize the tensor yields better results than using the tensor norm $\norm{\vcenter{\hbox{\includegraphics[page=69,scale=0.5]{TRGCFTGraphics.pdf}}}}=\norm{\vcenter{\hbox{\includegraphics[page=53,scale=0.5]{TRGCFTGraphics.pdf}}}}$.

\section{Fixing the degeneracy of conformal states}
\label{sec:fix-rotation}

From linearized TRG $M$ we obtain its 'raw' eigenvectors and eigenvalues:
\begin{equation}
    M \ket{u_a}= \lambda_a \ket{u_a}.
\end{equation}
Ideally, they should correspond to the conformal states:
\begin{equation}
    \ket{\mathcal{O}_\alpha} \in \{ \ket{\mathbb{1}}, \ket{\sigma}, \ket{\varepsilon}, \ket{\partial\sigma}, ...\}
\end{equation}
However, only from conformal dimensions one cannot different states which share the same conformal dimension, such as $\partial_x \sigma$ and $\partial_y \sigma$.
Moreover, in practice, the truncation error from finite bond dimensions, the anisotropy introduced by our HOTRG scheme, and the truncation from GILT introduced various numerical errors. We have observed that higher-level states with different conformal dimensions will also mix with each other. Furthermore, if the lTRG matrix is not strictly hermitian, the eigenvectors $\ket{u_\alpha}$ might not be orthogonal to each other.

To calibrate the above issues, we first orthonormalize the eigenvectors using the standard Gram-Schmidt process, where the eigenvectors with lower conformal dimensions have higher priorities.
The next step is to rotate the basis $\ket{u_a}$ to build the conformal states $\ket{\mathcal{O}_\alpha}$:
\begin{equation}
    \ket{\mathcal{O}_\alpha}=\sum_{\lambda_a=\Delta}S_{\alpha a}\ket{u_a}.
\end{equation}

We use the scheme described in \tabref{tab:OperatorInsertion} to build the implementation of conformal operators at the lattice level using the finite difference method and operator fusion. Note that $\ket{T_\alpha}$'s are the corresponding coarse-grained tensors and contain higher states in the conformal tower of $\mathcal{O}_\alpha$. For example:
\begin{equation}
    \label{eq:finiteDifferenceError}
    \begin{aligned}
        \ket{T_{\partial_x \sigma(x_0)}}&\sim\frac{1}{2\delta}(\sigma(x_0+\delta)-\sigma(x_0-\delta))\ket{0} \\
        &\sim\ket{\partial_x\sigma(x_0)} +\frac{1}{6}\ket{\partial^3_x\sigma(x_0)} +...
    \end{aligned}
\end{equation}

The Gram matrix $S'$ is the dot product between $\ket{u_a}$ and $\ket{T_\alpha}$:
\begin{equation}
    S'_{a\alpha}=\bra{u_a}\ket{T_\alpha}.
\end{equation}
We expect the block-diagonal part of $S'$ will cancel the rotation matrix, and the off-block-diagonal part is the higher terms that correspond to finite-difference error like \eqref{eq:finiteDifferenceError}:
\begin{equation}
    S^{(\Delta)}=S'^{(\Delta),-1},
\end{equation}
where $(\Delta)$ means restricted on eigenspace $\lambda_\alpha,\lambda_a\approx\Delta$.

However, in practice, some states with different scaling dimensions, such as $\Delta=2$ and $\Delta=2.125$, are mixed with each other, as well as with eigenvectors of higher conformal dimensions. It is shown in \figref{fig:FiniteDifference}. An easy fix is to use a larger block that contains both $\Delta=2$ and $\Delta=2.125$. One might also include more eigenvectors and replace the submatrix inverse with the submatrix pseudoinverse. We find that we can assume all the eigenvectors are somehow mixed and do the pseudo-inverse for the whole matrix regardless of the theoretical degeneracy, which will give the best result.
So, rather than ``fixing the degeneracy of the eigenvectors of lTRG using lattice operators'', what we actually did is more like projecting the lattice operators onto the low energy (conformal dimension) subspace of lTRG.

\section{Renormalization of fixed-point tensors and extraction of OPE coefficient}
\label{sec:Appendix_RenormalizeTensorStates}

To extract the CFT data from the fusion rules of coarse-grained tensors, it is tempting to identify the coarse-grained tensors with the ket vectors in \eqref{eq:CFTOPE}. From it, one may extract the OPE coefficient by comparing the coarse-grained tensor of two defect tensors with the third one. However, proper renormalization is required to get the correct result. It is due to the following reasons. 

1. We want the coarse-graining operation, which maps the states on a square of length $2 a_0$ to the states on the square of length $a_0$, to have the same behavior as the scaling operator in CFT, when acting on defects, as illustrated in \eqref{eq:CoarseGrainingScalingVacuum} and \eqref{eq:CoarseGrainingScaling}. It requires the vacuum tensor $T_\mathbb{1}$ to be properly normalized.

2. The size of the coarse-grained block $a_0$ is related to the lattice detail that we should be agnostic of when only the fixed-point tensor is provided. One needs to reassign $a_0$ with a finite number (e.g. $a_0=1$) and renormalize the spin operators according to \eqref{eq:CFTOperatorRenormalization}.

3. The inner product, compatible with the states in \eqref{eq:CFTOPE}, is defined using radial quantization, which is different from the inner product defined on contracting the corresponding external legs of two tensor networks. A simple fix is to rescale the bra tensors.

4. The spatial slice in our radial quantization is a square of length $2 a_0$. The coarse-graining process introduces an additional scale transformation to map it to a square of length $a_0$. We should scale the bra vector accordingly to compensate for the scaling.

For convenience, we use $\mathcal{M}\ket{\begin{array}{cc}T_a & T_b	\\T_c & T_d\end{array}}$ to denote the coarse-grained tensor of the four tensors listed inside the ket, where $\mathcal{M}$ is the coarse-grain operation that maps the tensor state from the larger square to the states at the original square. We also note that $\bra{T}$ has a different normalization factor than $\ket{T}$. The inner product $\bra{T_a}\ket{T_b}$ is the naive contraction of the corresponding legs of the two tensors (with complex conjugation on the $T_a$ tensor). Now, we list the whole procedure as follows.

1. Rescale $\ket{T_\mathbb{1}}$ such that 
\begin{equation}
    \label{eq:CoarseGrainingScalingVacuum}
    \mathcal{M}\ket{\begin{array}{cc}
    T_\mathbb{1} & T_\mathbb{1}	\\
    T_\mathbb{1} & T_\mathbb{1}
    \end{array}}=\ket{T_\mathbb{1}}.
\end{equation}

2. The scaling dimensions satisfies:
\begin{equation}
\begin{aligned}
    \label{eq:CoarseGrainingScaling}
    \mathcal{M}\ket{\begin{array}{cc}
    T_\sigma & T_\mathbb{1}	\\
    T_\mathbb{1} & T_\mathbb{1}
    \end{array}}=\frac{1}{2^{\Delta_\sigma}}\ket{T_\sigma}+...\\
    \mathcal{M}\ket{\begin{array}{cc}
    T_\varepsilon & T_\mathbb{1}	\\
    T_\mathbb{1} & T_\mathbb{1}
    \end{array}}=\frac{1}{2^{\Delta_\varepsilon}}\ket{T_\varepsilon}+...
\end{aligned}
\end{equation}

3. Choose the ``renormalized'' length scale $a_0$ as an arbitrary number. Here, we choose $a_0=1$.

4. Rescale (renormalize) $\ket{T_\sigma}$ such that 
\begin{equation}
    \mathcal{M}\ket{\begin{array}{cc}
    T_\sigma & T_\sigma	\\
    T_\mathbb{1} & T_\mathbb{1}
    \end{array}}=\frac{C_{\sigma\sigma\mathbb{1}}}{a_0^{2\Delta_\sigma}}\ket{T_\mathbb{1}}+...
\end{equation}
or equivalently,
\begin{equation}
    \mathcal{M}\ket{\begin{array}{cc}
    T_\sigma & T_\mathbb{1}	\\
    T_\mathbb{1} & T_\sigma
    \end{array}}=\frac{C_{\sigma\sigma\mathbb{1}}}{(\sqrt{2}a_0)^{2\Delta_\sigma}}\ket{T_\mathbb{1}}+...
\end{equation}

5. Repeat the same process to rescale $\ket{T_\varepsilon}$.

6. Rescale $\bra{T_\mathbb{1}}$, $\bra{T_\sigma}$, and $\bra{T_\varepsilon}$ such that 
\begin{equation}
    \bra{T_\mathbb{1}}\mathcal{M}\ket{\begin{array}{cc}
    T_\mathbb{1} & T_\mathbb{1}	\\
    T_\mathbb{1} & T_\mathbb{1}
    \end{array}}=1,
\end{equation}    
\begin{equation}
    \bra{T_\sigma}\mathcal{M}\ket{\begin{array}{cc}
    T_\sigma & T_\mathbb{1}	\\
    T_\mathbb{1} & T_\mathbb{1}
    \end{array}}=1,
\end{equation}  
\begin{equation}
    \bra{T_\varepsilon}\mathcal{M}\ket{\begin{array}{cc}
    T_\varepsilon & T_\mathbb{1}	\\
    T_\mathbb{1} & T_\mathbb{1}
    \end{array}}=1.
\end{equation}

7. Now one can read off the OPE coefficient:
\begin{equation}
    \bra{T_\varepsilon}\mathcal{M}\ket{\begin{array}{cc}
    T_\sigma & T_\sigma	\\
    T_\mathbb{1} & T_\mathbb{1}
    \end{array}}=\frac{C_{\sigma\sigma\varepsilon}}{a_0^{2\Delta_\sigma-\Delta_\varepsilon}},
\end{equation}
or equivalently,
\begin{equation}
    \bra{T_\varepsilon}\mathcal{M}\ket{\begin{array}{cc}
    T_\sigma & T_\mathbb{1}	\\
    T_\mathbb{1} & T_\sigma
    \end{array}}=\frac{C_{\sigma\sigma\varepsilon}}{(\sqrt{2}a_0)^{2\Delta_\sigma-\Delta_\varepsilon}},
\end{equation}
and similarly for $C_{\sigma\varepsilon\sigma}$.

\section{Determining Operator Families from the fixed-point tensor of an unknown CFT}

In the previous discussion, we already know the list of primary Ising CFT operators and their spectrum. When given an arbitrary fixed-point tensor, since one does not know the list of conformal families, it is unknown which of the eigenvectors of lTRG belong to which conformal family.

To determine which of the eigenvectors of lTRG corresponds to primaries, we can examine the two-point function. Only primaries and descendants within the same conformal family have a non-vanishing two-point function. However, for certain symmetric conditions, the correlation between the primaries and descendants may also be zero, so it is necessary to break the symmetry when placing the two points.

By examining the two-point function, we can classify the eigenvectors into different conformal families. The primary operator is the one within the family that has the smallest scaling dimension. If the primary has a symmetry structure, such as spins or internal symmetry, there may be degeneracy.
We also need to pay special attention to the descendants of the identity operator. The first two descendants are the stress energy tensors, which are easy to identify because they have a conformal dimension of $\Delta=2$.

\section{Huber Loss}
\label{sec:Appendix_HuberLoss}

\begin{figure}[h]
    \centering
    \includegraphics[width=.9\hsize]{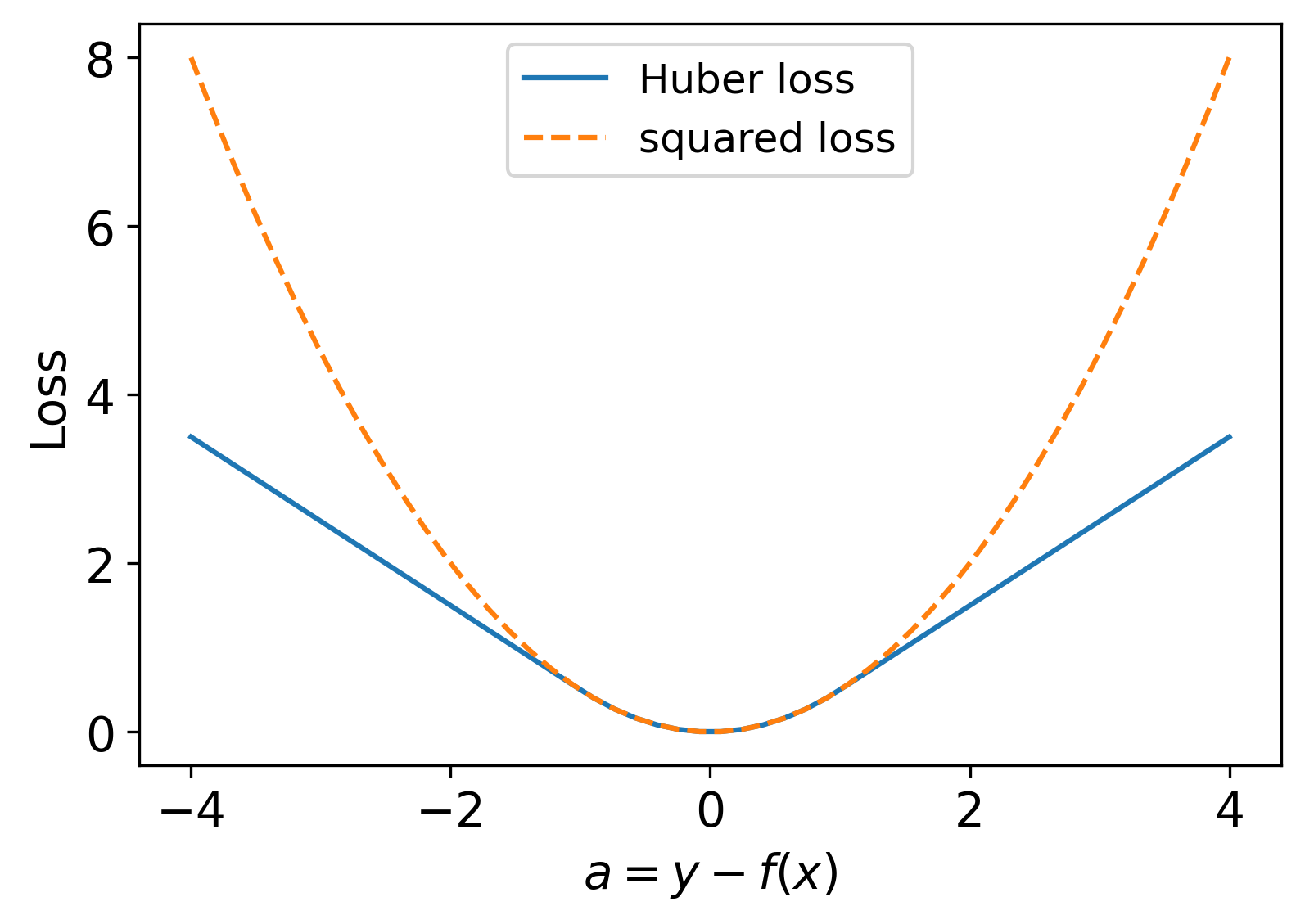}
    \caption{(color online) Huber loss ($\epsilon=1$) and squared error loss.}
    \label{fig:HuberLoss}
\end{figure}

When outliers are present in the data, they can introduce a significant error in the fitting, even in case the number of outliers takes only a small portion of the data. It is because in traditional least squares fitting $L=(y-f(x))^2$, the impact of a data point on the fitted function is proportional to its residual $y-f(x)$:

\begin{equation}
    \frac{\partial L}{\partial f(x)} = y-f(x)
\end{equation}

So, the outliers with a greater residual take a more significant weight in determining the fitting function.

The Huber loss\cite{huber1964HuberLoss} $L_\epsilon$ is supposed to suppress the excessive impact of outliers. Which is defined as:

\begin{equation}
    L_\epsilon(y,f(x))=\begin{cases}
        \frac{1}{2}(y-f(x))^2 & \abs{y-f(x)}\leq \epsilon \\
        \delta (\abs{y-f(x)}-\frac{1}{2}\epsilon, & \text{otherwise.}
    \end{cases}
    \label{eq:HuberLoss}
\end{equation}

 Where $x$, $y$ are the data to fit and $f$ is the fitted function. $\epsilon$ is an empirical parameter that determines the range in which a data point is considered an outlier. 
 
Huber loss results in a bounded impact of outliers on the fitting:

\begin{equation}
    \frac{\partial L_\epsilon}{\partial f(x)} =\begin{cases}
        y-f(x) & \abs{y-f(x)}\leq \epsilon \\
        \pm \epsilon, & \text{otherwise.}
    \end{cases}
\end{equation}

\section{Numerically finding the critical temperature $T_c(\chi)$}
\label{sec:Find_Tc}

Here we demonstrate how to find the optimal input inverse temperature $\beta_c$ which corresponds to the most stable HOTRG flow using the bisection algorithm. This method is modified from the code provided by \cite{lyu2021lTRG}.

\begin{algorithmnofloat}
    \caption{Finding the critical (inverse) temperature $\beta_c(\chi)$ using bisection method}\label{alg:Find_Tc}

    \textbf{Input:} The bisection search range $\beta_{c,\text{min}}$, $\beta_{c,\text{max}}$, bisection search cutoff $\epsilon$, the initial tensor $T^{(0)}(\beta)$, and the coarse-graining scheme $CG^{(l_M)}_\chi$ which returns properly normalized coarse-grained tensors after $1 ... l_M$ steps, truncated with maximal bond dimension $\chi$: $T^{(1)}, ... ,T^{(l_m)}=CG^{(l_M)}_\chi(T^{(0)})$.
    
    \textbf{Output:} $\beta_c(\chi)$, the ``critical (inverse) temperature'' which corresponds to the most stable RG flow in the specified numerical scheme and bond dimension $\chi$.
    
    \textbf{Steps:}
    
    \begin{algorithmic}
    \For{$\beta_{\text{max}}-\beta_{\text{min}}>\epsilon$}
        \vspace{.5em}
        \State $T^{(1)}_{\text{min}}, ... ,T^{(l_M)}_{\text{min}} \gets CG^{(l_M)}_\chi(T^{(0)}(\beta_{\text{min}}))$
        \vspace{.5em}
        \State $T^{(1)}_{\text{max}}, ... ,T^{(l_M)}_{\text{max}} \gets CG^{(l_M)}_\chi(T^{(0)}(\beta_{\text{max}}))$
        \vspace{.5em}
        \State $T^{(1)}_{\text{med}}, ... ,T^{(l_M)}_{\text{med}} \gets CG^{(l_M)}_\chi(T^{(0)}(\frac{\beta_{\text{min}}+\beta_{\text{max}}}{2}))$
        \vspace{.5em}
        \If {$\text{dist}({T^{(l_M)}_{\text{med}},T^{(l_M)}_{\text{min}}})<\text{dist}({T^{(l_M)}_{\text{med}},T^{(l_M)}_{\text{max}}})$}
            \vspace{.5em}
            \State $\beta_{\text{min}} \gets \frac{\beta_{\text{min}}+\beta_{\text{max}}}{2}$
        \Else
            \vspace{.5em}
            \State $\beta_{\text{max}} \gets \frac{\beta_{\text{min}}+\beta_{\text{max}}}{2}$
        \EndIf
        
    \EndFor
    \end{algorithmic}
\end{algorithmnofloat}

In the above procedure, $\text{dist}({T^{(i)}_{\text{1}},T^{(i)}_{\text{2}}})$ characterizes how the two systems are different from each other. One can also use other quantities (derived from the tensors), such as the magnetization $\expval{\sigma}$ to characterize the distance: $\text{dist}({T^{(i)}_{\text{1}},T^{(i)}_{\text{2}}})=| |\expval{\sigma}^{(i)}_{\text{1}}|-|\expval{\sigma}^{(i)}_{\text{2}}| |.$ We found that, in practice, using the tensors themselves may not be stable compared to using physical observables.



\section{Effects of maximal bond dimension $\chi$}
\label{sec:Discussion_Bond_dim}

\begin{figure*}[hbtp]
    \centering
    \includegraphics[width=.6\textwidth]{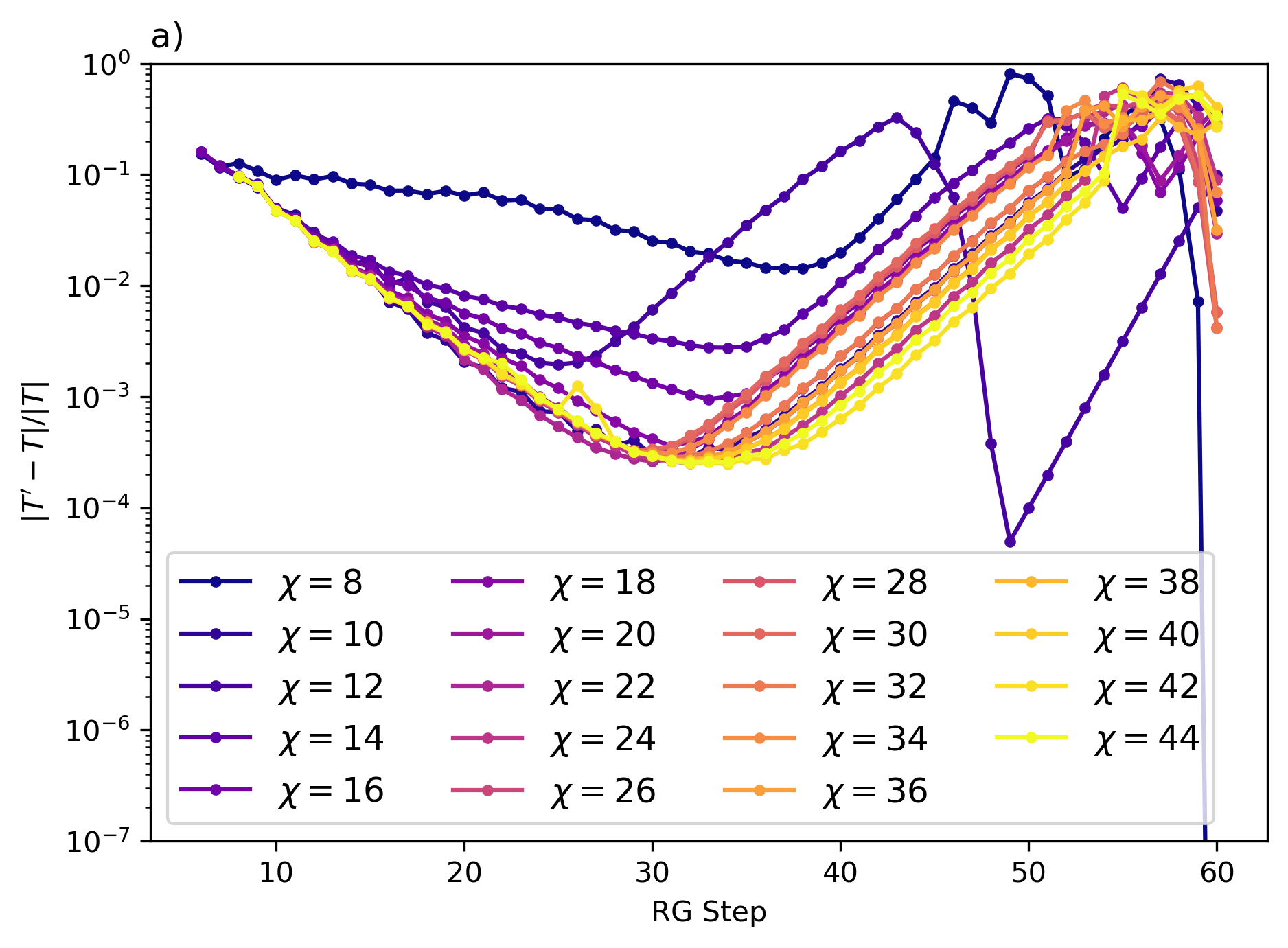}
    \includegraphics[width=.45\textwidth]{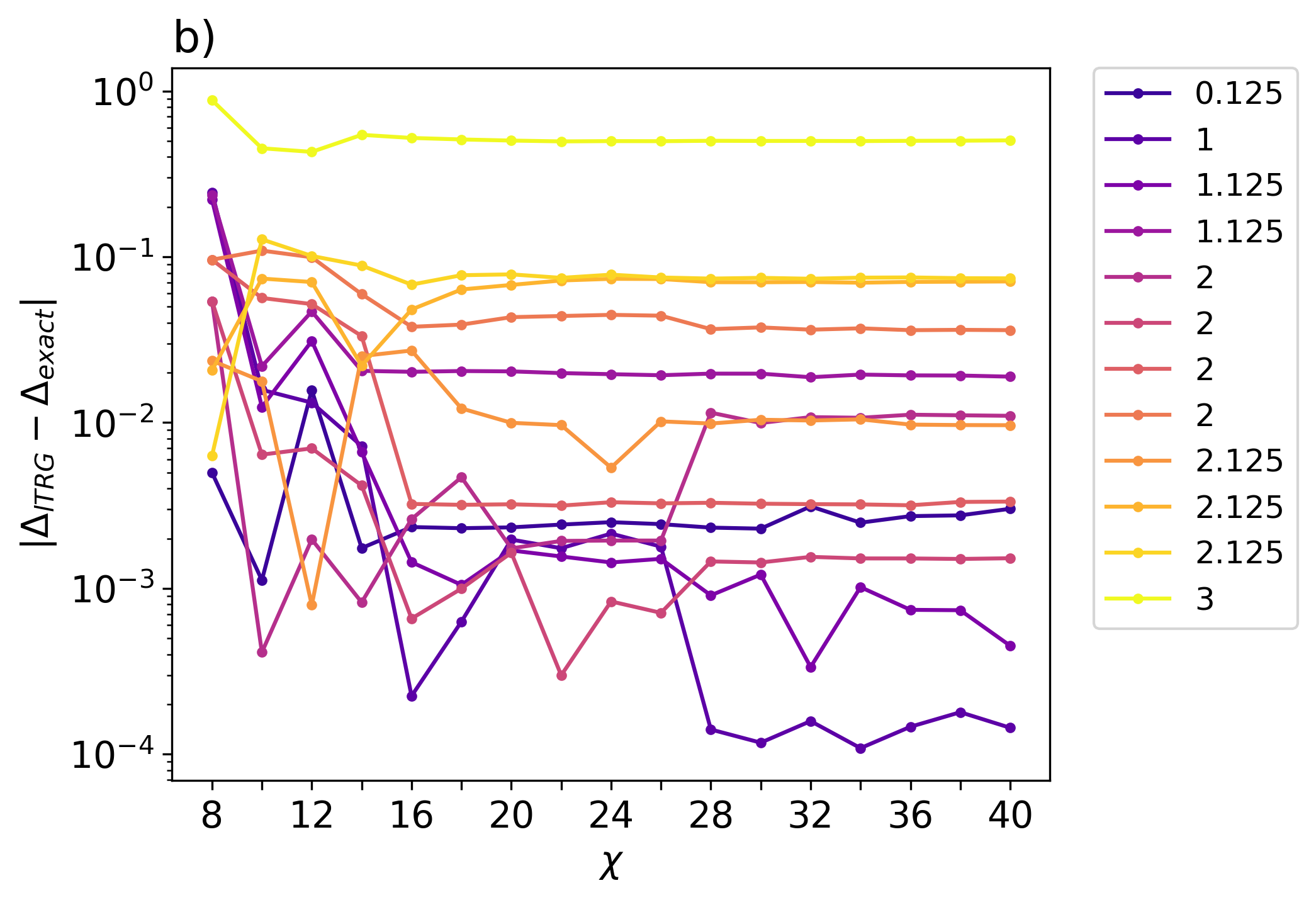}
    \includegraphics[width=.45\textwidth]{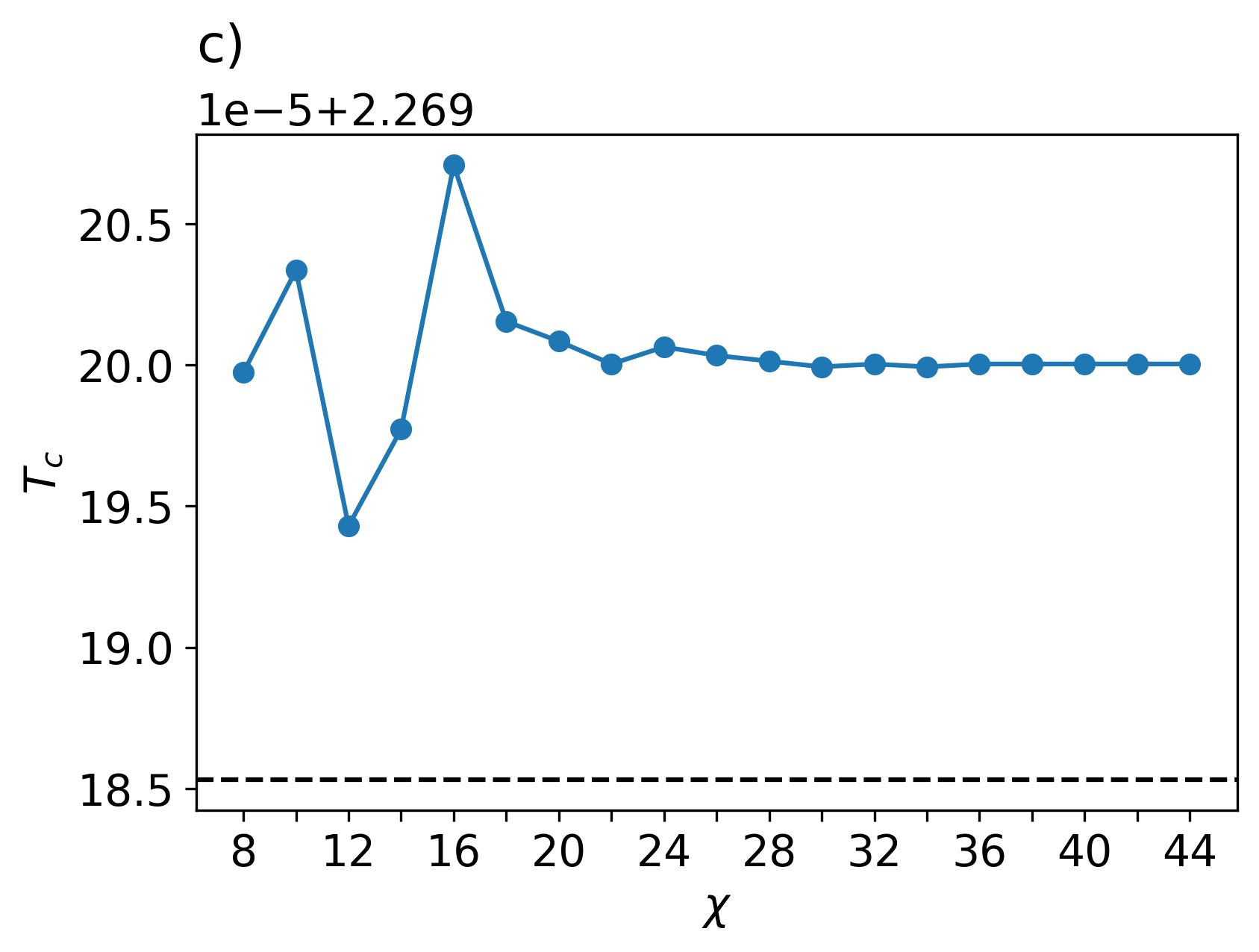}
    \caption{(color online) a) Per-component difference of coarse-grained tensors between every 2 RG steps, b) Error of scaling dimensions extracted through lTRG method at step 30, and c) Critical temperature estimated by inspecting the RG flow~\cite{lyu2021lTRG}, at different maximal bond dimensions $\chi$.}
    \label{fig:TdiffChi}
\end{figure*}

Here we demonstrate the effect of the maximal bond dimension on numerical errors. Regarding the stability of RG flow, the precision of scaling dimensions of higher-order operators, and the critical temperature, there is no significant improvement at $\chi>24$, as shown in \figref{fig:TdiffChi}. The only improvement from increasing $\chi$ from 24 to 40 is the scaling dimension of the first few operators. It indicates that the error of higher-order operators might be from the other part of the algorithm which we could not identify.
Thus, for convenience, we have used $\chi=24$ for the main part of our paper.

\end{document}